\documentclass[11pt]{article}
\usepackage{latexsym,amssymb,amstext,amsmath}
\usepackage{slashed}
\usepackage{mathrsfs}
\usepackage{hyperref}

\renewcommand{\baselinestretch}{1.2}
\allowdisplaybreaks
\def\slash#1{\rlap{\hbox{$\mskip 1 mu /$}}#1}      
\def\Slash#1{\rlap{\hbox{$\mskip 3 mu /$}}#1}      
\newcommand{\ft}[2]{{\textstyle\frac{#1}{#2}}}

\newcommand {\cD}{{\cal D}}
\newcommand {\cE}{{\cal E}}

\newcommand {\cH}{{\cal H}}

\newcommand {\cJ}{{\cal J}}

\newcommand {\cL}{{\cal L}}

\newcommand {\cN}{{\cal N}}

\newcommand {\cR}{{\cal R}}

\newcommand {\cV}{{\cal V}}
\newcommand {\cW}{{\cal W}}
\newcommand {\cX}{{\cal X}}

\def\q{\theta}

\def\L{\Lambda}

\def\S{\Sigma}

\def\ri{{\rm i}}

\newcommand{\dalpha}{{\dot{\alpha}}}
\newcommand{\dbeta}{{\dot{\beta}}}
\newcommand{\dgamma}{{\dot{\gamma}}}
\newcommand{\ddelta}{{\dot{\delta}}}

\newcommand{\dphi}{{\dot{\phi}}}

\newcommand{\1}{{\underline{1}}}
\newcommand{\2}{{\underline{2}}}

\newcommand{\pa}{\partial}

\newcommand{\be}{\begin{equation}}
\newcommand{\ee}{\end{equation}}
\newcommand{\bea}{\begin{eqnarray}}
\newcommand{\eea}{\end{eqnarray}}

\newcommand{\ba}{\begin{array}}
\newcommand{\ea}{\end{array}}

\def\double #1{#1{\hbox{\kern-2pt $#1$}}}

\newcommand{\bsubeq}{\begin{subequations}}
\newcommand{\esubeq}{\end{subequations}}

\newcommand{\ul}{\underline}

\newcommand{\rd}{\mathrm d}
\newcommand{\veps}{\varepsilon}

\newcommand{\bphi}{{\bar\phi}}
\newcommand{\bpsi}{{\bar\psi}}

\newcommand{\eps}{{\epsilon}}

\newcommand{\bsigma}{\bar{\sigma}}

\newcommand{\eol}{\nonumber \\}

\begin{document}
%
\begin{titlepage}
\begin{flushright} \small
  Nikhef-2013-024 \\  ITP-UU-13/17  
\end{flushright}
\bigskip

\begin{center}
  {\LARGE\bfseries New higher-derivative invariants in N=2
    supergravity\\[.2ex]
    and the Gauss-Bonnet term}
  \\[10mm]
\textbf{Daniel Butter$^{a}$, Bernard de Wit$^{a,b}$, Sergei M.
  Kuzenko$^{c}$ and Ivano Lodato$^{a}$}\\[5mm]
\vskip 6mm
$^a${\em Nikhef, Science Park 105, 1098 XG Amsterdam, The
  Netherlands}\\
$^b${\em Institute for Theoretical Physics, Utrecht
  University,} \\
  {\em Leuvenlaan 4, 3584 CE Utrecht, The Netherlands}\\
$^c${\em  School of Physics M012, The University of Western
  Australia, }\\
{\em 35 Stirling Highway, Crawley W.A. 6009, Australia}\\[3mm]
{\tt dbutter@nikhef.nl}\,,\;  {\tt B.deWit@uu.nl}\,,\; {\tt
  sergei.kuzenko@uwa.edu.au}\,,\;  {\tt ilodato@nikhef.nl} 
\end{center}

\vspace{3ex}

\begin{center}
{\bfseries Abstract}
\end{center}
\begin{quotation} \noindent A new class of $N\!=\!2$ locally
  supersymmetric higher-derivative invariants is constructed based on
  logarithms of conformal primary chiral superfields. They
  characteristically involve a coupling to $\mathcal{R}_{\mu\nu}{}^2
  -\tfrac13 \,\mathcal{R}^2$, which equals the non-conformal part of
  the Gauss-Bonnet term. Upon combining one such invariant with the
  known supersymmetric version of the square of the Weyl tensor one
  obtains the supersymmetric extension of the Gauss-Bonnet term. The
  construction is carried out in the context of both conformal
  superspace and the superconformal multiplet calculus. The new class
  of supersymmetric invariants resolves two open questions. The first
  concerns the proper identification of the $4D$ supersymmetric
  invariants that arise from dimensional reduction of the $5D$ mixed
  gauge-gravitational Chern-Simons term.  The second is why the pure
  Gauss-Bonnet term without supersymmetric completion has reproduced
  the correct result in calculations of the BPS black hole entropy in
  certain models.
\end{quotation}

\vfill

\flushleft{\today}
\end{titlepage}

\section{Introduction}
\label{sec:introduction}
\setcounter{equation}{0}
More detailed knowledge of supersymmetric higher-derivative terms is
becoming increasingly relevant. Although a substantial body of
research in supersymmetric field theories, supergravity and string
theory is based on supersymmetric invariants that are at most
quadratic in space-time derivatives, there are many questions that
require knowledge of supersymmetric invariants beyond the
two-derivative level. Originally the central question concerned the
issue of possible supersymmetric counterterms in the hope of
establishing the ultraviolet finiteness of certain supersymmetric
gauge and supergravity theories. Hence candidate counterterms were
studied whenever possible, motivated by the assumption that
supersymmetry must be the crucial element responsible for the
finiteness. However, there are also instances where one is actually interested in
finite effects corresponding to higher-derivative invariants, such as
encountered when determining subleading corrections to black hole
entropy.

This paper is directed to an extension of certain classes of
higher-derivative invariants in $N=2$ supergravity. From the technical
point of view, such a study is facilitated by the fact that there
exist formulations of $N=2$ supergravity where supersymmetry is
realized off-shell, i.e. without involving the equations of motion
associated with specific Lagrangians. In that case there exist
well-established methods such as superspace and component calculus
that enable a systematic study. There exists a healthy variety of
approaches: in this paper we will make use of conformal superspace
\cite{Butter:2011sr} which is closely related to the superconformal
multiplet calculus \cite{deWit:1984pk,deWit:1984px} that is carried
out in component form.\footnote{Other off-shell methods
include the $N=2$ harmonic \cite{GIOS} and projective \cite{KLRT,KLRT-M2}
superspace approaches, which make it possible to realize the most general
off-shell supergravity-matter couplings.}
We will be using these methods in parallel. For
higher-extended supersymmetry the application of methods such as these
becomes problematic for the simple reason that off-shellness is not
realized, up to a few notable exceptions such as the Weyl
multiplet in $N=4$ supergravity.

Some higher-derivative invariants in $N=2$ supersymmetry and
supergravity have been known for some time, such as those
involving functions of the field strengths for supersymmetric gauge theories
\cite{Henningson:1995eh,deWit:1996kc,Buchbinder:1999jn,Banin:2002mf,Argyres:2003tg},
the chiral invariant containing the square of the Weyl tensor
(possibly coupled to matter chiral multiplets)
\cite{Bergshoeff:1980is} and invariants for tensor multiplets
\cite{deWit:2006gn}. A full superspace integral has also been used
to generate an $\cR^4$ term in the context of ``minimal'' Poincar\'e
supergravity \cite{Moura:2002ip}.
More recently, a large class of
higher-derivative supersymmetric invariants was constructed using the
superconformal multiplet calculus, corresponding to integrals
over the full $N=2$ superspace \cite{deWit:2010za}.\footnote{The action considered
in \cite{Moura:2002ip} can be interpreted within the conformal framework of
\cite{deWit:2010za} as the full superspace integral of
$\cH = (T_{ab\, ij})^2 (T^{cd\, kl})^2 / (X_0 \bar X_0)^2$
where $X_0$ is a compensating vector multiplet, in the presence
of an additional non-linear multiplet.}
This action involved arbitrary chiral multiplets, which could play the role
of composite fields consisting of homogeneous functions of
vector multiplets. This entire class had the remarkable property that
the corresponding invariants and their first derivatives (with respect
to the fields or to coupling constants) vanish in a fully
supersymmetric background. This result ensures that these invariants
do not contribute to either the entropy or the electric charge of BPS
black holes. Actions of this class have also been used recently to
study supergravity counterterms and the relation between off-shell and
on-shell results \cite{Chemissany:2012pf}. Furthermore, in
\cite{Butter:2010jm}, higher-derivative actions were constructed in
projective superspace by allowing vector multiplets and/or tensor
multiplets to be contained in similar homogeneous functions
of other multiplets. Because the invariants derived in
\cite{deWit:2006gn,deWit:2010za,Butter:2010jm} can involve several
independent homogeneous functions at the same time, they cannot be
classified concisely, although this forms no obstacle when considering
applications.

Nevertheless, these broad classes do not exhaust the possibilities for
higher-derivative invariants. A previously unknown $4D$ higher-derivative
term was identified recently in \cite{Banerjee:2011ts} when applying
off-shell dimensional reduction to the $5D$ mixed gauge-gravitational
Chern-Simons term \cite{Hanaki:2006pj}. It turned out to involve a
Ricci-squared term $\cR^{ab} \cR_{ab}$ multiplied by the ratio
of vector multiplets. This curvature combination does not appear in
the previous known invariants and is suggestive of
the Gauss-Bonnet term, whose $N=2$ extension has, remarkably,
never been constructed before.

A related issue, also involving the Gauss-Bonnet term, arose
several years ago in a different context:
the calculation of black hole entropy from higher-derivative couplings in an
effective supergravity action. It was observed in a certain model \cite{Sen:2005iz}
that one could calculate the entropy of a BPS black hole by considering
the effective action involving the product of a dilaton field with
the Gauss-Bonnet term \emph{without} supersymmetrization.
This result agreed with the original calculation based on
the square of the Weyl tensor,
which depended critically on its full
supersymmetrization \cite{LopesCardoso:1998wt,LopesCardoso:2000qm},
but it remained unclear why the non-supersymmetric approach
of \cite{Sen:2005iz} would yield the same answer and whether
the outcome was indicative of some deeper result.

Both of these issues would be resolved by a full knowledge of the
$N=2$ Gauss-Bonnet invariant and the broader class of higher-derivative
supersymmetric invariants to which it belongs.
The goal of this paper is to present this class and 
to discuss whether it shares the same properties with the previously
explored classes of invariants.

Let us first briefly recall some features of the Gauss-Bonnet
invariant as well as other invariants quadratic in the Riemann tensor.
In this introductory text we restrict ourselves to
bosonic fields; the supersymmetric extension will be discussed in the
subsequent sections.  In four space-time dimensions there are two
terms quadratic in the Riemann tensor whose space-time integral
defines topological invariants: these are the Pontryagin density,
\begin{align}
  \label{eq:Pontryagin}
  \mathcal{L}_\mathrm{P} = \tfrac12
  \varepsilon^{\mu\nu\rho\sigma}\,\mathcal{R}_{\mu\nu}{}^{\lambda\tau}\,
  \mathcal{R}_{\rho\sigma\lambda\tau} \,,
\end{align}
and the Euler density,
\begin{align}
  \label{eq:Euler}
  e^{-1} \mathcal{L}_\chi = \tfrac{1}{4}\,\varepsilon^{\mu\nu\rho\sigma}\,
  \mathcal{R}_{\mu\nu}{}^{\lambda\tau}\, 
  \mathcal{R}_{\rho\sigma}{}^{\delta\epsilon}\,\varepsilon_{\lambda\tau\delta\epsilon} =
  \mathcal{R}^{\mu\nu\rho\sigma} \mathcal{R}_{\mu\nu\rho\sigma} - 4 
  \mathcal{R}^{\mu\nu} \mathcal{R}_{\mu\nu} + \mathcal{R}^2~.
\end{align}
The integral of the Euler density is the Gauss-Bonnet invariant. 
Their difference can be made more apparent by trading the Riemann
tensor for the Weyl tensor, $C_{\mu\nu}{}^{\rho\sigma} =
\mathcal{R}_{\mu\nu}{}^{\rho\sigma} - 2\, \delta_{[\mu}{\!}^{[\rho}
\,\mathcal{R}_{\nu]}{}^{\sigma]} + \tfrac13
\delta_{\mu}{}^{[\rho} \delta_{\nu}{}^{\sigma]}\,\mathcal{R}$,
\begin{align}
  \label{eq:Pontryagin-GB-2}
  \mathcal{L}_\mathrm{P} = \tfrac12 \varepsilon^{\mu\nu\rho\sigma}\, C_{\mu\nu}{}^{\lambda \tau}\,
  C_{\rho\sigma \lambda \tau} , \qquad e^{-1} \mathcal{L}_\chi =
  C^{\mu\nu\rho\sigma} \,C_{\mu\nu\rho\sigma} - 2\, 
  \mathcal{R}^{\mu\nu} \,\mathcal{R}_{\mu\nu} + \tfrac{2}{3}
  \mathcal{R}^2 ~.
\end{align}
From the perspective of supersymmetry \eqref{eq:Pontryagin-GB-2} is
not a good basis for discussing supersymmetric extensions. Rather, it
turns out that the following combinations are more natural,
\begin{align} 
  \label{eq:susy-R-squared}
  e^{-1} \mathcal{L}_\mathrm{W} {}^\pm = &\,
  \tfrac{1}{2} C_{\mu\nu}{}^{ab}\,C^{\mu\nu cd}
  \big[\eta_{ac}\eta_{bd}
  \pm \tfrac12 \varepsilon_{abcd}\big] =  C_{\mu
    \nu}{}^{ab\pm} C^{\mu\nu}{}^\pm_{ab} \,, \nonumber\\ 
  e^{-1} \mathcal{L}_\mathrm{NL} =&\, -\mathcal{R}^{\mu\nu}
  \,\mathcal{R}_{\mu\nu} + \tfrac{1}{3} \mathcal{R}^2 \,.
\end{align}
The first expression is the square of the anti-selfdual (selfdual)
Weyl tensor, which belongs to a chiral (anti-chiral) multiplet, and
whose superextension has been known for a long time
\cite{Bergshoeff:1980is}. The supersymmetric extension of the second
term will be one of the results of this paper.

In the expressions \eqref{eq:susy-R-squared} we made use of tangent-space indices,
$a,b,\ldots$, because the metric formulation is not suitable for
supersymmetric theories, which necessarily contain fermions and
therefore require vierbein fields $e_\mu{}^a$. 
In this paper, we will employ a superconformal description
in which the tangent space will be
subject to Lorentz transformations (M), dilatations (D), and conformal
boosts (K). This implies that we will be dealing with three tangent
space connections, namely the spin connection $\omega_\mu{}^{ab}$,
the dilatation connection $b_\mu$, and the connection associated with
conformal boosts $f_\mu{}^a$. The connections $\omega_\mu{}^{ab}$ and
$f_\mu{}^a$ will turn out to be composite, as we will explain
momentarily. As our goal will be to construct the supersymmetric
extension of the second invariant in \eqref{eq:susy-R-squared},
we must first discuss how this invariant can arise in the
framework of conformal gravity.

Under dilations and conformal boosts, the vierbein fields and the
various connections transform as follows,
\begin{alignat}{2}
  \label{eq:D-K-transform}
   \delta e_\mu{}^a &= -\Lambda_\mathrm{D}\, e_\mu{}^a\,, &\qquad
   \delta \omega_\mu{}^{ab} &= 2\, \Lambda_\mathrm{K} {}^{[a}
   \,e_\mu{}^{b]} \,, \nonumber\\ 
   \delta b_\mu &= \partial_\mu\Lambda_\mathrm{D} + \Lambda_\mathrm{K}{}^a\,e_{\mu a}\,,  &\qquad
   \delta f_\mu{}^a &=\, \mathcal{D}_\mu\Lambda_\mathrm{K}{}^a +
   \Lambda_\mathrm{D} \,f_\mu{}^a \,,
\end{alignat}
where we use Lorentz and dilatationally covariant derivatives
$\mathcal{D}_\mu$, such as in 
\begin{equation}
  \label{eq:cov-cal-der}
  \mathcal{D}_\mu\Lambda_\mathrm{K}{}^a =
  (\partial_\mu-b_\mu)\Lambda_\mathrm{K}{}^a -\omega_\mu{}^{ab}\,
  \Lambda_{\mathrm{K} b} \,. 
\end{equation}
The corresponding curvatures take the following form,
\begin{align}
  \label{eq:PMDK-curv}
  R(P)_{\mu\nu}{}^a =&\, 2\,\mathcal{D}_{[\mu} e_{\nu]}{}^a  \,, \nonumber\\
  R(M)_{\mu\nu}{}^{ab} =&\, 2\,\partial_{[\mu}  \omega_{\nu]}{}^{ab}
  -2\, \omega_{[\mu}{}^{ac} \,\omega_{\nu] c}{}^b -4\, f_{[\mu}{}^{[a}\,
  e_{\nu]}{}^{b]} \,, \nonumber\\
  R(D)_{\mu\nu}   =&\, 2\,\partial_{[\mu}  b_{\nu]} -2\, f_{[\mu}{}^a\,
  e_{\nu] a} \,, \nonumber\\
  R(K)_{\mu\nu}{}^a =&\, 2\,\mathcal{D}_{[\mu}  f_{\nu]}{}^a\,. 
\end{align}
In terms of these curvatures one imposes the following conventional
constraints,
\begin{equation}
  \label{eq:constraints}
  R(P)_{\mu\nu}{}^a =0\,,\qquad R(M)_{\mu\nu}{}^{ab} \, e_b{}^\nu
  =0\,. 
\end{equation}
Because the constraints~\eqref{eq:constraints} are invariant under
Lorentz transformations, dilatations and conformal boosts, the
transformation rules \eqref{eq:D-K-transform} remain unaffected. For
the supersymmetric extension this will no longer be the case and
additional terms will emerge.  The Bianchi identities together
with the constraints \eqref{eq:constraints} imply the following
relations,\footnote{Here and below we take $\mathcal{D}_\mu$ and $D_\mu$
to contain also the affine connection
$\Gamma_{\mu\nu}{}^\rho= e_a{}^\rho\,\mathcal{D}_\mu e_\nu{}^a$
when acting on quantities with world indices. $D_\mu$
is the conformally covariant derivative and contains the connection
$f_\mu{}^a$ in addition to the spin and dilatation connections. In later sections,
we will use the same symbol for the supercovariant derivative.}
\begin{align}
  \label{eq:bianchi}
  R(D)_{\mu\nu}= 0\,,\qquad 
   R(M)_{[\mu\nu}{}^{ab} \,e_{\rho]b}= 0 \,,\qquad
  R(K)_{\mu\nu}{}^{a}  = D_b R(M)_{\mu\nu}{}^{ba}\,. 
\end{align}
The constraints \eqref{eq:constraints} express the spin connection
field $\omega_\mu{}^{ab}$ and the $\mathrm{K}$-connection field
$f_\mu{}^a$ in terms of $e_\mu{}^a$ and $b_\mu$. The resulting
expression for $f_\mu{}^a$ reads as follows,
\begin{align}
  \label{eq:f-field}
  f_\mu{}^a= \tfrac12 \mathcal{R}(e,b) _\mu{}^a-\tfrac1{12}
  e_\mu{}^a\,\mathcal{R}(e,b) \,, \qquad
  f_\mu{}^\mu \equiv f= \tfrac1{6}\, \mathcal{R}(e,b)\,, 
\end{align}
where $\mathcal{R}(e,b)_{\mu\nu}{}^{ab}$ denotes the curvature
associated with the spin connection,
\begin{equation}
  \label{eq:spin-conn-curv}
  \mathcal{R}(e,b)_{\mu\nu}{}^{ab} = 2\,\partial_{[\mu}  \omega(e,b)_{\nu]}{}^{ab}
  -2\, \omega(e,b)_{[\mu}{}^{ac} \,\omega(e,b)_{\nu] c}{}^b \,. 
\end{equation}
Note that it is possible to impose the gauge $b_\mu=0$, so that only
the vierbein remains as an independent field. The spin connection is
then the standard torsion-free connection, the curvature
$\mathcal{R}(e,b)_{\mu\nu}{}^{ab}$ corresponds to the standard Riemann
tensor with a symmetric Ricci tensor, while $f_\mu{}^a e_{\nu a}$ is
symmetric. However, it is advantageous to not impose such a gauge at
this stage. The curvature \eqref{eq:spin-conn-curv} satisfies the
Bianchi identity $\mathcal{D}_{[\mu}
\mathcal{R}(e,b)_{\nu\rho]}{}^{ab} = 0$.  From this identity it follows
that $\mathcal{D}_a\big[ 2\, \mathcal{R}(e,b)_\mu{}^a - e_\mu{}^a
\,\mathcal{R}(e,b) \big]=0$, where the Ricci tensor is not
symmetric. This equation is equivalent to
\begin{equation}
  \label{eq:bianchi-f}
  \mathcal{D}_a \big[ f_\mu{}^a -f\,e_\mu{}^a\big] = 0\,. 
\end{equation}

To exhibit some salient features of the above formalism and to give an
early demonstration of the strategy we intend to follow in this paper,
let us consider a scalar field $\phi$ transforming under dilatations
as
\begin{equation}
  \label{eq:scalar-phi-w}
  \delta_{\rm D} \phi = w\,\Lambda_\mathrm{D} \phi\,,
\end{equation}
where the constant $w$ is known as the Weyl weight. We stress that
$\phi$ does not have to be an elementary field; it could also be a
composite field, as long as it transforms in the prescribed way under
dilatations. It is now straightforward (but more and more tedious) to
determine explicit expressions for multiple conformally covariant derivatives of
$\phi$ and their transformation behaviour under
$\mathrm{K}$-transformations (c.f. appendix B of
\cite{deWit:2010za}),
\begin{align}
  \label{eq:mutiple-der-phi}
  D_\mu\phi =&\,  \mathcal{D}_\mu \phi = \pa_\mu \phi - w b_\mu \phi \,, \nonumber\\
  D_\mu D_a\phi =&\, \mathcal{D}_\mu D_a\phi +w\,f_{\mu  a} \,\phi
  \,, \nonumber\\
    D_\mu\Box_\mathrm{c}\,\phi =&\,  \mathcal{D}_\mu\Box_\mathrm{c} \phi
  +2(w-1)f_\mu{}^a D_a \phi\,,  \nonumber\\ 
  \Box_\mathrm{c} \Box_\mathrm{c}\, \phi =&\,  \mathcal{D}_a D^a
  \Box_\mathrm{c} \phi +(w+2)f \,\Box_\mathrm{c}  \phi + 2 (w-1)
  f_{\mu a}\, D^\mu D^a\phi \,,   
\end{align}
whose variations under $\mathrm{K}$-transformations read,
\begin{align}
  \label{eq:K-var-multiple-der-phi}
  \delta_\mathrm{K} D_a \phi= &\, -w\,\Lambda_{\mathrm{K}a}\,
  \phi\,,  \nonumber \\
  \delta_\mathrm{K} D_\mu D_{a} \phi=&\, -(w+1)
  \big[\Lambda_{\mathrm{K}\mu}\, D_a+ \Lambda_{\mathrm{K}a} \, D_\mu\big]
  \phi + e_{\mu a} \Lambda_\mathrm{K}{}^b \,D_b \phi
  \,,\nonumber \\
  \delta_\mathrm{K} \Box_\mathrm{c} \phi=&\,
  - 2(w-1)\Lambda_\mathrm{K}{}^a \, D_a\phi\,,  \nonumber\\
   \delta_\mathrm{K} D_\mu \Box_\mathrm{c} \phi=&\,
  -(w+2) \Lambda_{\mathrm{K}\mu}\,\Box_\mathrm{c} \phi 
  -2(w-1)\Lambda_\mathrm{K}{}^a \, D_\mu D_a\phi\,,  \nonumber\\
    \delta_\mathrm{K} \Box_\mathrm{c} \Box_\mathrm{c} \phi=&\, -2(w-1)
  \Lambda_{\mathrm{K}}{}^a \,\Box_\mathrm{c}\, D_a\phi - 2(w+1)
  \Lambda_\mathrm{K}{}^a D_a \Box_\mathrm{c} \phi \,.
\end{align}
It turns out that, for specific Weyl weights, $\Box_\mathrm{c}\phi$ and
$\Box_\mathrm{c} \Box_\mathrm{c}\phi$ are $\mathrm{K}$-invariant,
\begin{alignat}{2}
  \label{eq:Box-Box-squared}
  \delta_\mathrm{K} \Box_\mathrm{c} \phi &=\, 0\,, &\quad
  &(\mathrm{for}\;w=1)\,, \nonumber \\ 
  \delta_\mathrm{K} \Box_\mathrm{c} \Box_\mathrm{c} \phi&=\,
  2\,\Lambda_\mathrm{K}{}^a \big( \Box_\mathrm{c} \,D_a -D_a
  \Box_\mathrm{c}\big) \phi = 0\,,  &\quad &(\mathrm{for}\; w=0)\,,
\end{alignat}
where, to prove the last part of the second equation, we rewrote $
\Box_\mathrm{c} \,D_a \phi -D_a \Box_\mathrm{c} \phi =
D^b\big[D_b,D_a\big] \phi +\big[D_b,D_a\big] D^b\phi$ and made use of
the Ricci identity and the curvature constraints.  From
\eqref{eq:Box-Box-squared} one derives two conformally invariant
Lagrangians by multiplying with a similar scalar field $\phi^\prime$
of the same Weyl weight as $\phi$,
\begin{alignat}{2}
  \label{eq:lagrangians-1-2}
  e^{-1} \mathcal{L} \propto \phi^\prime \,\Box_\mathrm{c}\phi
  &= -\mathcal{D}^\mu\phi^\prime  \,\mathcal{D}_\mu\phi +
  f\,\phi^\prime \phi \,, &\quad
  &(\mathrm{for}\;w=1) \nonumber\\[.2ex]
  e^{-1} \mathcal{L} \propto \phi^\prime \,\Box_\mathrm{c}
  \Box_\mathrm{c} \phi &=\,
  \,\mathcal{D}^2\phi^\prime\,\mathcal{D}^2\phi  
  +2\, \mathcal{D}^\mu\phi^\prime \big[2\, f_{(\mu}{}^a e_{\nu)a}
    -  f \,g_{\mu\nu}\big] \mathcal{D}^\nu\phi \,,
	&\quad &(\mathrm{for}\;w=0)
\end{alignat}
up to total derivatives. Note that we have made use here of
\eqref{eq:bianchi-f}. Both the above expressions are symmetric in
$\phi$ and $\phi^\prime$. 

Let us comment on the two Lagrangians~\eqref{eq:lagrangians-1-2}. In
both Lagrangians the dependence on $b_\mu$ will cancel as a result of
the invariance under conformal boosts. In the first Lagrangian one may
then adjust the product $\phi^\prime \phi$ to a constant by means of a
local dilatation. In that case the second term of the Lagrangian is
just proportional to the Ricci scalar, so that one obtains the
Einstein-Hilbert term. The kinetic term for the scalars depends on the
choice made for $\phi^\prime$ and $\phi$. For instance, when the two
fields are the same, then $\phi$ equals a constant; when they are not
the same (elementary or composite) fields, the kinetic term can be
exclusively written in terms of $\phi$ and will be proportional to
$\phi^{-2}\big(\partial_\mu\phi\big)^2$. In that case the first
Lagrangian describes an elementary or a composite scalar field coupled
to Einstein gravity.

The situation regarding the second Lagrangian is fundamentally
different, because one cannot adjust the scalar fields to any
particular value by local dilatations in view of the vanishing Weyl
weight. The scalar fields may be equal to constants (in which case the
Lagrangian vanishes) or to homogeneous functions of other fields such
that the combined Weyl weight remains zero, without affecting the
invariance under local dilatations.  We should also mention that the
operator $\Box_{\rm c} \Box_{\rm c}$ appearing in this Lagrangian,
when acting on a scalar field with $w=0$, is the same
operator $\Delta_0$ given in \cite{FT1982}
and has an interesting history
in its own right.\footnote{
This operator was discovered by Fradkin and Tseytlin in 1981 \cite{FT1982}
and re-discovered by Paneitz in 1983 \cite{Paneitz}.
In the mathematics literature, it is known as the Paneitz operator. 
 The same operator along with the second Lagrangian in
  \eqref{eq:lagrangians-1-2} was used by Riegert \cite{Riegert:1984kt} for the
  purpose of integrating the conformal anomaly. There is a unique
  generalization to higher dimensions, see e.g. \cite{Gover:2002ay}
  and references therein. } 

It is, of course, possible to construct invariants which also involve
the Weyl tensor. For instance, any scalar field of zero Weyl weight
times the square of the Weyl tensor will define a conformally
invariant Lagrangian. But how to include invariants such as the
four-dimensional Gauss-Bonnet term is less obvious.  As it turns out,
the crucial assumption made in the examples above is that the scalar
fields transform linearly under dilatations. To demonstrate how the
situation changes when this is not the case, let us repeat the
previous construction for $\ln\phi$, which transforms inhomogeneously
under dilatations, $\delta_\mathrm{D}\ln\phi=
w\,\Lambda_\mathrm{D}$. In the same way as above, we derive the
following definitions,
\begin{align}
  \label{eq:DD-ln-phi}
  D_\mu\ln\phi=&\, \mathcal{D}_\mu \ln\phi = \partial_\mu\ln\phi -w\,b_\mu\,,\nonumber\\
  D_\mu D_a\ln\phi=&\,\mathcal{D}_\mu D_a\ln\phi + w\, f_{\mu a}\,,
  \nonumber\\
    D_\mu\Box_\mathrm{c}\,\ln \phi =&\,  \mathcal{D}_\mu\Box_\mathrm{c} \ln\phi
  -2\,f_\mu{}^a D_a \ln\phi\,,  \nonumber\\ 
  \Box_\mathrm{c} \Box_\mathrm{c}\, \ln\phi =&\,  \mathcal{D}_a D^a 
  \Box_\mathrm{c} \ln \phi +2f\,\Box_\mathrm{c}  \ln\phi
  -2\,  f_{\mu a}\, D^\mu D^a\ln\phi \,.
\end{align}
The equations above show an interesting systematics, namely that,
after applying a certain number of covariant derivatives on $\ln\phi$,
these expressions take the same form as in
\eqref{eq:K-var-multiple-der-phi} with $w=0$. However, it is important
to realize that the details implicit in the multiple covariant
derivatives will still depend on the characteristic features
associated with the logarithm. The same observation can be made for
the $\mathrm{K}$-transformations of multiple derivatives which also
transform as if one were dealing with a $w=0$ scalar field,
\begin{align}
  \label{eq:K-var-multiple-der-ln-phi}
  \delta_\mathrm{K}D_a\ln\phi =&\, -w \,\Lambda_{\mathrm{K}a}\,,
  \qquad \delta_\mathrm{D} D_a\ln\phi = \Lambda_\mathrm{D}
  \,D_a\ln\phi\,, \nonumber\\
  \delta_\mathrm{K} D_\mu D_{a} \ln\phi=&\, -
  \big[\Lambda_{\mathrm{K}\mu}\, D_a+ \Lambda_{\mathrm{K}a} \,
  D_\mu\big] \ln \phi + e_{\mu a} \Lambda_\mathrm{K}{}^b \,D_b \ln
  \phi
  \,,\nonumber \\
  \delta_\mathrm{K} \Box_\mathrm{c} \ln\phi=&\,
  2\,\Lambda_\mathrm{K}{}^a \, D_a\ln\phi\,,  \nonumber\\
  \delta_\mathrm{K} D_\mu \Box_\mathrm{c} \ln\phi=&\, - 2\,
  \Lambda_{\mathrm{K}\mu}\,\Box_\mathrm{c} \ln\phi +
  2\,\Lambda_\mathrm{K}{}^a \, D_\mu D_a\ln\phi\,,  \nonumber\\
  \delta_\mathrm{K} \Box_\mathrm{c} \Box_\mathrm{c} \ln\phi=&\, 2
  \Lambda_{\mathrm{K}}{}^a \,\Box_\mathrm{c}\, D_a\ln\phi -2\,
  \Lambda_\mathrm{K}{}^a D_a \Box_\mathrm{c} \ln\phi = 0\,.
\end{align}

In four space-time dimensions the only conformally invariant
Lagrangian based on the above expression must be equal to
$\Box_\mathrm{c} \Box_\mathrm{c} \ln\phi$, possibly multiplied with a
scalar field of zero Weyl weight. This constitutes the non-linear
version of the second Lagrangian in \eqref{eq:lagrangians-1-2}, namely
$\sqrt{g}\, \phi^\prime \Box_\mathrm{c} \Box_\mathrm{c} \ln\phi$, where
$\phi$ has a non-vanishing, but arbitrary Weyl weight $w$ and
$\phi^\prime$ has zero Weyl weight. Taking the explicit form of $\Box_\mathrm{c}
\Box_\mathrm{c} \ln\phi$ this Lagrangian is given by
\begin{align}
  \label{eq:box-box-log}
\sqrt g\, \phi' \Box_\mathrm{c} \Box_\mathrm{c} \ln\phi =&\,
  \sqrt g\, \phi' \Big\{\big(\mathcal{D}^2\big){}^2\ln\phi - 2\, \mathcal{D}^\mu \big[\big(2\,
  f_{(\mu}{}^a e_{\nu)a} - f\,g_{\mu\nu}\big)
  \mathcal{D}^\nu\ln\phi\big]
	\nonumber \\ & \qquad\qquad
  +  w \big[\mathcal{D}^2 f + 2\, f^2 -2\, (f_\mu{}^a)^2 \big] \Big\}.
\end{align}
There are two features to note about this Lagrangian
The first is that its dependence on $\ln\phi$ is isolated in the first line on
the right-hand side, which is a total derivative when $\phi'$ is constant.
In other words, the action is independent of the choice of $\ln\phi$ when
$\phi'$ is constant.
The second feature is that the Lagrangian is K-invariant, so all the
$b_\mu$ terms must drop out. Equivalently, one can adopt a K-gauge
where $b_\mu=0$. Using \eqref{eq:f-field}, one finds
\begin{equation}
    \label{eq:w-terms-in-R}
    \mathcal{D}^2 f + 2\, f^2 -2\, (f_\mu{}^a)^2 = 
     \tfrac16\mathcal{D}^2 \mathcal{R}  
    -\tfrac12  \mathcal{R}^{ab} \,\mathcal{R}_{ab} +\tfrac16 \mathcal{R}^2~,
\end{equation}
which is proportional to $\mathcal{L}_\mathrm{NL}$
(c.f. \ref{eq:susy-R-squared}) up to a total covariant derivative.
When combined with the square of the Weyl tensor with an appropriate
relative normalization one obtains the Gauss-Bonnet invariant up to a
total covariant derivative
\begin{align}
  \label{eq:Lchi'a}
  e^{-1}\mathcal{L}_\chi =&\, C^{abcd} C_{abcd} +
  4\,w^{-1} \,\Box_{\rm c} \Box_{\rm c} \ln\phi \nonumber \\
	=&\, C^{abcd} C_{abcd}
	- 2\,\mathcal{R}^{ab}\mathcal{R}_{ab}
	+ \tfrac{2}{3} \mathcal{R}^2
	+ \tfrac{2}{3} \mathcal{D}^2 \mathcal{R}
	\nonumber \\ 
        &\, + 4 w^{-1} \Big\{\big(\mathcal{D}^2\big){}^2 \ln \phi
	+  \cD^a \Big(
		\tfrac{2}{3} \mathcal{R} \,\mathcal{D}_a \ln\phi - 2\,
                \mathcal{R}_{ab} \,\mathcal{D}^b \ln\phi\big)\Big\} ~,
\end{align}
where we have taken the gauge $b_\mu=0$ in the second equality.
Discarding the (explicit) total derivatives, this result reduces to the
Euler density. Alternatively the dilatation gauge $\phi=1$
reduces it to
\begin{align}
  \label{eq:Lchi'b}
        e^{-1} \mathcal{L}_\chi =&\, C^{abcd} C_{abcd} - 2\,
  \mathcal{R}^{ab}\mathcal{R}_{ab} + \tfrac{2}{3} \mathcal{R}^2 +
  \tfrac{2}{3} \cD^2 \mathcal{R}\,.
\end{align}
This differs from the usual Euler density \eqref{eq:Euler} by an explicit total
derivative. Obviously additional invariants are obtained by
multiplying this result with a $w=0$ independent (composite or
elementary) scalar field $\phi'$.

The above relatively simple bosonic Lagrangians indicate how
higher-derivative couplings will be characterized in this paper. As we
shall argue in the next section, all these Lagrangians have an $N=2$
supersymmetric counterpart based on {\it chiral superfields}. These include
the well-known Lagrangians quadratic in derivatives, the class of
higher-derivative Lagrangians discussed in \cite{deWit:2010za}, and a
new class of Lagrangians based on $\sqrt{g}\,\phi^\prime
\,\Box_\mathrm{c}\Box_\mathrm{c} \ln\phi$, where $\phi^\prime$ and
$\phi$ are the lowest components of
chiral multiplets with $w^\prime=0$ and $w\not=0$. This
last class must contain the $N=2$ supersymmetric higher-derivative
invariant that was found upon reducing the $5D$ higher-derivative
invariant coupling to four dimensions \cite{Banerjee:2011ts}. The main
purpose of this paper is to study this new class of invariants.

This paper is organized as follows. In section
\ref{sec:ext-chiral-super} we explain how to extend the present
results to $N=2$ supersymmetry by assigning the various fields to
chiral multiplets.  This discussion will be at the level of flat
superspace. We introduce the so-called kinetic multiplet, which
supersymmetrizes $\Box_{\rm c} \Box_{\rm c} \phi$, and its non-linear
version, corresponding to $\Box_{\rm c} \Box_{\rm c} \ln\phi$. In the
subsequent section \ref{sec:curved-kinetic} we extend these results to
curved superspace. Then, in section
\ref{sec:component-kinetic-multiplet}, we exhibit the component
structure of the kinetic multiplet, both in the linear and in the
non-linear case. Explicit results are given for a new class of
higher-derivative supersymmetric invariants based on the
supersymmetrization of $\Box_{\rm c} \Box_{\rm c} \ln\phi$. The result
here is the direct extension of the result presented in
\cite{deWit:2010za} and it can be used for similar purposes. One
application that is typical for this class concerns the supersymmetric
Gauss-Bonnet term. Therefore section \ref{sec:GaussBonnet} deals with
a number of characteristic features of this term. Conclusions and
implications of our results are discussed in section
\ref{sec:summary-conclusions}.  A number of appendices has been
included with additional material.

\section{The extension to chiral superfields in flat N=2 superspace}
\label{sec:ext-chiral-super}
\setcounter{equation}{0}
In the introduction we noted the existence of four different types of
conformally invariant Lagrangians and we pointed out that those can
rather easily be embedded into $N=2$ supersymmetric invariants on the
basis of chiral superfields. Just as conformal transformations are an
invariance in flat space-time, defined by a constant vierbein and
vanishing connections $\omega_\mu{}^{ab}$, $b_\mu$, $f_\mu{}^a$,
superconformal transformations leave a flat {\it superspace}
invariant. Furthermore, almost every statement we will make about flat
superspace can transparently be lifted to curved superspace although
the required calculations are considerably more involved. Therefore we
will first discuss flat superspace in this section. Since chiral
multiplets are intrinsically complex, the superfields and
corresponding invariants involving them are complex as well. We
subsequently describe the systematics of these superfields, discuss
the notion of an $N=2$ superconformal \emph{kinetic multiplet}, and
present the four types of invariants. In the next section
\ref{sec:curved-kinetic} we will extend this analysis to curved
superspace.

Superfields can be defined as functions of the flat superspace
coordinates $z^A= (x^a, \theta^{\alpha i}, \bar\theta_{\dot\alpha
  i})$. Here  our notation will reflect the fact that in flat superspace world
and tangent-space indices can be identified. The tangent space
derivatives are
\begin{align}
  \label{eq:D-derivaties}
  \partial_a = \frac{\pa}{\pa x^a}~,\qquad D_{\alpha i} =
  \frac{\partial}{\partial \theta^{\alpha i}} + 
  \mathrm{i} (\sigma^a)_{\alpha \dot\alpha} \,\bar
  \theta^{\dot\alpha}{}_i \frac{\partial}{\partial x^a}~, \quad \bar
  D^{\dot\alpha i}= \frac{\partial}{\partial \bar\theta_{\dot\alpha
      i}} + \mathrm{i} (\bar\sigma^a)^{\dot\alpha \alpha} \,\theta_\alpha{}^i
  \frac{\partial}{\partial x^a}~.
\end{align}
In the context of curved superspace, we will be employing a vector
tangent-space derivative $\nabla_a$ and spinor tangent-space
derivatives $\nabla_{\alpha i}$ and $\bar\nabla^{\dot \alpha i}$ which
are the direct extension of the derivatives in
\eqref{eq:D-derivaties}. We remind the reader that we use
two-component spinor notation in the context of superspace where
spinor indices are raised and lowered with the antisymmetric epsilon
tensor (see Appendix \ref{App:NC}).

Chiral superfields satisfy the differential superspace constraint
$\bar D^{\dalpha i} \Phi = 0$.  We will
denote the components of a general chiral multiplet $\Phi$ by
\cite{deRoo:1980mm,deWit:1980tn},
\begin{alignat}{3}
  \label{eq:flatPhi_components}
  A &:= \Phi \vert_{\theta=0}~, \qquad&
  \Psi_{\alpha i} &:= D_{\alpha i} \Phi\vert_{\theta=0}~, \qquad&
  B_{ij} &:= -\tfrac{1}{2} D_{ij} \Phi\vert_{\theta=0}~, \nonumber \\
  F_{ab}^- &:= -\tfrac{1}{4} (\sigma_{ab})_\alpha{}^\beta
  D_\beta{}^\alpha \Phi\vert_{\theta=0} ~,\qquad& 
  \Lambda_{\alpha i} &:= \tfrac{1}{6} \varepsilon^{jk} D_{\alpha k}
  D_{ji} \Phi\vert_{\theta=0}~, \qquad& 
  C &:= -2 D^4 \Phi \vert_{\theta=0}~, 
\end{alignat}
where 
\begin{equation}
  \label{eq:double-D}
  D_{ij} := - D_{\alpha (i}  D^\alpha{}_{j)}  \,,\qquad 
    D_{\alpha\beta} := -\varepsilon^{ij} D_{(\alpha i} D_{\beta)j}\,.
\end{equation}
Hence a chiral multiplet comprises a $16+16$ bosonic and fermionic
components, consisting of a complex scalar $A$, a chiral
spinor doublet $\Psi_i$, a complex symmetric scalar $B_{ij}$, an
anti-selfdual tensor $F_{ab}^-$, a chiral spinor doublet $\Lambda_i$,
and a complex scalar $C$. 

Under dilatations and chiral $\mathrm{U}(1)$ transformations (with
constant parameters $\Lambda_\mathrm{D}$ and $\Lambda_\mathrm{A}$ in
flat superspace) the superspace coordinates change according to
\begin{equation}
  \label{eq:dil-chiral-coord}
  x^\prime = \exp\big[-\Lambda_\mathrm{D}\big]\,x\,, \qquad
  \theta^\prime =  \exp\big[-\tfrac12(\Lambda_\mathrm{D} +
  \mathrm{i}\Lambda_\mathrm{A} ) \big]\,\theta\,, \qquad
  \bar\theta^\prime =  \exp\big[-\tfrac12(\Lambda_\mathrm{D} -
  \mathrm{i}\Lambda_\mathrm{A} ) \big]\,\bar\theta\,, 
\end{equation}
and superfields $\Psi(x,\theta,\bar\theta)$ are usually assigned to
transform as
\begin{equation}
  \label{eq:dilatation-chiral}
  \Psi^\prime(x^\prime,\theta^\prime,\bar\theta^\prime) = \exp\big[
  w\,\Lambda_\mathrm{D} +\mathrm{i}   c\,\Lambda_\mathrm{A}\big]\,
  \Psi(x,\theta,\bar\theta) \,,
\end{equation}
where $w$ and $c$ are called the Weyl and the chiral weight. For chiral
multiplets these weights are related by $c=-w$. In that case the Weyl
weight of $A$ equals $w$ and the highest-$\theta$ component $C$ has
weight $w+2$. All the components scale homogeneously and since there
are no chiral superfield components with Weyl weight less than $w$ it
follows that $A$ must be invariant under S-supersymmetry. This implies
that it is also invariant under $\mathrm{K}$ transformations. Such a
chiral superfield is called a {\it conformal primary} field. All
these properties can be derived systematically on the basis of the
superconformal algebra using the chiral constraint.

Just as in $N=1$ superspace one can integrate the product
$\Phi^\prime\,\bar\Phi$ of a chiral and an anti-chiral superfield,
respectively, to obtain an expression involving {\it four} space-time
derivatives (discarding total derivatives in the equalities),
\begin{align}
  \label{eq:phi-phi-bar}
  \int \mathrm{d}^4\theta\,\mathrm{d}^4\bar\theta\,
  \Phi' \,\bar\Phi = \int \mathrm{d}^4\theta\, \Phi' \,\big(\bar D^4
  \bar\Phi\big) = A' \,\Box \Box \bar A + \cdots~,
\end{align}
where $\bar D^4= \tfrac1{48}\varepsilon_{ik} \varepsilon_{jl}\, \bar D{}^{ij}
\,\bar D{}^{kl}$ and $A$ and $A^\prime$ are the lowest-$\theta$
components of $\Phi$ and $\Phi^\prime$, respectively.  Obviously this
class of Lagrangians defines a superconformal extension of the second
Lagrangian in \eqref{eq:lagrangians-1-2}. In order for the action to
be superconformally invariant, the chiral superfields $\Phi$ and
$\Phi'$ must both have vanishing Weyl weights, implying that $A$ and
$A^\prime$ are scale invariant. 

The intermediate equality in \eqref{eq:phi-phi-bar} involves the
so-called $N=2$ \emph{kinetic multiplet} $\mathbb T(\bar\Phi)$
\cite{deWit:1980tn}, conventionally normalized as $\mathbb T(\bar\Phi)
:= -2 \,\bar D^4 \bar\Phi$. When $\Phi$ has zero Weyl weight the
highest-$\theta$ component of the chiral superfield $\Phi$, denoted by
$C$, is S-supersymmetric. Since $\bar C$ equals the lowest-$\theta$
component of $\mathbb T(\bar\Phi)$, the kinetic multiplet is
therefored a conformal primary chiral superfield. The kinetic multiplet
itself thus has Weyl weight $w=2$.\footnote{
  Some of these properties will be more obvious once we present the
  general transformation rules under Q- and S-supersymmetry for a
  generic chiral multiplet of arbitrary Weyl weight. Those will be
  given in \eqref{eq:conformal-chiral} for a general curved
  superspace. } 
Its flat-space components are
\begin{alignat}{2}
  \label{eq:flatT-components}
  A\vert_{\mathbb{T}(\bar\Phi)} &= \bar C~, \qquad&
  \Psi_i\vert_{\mathbb{T}(\bar\Phi)} &=
  - 2\,\varepsilon_{ij} \,\slash{\pa}\Lambda^j~, \nonumber \\[.6ex] 
  B_{ij}\vert_{\mathbb{T}(\bar\Phi)} &=\,
	- 2\,\varepsilon_{ik}\varepsilon_{jl} \,\Box B^{kl}~,
	\qquad&
  F_{ab}^-\vert_{\mathbb{T}(\bar\Phi)} &=\,
	- 4 \big(\delta_a{}^{[c} \delta_b{}^{d]}
    - \tfrac12\varepsilon_{ab}{}^{cd}\big)
		\pa_c \pa^e F^+_{ed} \,,\nonumber\\[.6ex] 
  \Lambda_i\vert_{\mathbb{T}(\bar\Phi)} &=\,
	2\,\Box \,\slash{\pa} \Psi^{j}\varepsilon_{ij}
	\,,  \qquad&
        C\vert_{\mathbb{T}(\bar\Phi)} &=\, 
	4 \,\Box \Box \bar A~.
\end{alignat}
They transform as a chiral multiplet, while depending on the
components of the anti-chiral multiplet $\bar\Phi$.  

An obvious question concerns the derivation of the supersymmetric
extension of the first Lagrangian in \eqref{eq:lagrangians-1-2}, which
is only quadratic in space-time derivatives. As it turns out this
Lagrangian is associated with a {\it reduced} chiral superfield $\cX$.
Besides the chiral constraint, reduced superfields obey the additional
constraint $D_{ij} \cX = \varepsilon_{ik} \varepsilon_{jl} \,\bar D^{kl}
\bar \cX$ that halves the number of independent field components by
expressing the higher-$\theta$ components in terms of space-time
derivatives of the lower-$\theta$ components. The independent
components of the reduced chiral superfield are a complex scalar $X$,
a chiral spinor doublet $\Omega_{\alpha i}$, an anti-selfdual tensor
$F^-_{ab}$ and a triplet of auxiliary fields $Y_{ij}$, conventionally
normalized as
\begin{align}
X := A \vert_{\cX}~, \qquad \Omega_{\alpha i} := \Psi_{\alpha
  i}\vert_{\cX}~, \qquad F_{ab}^- := F_{ab}^-\vert_\cX~,\qquad
	Y_{ij} := B_{ij}\vert_{\cX}~,
\end{align}
using the definitions for the components of a chiral multiplet.  The
reducibility constraint on the superfield $\cX$ requires that $Y_{ij}$
is real, $(Y_{ij})^*=Y^{ij} = \varepsilon^{ik} \varepsilon^{jl}\,
Y_{kl}$, whereas the tensor obeys a Bianchi identity implying that
$F_{ab} = F_{ab}^- + F_{ab}^+$ equals $F_{ab} = 2 e_a{}^\mu e_b{}^\nu\,
\partial_{[\mu} W_{\nu]}$ where $W_\mu$ is a vector gauge
field. Therefore this multiplet is known as the {\it vector
  multiplet}. It comprises $8+8$ bosonic and fermionic components.  In view of
the reducibility constraint the vector multiplet carries Weyl weight
$w=1$.  It is now straightforward to verify that
\begin{align} \label{eq:vector-action}
  \frac{1}{2} \int \rd^4\theta\, \cX^2 = X \Box \bar X + \cdots\,,
\end{align}
where the D'Alembertian arises from the fact that the chiral
superfield is reduced. This example demonstrates
how supersymmetric versions of actions such as the first one in
\eqref{eq:lagrangians-1-2} arise in the context of $N=2$ chiral
superspace. 

Incorporating the third type of Lagrangian \eqref{eq:box-box-log} 
in the context of chiral multiplets seems rather obvious.
Taking $\bar\Phi$ to be an
anti-chiral multiplet of weight $w$, we consider the chiral integral
\begin{align}\label{eq:flat-nlk}
  \int \mathrm{d}^4\theta\,\Phi^\prime \,\big(\bar D^4 \ln\bar\Phi\big)
	= A^\prime \,\Box \Box \ln\bar A + \cdots~,
\end{align}
where $\Phi^\prime$ is a $w=0$ chiral superfield and $A^\prime$
denotes its lowest component.  Naively, this resembles the previous
action \eqref{eq:phi-phi-bar}, but there is a crucial difference: the
anti-chiral multiplet $\bar\Phi$ has arbitrary Weyl weight $w$ and so
$\ln\bar\Phi$ transforms non-linearly under dilatations. Remarkably,
the corresponding kinetic multiplet $\mathbb T(\ln\bar\Phi) := -2 \,
\bar D^4 \ln\bar\Phi$ is nevertheless a conformal primary chiral
multiplet in flat superspace.\footnote{The multiplet $\mathbb
  T(\ln\bar \cX) / \cX^2$ was considered in \cite{Buchbinder:1999jn}
  with $\cX$ a reduced chiral superfield, and shown to be a $w=0$
  conformal primary. The extension of that analysis to $\mathbb
  T(\ln\bar\Phi)$ for an arbitrary anti-chiral multiplet $\bar\Phi$ is
  completely straightforward.}  In other words, it transforms linearly
under dilatations with $w=2$ and its lowest component is invariant
under S-supersymmetry.

We should stress that the non-linearities in $\mathbb{T}(\ln\bar\Phi)$
are of two different types. First of all, the logarithm leads to an
anti-chiral superfield that will depend non-linearly on the components
of $\bar\Phi$. Because of this behaviour, the superconformal
transformations will also be realized in a non-linear fashion, and as
a result the covariantizations that are required in curved superspace
will involve non-linearities depending on the Weyl weight $w$.  In
spite of all these complications, there is a rather systematic way of
writing the various components of $\mathbb{T}(\ln\bar\Phi)$, although
the various explicit expressions tend to become rather complicated,
expecially because they involve higher space-time derivatives. These
non-linearities are the reason why the kinetic multiplet
$\mathbb{T}(\ln\bar\Phi)$ differs in a crucial way from the original
one $\mathbb{T}(\bar\Phi)$.

As a first step in constructing the components of  $\mathbb T(\ln\bar\Phi)$, we must
replace the components of $\bar\Phi$ in \eqref{eq:flatT-components}
with those of $\ln\bar\Phi$. This will simply involve replacing $\bar
A \rightarrow \bar A\vert_{\ln \bar\Phi}$, \ldots, $\bar C
\rightarrow \bar C\vert_{\ln \bar\Phi}$, 
where the components of the multiplet $\ln\Phi$ are identified as 
\begin{align}
  \label{eq:conformal-nonlinear-chiral}
  {A}\vert_{\ln\Phi} &=\, \ln A~, \qquad\qquad\qquad\qquad\qquad\qquad
  {\Psi}_i\vert_{\ln\Phi}=\,\frac{\Psi_i}{A}~, \nonumber\\[.2ex]
  {B}_{ij}\vert_{\ln\Phi}&=\,\frac{B_{ij}}{A}
  +\frac{1}{2A^2}\bar\Psi_{(i}\Psi_{j)}~,\qquad\qquad\qquad
  {F}_{ab}^-\vert_{\ln\Phi}=\,\frac{F_{ab}^-}{A}+ 
  \frac{1}{8A^2}\,\varepsilon^{ij}\bar\Psi_i\gamma_{ab}\Psi_j\,,\nonumber\\[.2ex]  
  {\Lambda}_i\vert_{\ln\Phi}&=\,\frac{\Lambda_i}{A}+
  \frac{1}{2A^2}\big(B_{ij}\varepsilon^{jk}\Psi_k  
  +\tfrac12\,F_{ab}^-\gamma^{ab}\Psi_i\big)+
  \frac{1}{24A^3}\gamma^{ab}\Psi_i\varepsilon^{jk}\bar\Psi_j\gamma_{ab}\Psi_k\,,
  \nonumber\\[.2ex]
  {C}\vert_{\ln\Phi}&=\,\frac{C}{A}+\frac{1}{4A^2}\big(\varepsilon^{ik}\varepsilon^{jl}B_{ij}B_{kl}  
  -2F^{-ab}F_{ab}^-+4\varepsilon^{ij}\bar\Lambda_i\Psi_j\big) 
	\eol & \quad
        \,+\frac{1}{2A^3}\big(\varepsilon^{ik}\varepsilon^{jl}B_{ij}\bar\Psi_k\Psi_l- 
        \tfrac12\varepsilon^{kl}F_{ab}^-\bar\Psi_k\,\gamma^{ab}\Psi_l\big)
        -\frac{1}{32A^4}\varepsilon^{ij}\bar\Psi_i\,\gamma_{ab}\Psi_j
        \varepsilon^{kl}\bar\Psi_k\gamma^{ab}\Psi_l~.
\end{align}
When the chiral superfield $\Phi$ has zero Weyl weight, the logarithm
is merely a field redefinition in superspace, which has no direct
consequences.  However, in the superconformal setting that we are
considering, this is no longer the case for non-zero Weyl weight and
the two chiral multiplets $\Phi$ and $\ln \Phi$ are very different. In
particular $\ln\Phi$ does not satisfy the assignment
\eqref{eq:dilatation-chiral} as it transforms {\it inhomogeneously}
under (constant) dilatations and chiral $\mathrm{U}(1)$
transformations,
\begin{align}
  \label{eq:D-A-shift} 
\delta A\vert_{\ln \Phi} = w \,\big(\Lambda_{\rm D} -\mathrm{i} \Lambda_{\rm A}\big)~.
\end{align}
There are further inhomogeneous transformations, such as
S-supersymmetry that acts inhomogeneously on
$\Psi_i\vert_{\ln\Phi}$. However, the higher-$\theta$ components all
scale consistently as if they belong to a $w=0$ chiral multiplet.  In
flat superspace this phenomenon also extends to the Q- and
S-supersymmetry transformations, although, as we shall see later,
there are some minor exceptions in curved superspace.  The explicit
components in $\mathbb{T}(\ln\bar\Phi)$ will take a rather different
form than in $\mathbb{T}(\bar\Phi)$, but much of the global structure
of $\mathbb{T}(\ln\bar\Phi)$ will still match that of
$\mathbb{T}(\bar\Phi)$.  In particular, the highest
$\theta$-component, ${C}\vert_{\ln\Phi}$ will remain {\it invariant}
under S-supersymmetry, irrespective of the value of the Weyl weight of
$\Phi$. As explained earlier, the latter implies that the kinetic
multiplet $\mathbb{T}(\ln\bar\Phi)$, defined from a generic chiral
multiplet $\Phi$ of arbitrary Weyl weight $w$, will constitute a
conformal primary $w=2$ chiral multiplet.  This observation is
essential as it forms the basis for the approach followed in this
paper.  We will be more explicit in section
\ref{sec:component-kinetic-multiplet}.

The last quantity of interest is the Weyl tensor, which turns out to
be one of the components of the Weyl multiplet. This multiplet is a
reduced chiral {\it tensor} superfield $W_{\alpha\beta}$, symmetric in
$(\alpha\beta)$ with Weyl weight $w=1$. It obeys the constraint
$D^{\alpha \beta} W_{\alpha\beta} = \bar D_{\dalpha \dbeta} \bar
W^{\dalpha \dbeta}$, which reduces it to $24+24$ degrees of
freedom. Those are captured by the field strengths for the
independent gauge fields, namely the vierbein $e_\mu{}^a$, the doublet
of gravitini $\psi_\mu{}^i$, the gauge fields of the
$\mathrm{SU}(2)\times\mathrm{U}(1)$ R-symmetry,
$\mathcal{V}_\mu{}^i{}_j$ and $A_\mu$, as well as three matter fields,
an anti-selfdual tensor $T_{ab}{}^{ij}$, a chiral spinor doublet
$\chi^i$, and a scalar $D$.  Its lowest independent components are
given by
\begin{alignat}{2}
  \label{eq:flatWcomp}
  W_{\alpha\beta}\vert_{\q=0} &= -\tfrac{1}{8}
  (\sigma^{ab})_{\alpha\beta}\, T_{ab}{}^{ij} \varepsilon_{ij}~,
  &\qquad D_{\gamma j} W_{\alpha\beta}\vert_{\q=0} &=
  -\varepsilon_{jk} \, (\sigma^{ab})_{\alpha\beta}\, R(Q)_{ab\, \gamma}{}^k~, \nonumber \\[.4ex]
  D_{ij} W_{\alpha\beta} \vert_{\q=0} &= 2 \,\varepsilon_{ik}
  (\sigma^{ab})_{\alpha\beta}\, R(\cV)_{ab}{}^k{}_j~, &\qquad
  D_{\gamma \delta} W_{\alpha\beta} \vert_{\q=0} &= 2
  \,(\sigma^{ab})_{\alpha\beta}\, (\sigma_{cd})_{\gamma\delta}
  \,R(M)_{ab}{}^{cd}~,
\end{alignat}
where $R(Q)$, $R(\cV)$ and $R(M)$ are the (linearized) curvatures of
conformal supergravity. The usual Weyl tensor as well as the field $D$
are contained within $R(M)$, while $\chi^i$ is contained within
$R(Q)^i$.  The chiral superspace integral of the square of
$W_{\alpha\beta}$ contains therefore the square of the anti-selfdual
component of the Weyl tensor. At the linearized level, we can work
with flat superspace, and we find
\begin{equation} 
  \label{eq:flatW2}
  \mathcal{L}_{\rm W}^- = -\int \mathrm{d}^4\theta \, W_{\alpha\beta}
  W^{\alpha\beta} = C^{abcd -} C_{abcd}^- + \cdots\,.  
\end{equation}
From these results we can now define characteristic terms of the
(linearized and complex) expression for the Gauss-Bonnet density in
flat superspace,
\begin{align}
  \label{align:su-GB}
	\mathcal{L}_\chi^- =&\,   -\int
  \mathrm{d}^4\theta 
  \,\big\{
	W_{\alpha\beta}\, W^{\alpha\beta} 
        +w^{-1} \, \mathbb{T}(\ln \bar\Phi)\big\}  \nonumber\\ 
	=&\,
	\tfrac{1}{2} C^{abcd} C_{abcd} - \tfrac{1}{2} C^{abcd} \tilde C_{abcd}
	+ 2 w^{-1} \Box \Box \ln \bar A + \cdots\,,
\end{align}
where the additional terms depend on the remaining components of the
linearized Weyl multiplet.

These observations are in principle restricted to flat superspace and
to the linearized Weyl multiplet action.  Nevertheless, all of the
Lagrangians above exist in curved superspace.  At the component level,
this is due to the existence of an off-shell conformal supergravity
multiplet which can be used to extend the global supersymmetry algebra
to a local one and impose it on the matter multiplets.  Of particular
use is the chiral density formula (whose explicit form we give in the
next section), which allows the construction of a locally
supersymmetric invariant from a generic weight-two chiral multiplet,
analogous to chiral superspace integrals. The full Lagrangian
corresponding to the Weyl multiplet action \eqref{eq:flatW2}, given
long ago in \cite{Bergshoeff:1980is}, falls into this class, as does
the action \eqref{eq:phi-phi-bar} built upon the kinetic multiplet
$\mathbb T(\bar\Phi)$, whose locally supersymmetric version was shown
to be a conformal primary chiral multiplet in \cite{deWit:2010za}.
For the more complicated Lagrangian \eqref{eq:flat-nlk}, the key
property to determine is similarly whether $\mathbb T(\ln\bar\Phi)$
similarly exists as a proper chiral multiplet; once that is
established, the locally supersymmetric extension follows.  One can
then, as a simple application, construct the $N=2$ Gauss-Bonnet
invariant using the non-linear version of \eqref{align:su-GB}, which
we can immediately deduce must look like
\begin{align}
  \label{eq:Lchi'c}
  e^{-1}\mathcal{L}_\chi^- =&\, \tfrac{1}{2} C^{abcd} C_{abcd} -
  \tfrac{1}{2} C^{abcd} \tilde C_{abcd}
  + 2\,w^{-1} \,\Box_{\rm c} \Box_{\rm c} \ln \bar A + \cdots \nonumber \\
  =&\, \tfrac{1}{2} C^{abcd} C_{abcd} - \tfrac{1}{2} C^{abcd} \tilde
  C_{abcd} - \mathcal{R}^{ab}\mathcal{R}_{ab} + \tfrac{1}{3}
  \mathcal{R}^2
  + \tfrac{1}{3} \mathcal{D}^2 \mathcal{R} \nonumber \\
  & \, + 2 w^{-1} \Big\{ (\mathcal{D}^2)^2  \ln \bar A +  \mathcal{D}^a \big(
  \tfrac{2}{3} \mathcal{R} \,\mathcal{D}_a \ln \bar A - 2 \mathcal{R}_{ab} \,\mathcal{D}^b
  \ln \bar A\big) \Big\} + \cdots
\end{align}
where the missing terms depend on the rest of the Weyl and chiral multiplets.

To extend flat superspace to curved superspace has the advantage that
local supersymmetry will be manifest from the start.  A consistent
definition of curved superspace requires a suitable structure group
and corresponding constraints on the superspace geometry. Subsequently
one can replace the flat spinor derivatives $D_{\alpha i}$ by curved
tangent space derivatives $\nabla_{\alpha i}$ in the explicit
superspace actions as well as in the definitions of the superfield
components and the superfield constraints; these curved derivatives
contain the relevant connection fields whereas the gravitino fields
are introduced as fermionic components of the superspace vielbein. The
superspace formulation that is used here \cite{Butter:2011sr} shares
the same structure group with the conformal multiplet calculus,
encompassing it in a more geometric setting.

\section{Curved superspace, chiral superfields and the kinetic
  multiplet} 
\label{sec:curved-kinetic}
\setcounter{equation}{0}
In this section we first introduce the extension of flat superspace to
the $N=2$ conformal superspace \cite{Butter:2011sr}, which is closely
related to the $N=2$ superconformal multiplet
calculus.\footnote{
  This is not the only way to formulate conformal supergravity in
  superspace.  The most well-known formulations involve either the
  structure group $\rm SO(3,1)\times U(2)$ \cite{Howe}, or the simpler
  structure group $\rm SO(3,1)\times SU(2)$ \cite{KLRT}.  Both realize
  the superconformal symmetries as a super-Weyl transformation, so the
  connection with superconformal multiplet calculus is less direct.
  The relation between the two is spelled out in \cite{KLRT-M2}, and
  their relation to conformal superspace is described in
  \cite{Butter:2011sr}.  } 
Subsequently, we will discuss the chiral multiplet Lagrangians and the
kinetic multiplet in curved superspace.

\subsection{Some details of curved superspace}
Our starting point is a supermanifold parametrized by local
coordinates $z^M = (x^\mu, \theta^{\mathfrak{m} \,\imath},
\bar\theta_{\dot{\mathfrak{m}}\, \imath})$.~\footnote{
  The index $\imath$ on the Grassmann coordinates is a world index
  rather than a tangent space $\mathrm{SU}(2)$ index.}  
The coordinates $x^\mu$ parametrize the bosonic part of the manifold
while the eight Grassmann (anticommuting) coordinates
$\theta^{\mathfrak{m} \,\imath}$ and $\bar\theta_{\dot{\mathfrak{m}}
  \,\imath}$, with $\mathfrak{m} = 1,2$, $\dot{\mathfrak{m}} = \dot 1,
\dot 2$ and $\imath=\ul{1}, \ul{2}$, are associated with the eight
supersymmetries. In addition to (super)diffeomorphisms, we equip the
superspace with the following symmetry generators: Lorentz
transformations, $M_{ab}$; Weyl dilatations, $\mathbb{D}$; chiral $\rm
U(1)$ rotations, $\mathbb{A}$; $\rm SU(2)$ transformations, $I^i{}_j$;
special conformal transformations, $K_a$; and the S-supersymmetries,
$S_\alpha{}^i$ and $\bar S^\dalpha{}_i$.  We introduce a connection
associated with each of these: the spin connection $\Omega_M{}^{ab}$;
the dilatation connection $B_M$; the $\rm U(1)$ and $\rm SU(2)$
connections $A_M$ and $\cV_M{}^i{}_j$; and the K and S-supersymmetry
connections $F_M{}^a$, $\Phi_M{}^\alpha{}_i$ and
$\bar\Phi_M{}_\dalpha{}^i$.  In addition, we introduce the superspace
vielbein $E_M{}^A$, which relates the world index $M$ to the tangent
space index $A$.  In terms of the connections, we can construct the
covariant derivative $\nabla_A = (\nabla_a, \nabla_{\alpha i},
\bar\nabla^{\dalpha i})$ implicitly via the equation
\begin{align}
  \label{eq:defCD}
  E_M{}^A \nabla_A =&\, \partial_M
	- \tfrac{1}{2} \Omega_M{}^{ab} M_{ab}
	- B_M \mathbb D - A_M \mathbb A - \tfrac{1}{2} \cV_M{}^i{}_j \,I^j{}_i
	\eol & \,	- \tfrac{1}{2} \Phi_M{}^{\alpha}{}_i \, S_{\alpha}{}^i
	- \tfrac{1}{2} \bar\Phi_M{}_{\dalpha}{}^i\, \bar S^{\dalpha}{}_i
	- F_M{}^a K_a~,
\end{align}
from which $\nabla_A$ can be solved using the inverse vielbein
$E_A{}^M$.  The supergravity gauge group consists of covariant
diffeomorphisms generated by $\nabla_A$ and the additional
superconformal gauge transformations. A \emph{covariant} (scalar)
superfield $\Psi(x,\theta,\bar\theta)$ transforms as
\begin{align}
\delta \Psi = \Big(\xi^A \nabla_A + \tfrac{1}{2} \L^{ab} M_{ab}
	+ \Lambda_{\rm D} \mathbb D + \L_{\rm A} \mathbb A +
        \tfrac{1}{2} \Lambda^i{}_j \,I^j{}_i 
	+ \eta^\alpha{}_i S_\alpha{}^i
	+ \bar\eta_\dalpha{}^i \bar S^\dalpha{}_i
	+ \Lambda_{\rm K}{}^a K_a\Big) \Psi\,,
\end{align}
without any derivative on the parameters.  $\Psi$ has Weyl weight $w$
and chiral weight $c$ if $\mathbb{D} \Psi = w\, \Psi$ and $\mathbb A
\Psi = \mathrm{i} c\, \Psi$. Note that the space-time diffeomorphisms
and the Q-supersymmetry transformations comprise the
superspace diffeomorphisms generated by $\xi^A\nabla_A$.

Just as in flat superspace, invariant actions are constructed in two
ways.  A \emph{full superspace} integral involves an integral over the
eight Grassmann coordinates of some superspace Lagrangian, which we
denote using the symbol $\mathscr{L}$ (to distinguish it from a
component Lagrangian $\mathcal{L}$),
\begin{align}
  \label{eq:Dterm}
  \int \rd^4x\, \rd^4\theta\, \rd^4\bar\theta\, E\, \mathscr{L}~.
\end{align}
The measure factor $E = \textrm{Ber}(E_M{}^A)$ is the Berezinian (or
superdeterminant) of the superspace vielbein and plays the same role
as the vierbein determinant $e$ on a bosonic manifold. In order for
the action to be invariant under the supergravity gauge group, the
superspace Lagrangian $\mathscr{L}$ must be a conformal primary scalar
with Weyl and chiral weight zero.

A \emph{chiral superspace} integral can be written as
\begin{align}
  \label{eq:Fterm}
  \int \rd^4x\, \rd^4\theta\, \mathcal{E}\, \mathscr{L}_\mathrm{ch}\,,
\end{align}
where $\mathcal{E}$ is the appropriate chiral measure and the
Lagrangian $\mathscr L_\mathrm{ch}$ must be covariantly chiral
(i.e. subject to $\bar\nabla^{\dalpha i} \mathscr L_\mathrm{ch} = 0$) and
a conformal primary with Weyl weight $2$ and chiral weight $-2$.
Generally, any integral over the full superspace can be rewritten (up
to a total derivative) as an integral over chiral superspace,
\begin{align} 
  \label{eq:DtoFterm}
  \int \rd^4x\, \rd^4\theta\, \rd^4\bar\theta\, E\, \mathscr{L}
  = \int \rd^4x\, \rd^4\theta\, \mathcal{E}\, \bar\nabla^4 \mathscr{L}
\end{align}
using the chiral projection operator $\bar\nabla^4$,
\begin{align}
\bar\nabla^4 = \tfrac{1}{48} \varepsilon_{ik} \varepsilon_{jl}
\bar\nabla^{kl} \bar\nabla^{ij}~, \quad 
\bar\nabla^{ij} := \bar\nabla_\dalpha{}^{(i} \bar\nabla^{\dalpha j)}~.
\end{align}
This is a non-trivial statement in curved superspace: one must check that
$\bar\nabla^4 \mathscr{L}$ is indeed chiral and annihilated by S-supersymmetry.

One must have a method to relate superspace integrals to the
usual integrals over the bosonic manifold. Performing the
$\theta$ integrals in \eqref{eq:Fterm} leads to \cite{Butter:2011sr}
\begin{align}
  \int \rd^4x\, \rd^4\theta\, \mathcal{E}\, \mathscr{L}_\mathrm{ch} =
  \int \rd^4x\, \mathcal L_\mathrm{ch} 
\end{align}
where, in two-component notation,
\begin{align}
  \label{eq:chiral-densitySuper}
  e^{-1} \mathcal L_\mathrm{ch} =&\, \Big[\nabla^4
  \mathscr{L}_\mathrm{ch}  - \tfrac{1}{12} \ri\, \varepsilon^{ik}
  \varepsilon^{jl} \,(\bar \psi_{\mu i} \,\bsigma^\mu)^\alpha 
  (\nabla_{\alpha j} \nabla_{kl} \mathscr{L}_\mathrm{ch})
  - \tfrac{1}{2} \ri\,\varepsilon^{ij} \bar\psi_{\mu\,\dgamma i}
  \,\bar W^\dgamma{}_\dbeta 
  \,(\bsigma^\mu){}^{\dbeta \alpha} (\nabla_{\alpha j}
  \,\mathscr{L}_\mathrm{ch}) 
  \nonumber \\ 
  &\quad
  + \bar W^{\dalpha \dbeta} \bar W_{\dalpha \dbeta} \,\mathscr{L}_\mathrm{ch}
  + \tfrac{1}{4} \varepsilon^{ik} \varepsilon^{jl}\, (\bar\psi_{\mu i}
  \bsigma^{\mu \nu} \bar\psi_{\nu j}) \, 
		(\nabla_{kl} \mathscr{L}_\mathrm{ch})
  \nonumber \\
  & \quad
  + \varepsilon^{ij} (\bar\psi_{\mu i} \bar\psi_{\nu j})\,
	\Big( \,\tfrac{1}{8} (\sigma^{\mu\nu})_\alpha{}^\beta
        \,(\nabla_{\beta}{}^\alpha \mathscr{L}_\mathrm{ch}) 
		+ (\bsigma^{\mu \nu})^\dalpha{}_\dbeta \bar
                W^\dbeta{}_{\dalpha} \,\mathscr{L}_\mathrm{ch}\Big)
                \nonumber \\ 
      & \quad 
     + \tfrac{1}{4} e^{-1} \varepsilon^{\mu\nu\rho\tau}
     \varepsilon^{ij} \varepsilon^{kl} (\bar\psi_{\mu i} \bar\psi_{\nu j}) 
          \Big(i (\bar\psi_{\rho\,k}\bsigma_\tau)^\alpha
          (\nabla_{\alpha l} \mathscr{L}_\mathrm{ch}) 	
          + (\bar\psi_{\rho k} \bar\psi_{\tau l})
          \,\mathscr{L}_\mathrm{ch}\Big)\Big]_{\theta=0}~. 
\end{align}
Provided one defines the components of the chiral multiplet
$\mathscr{L}_\mathrm{ch}$ as in \eqref{eq:flatPhi_components},
replacing $D_{\alpha i}$ with $\nabla_{\alpha i}$, one recovers the
usual chiral density rule of the conformal multiplet calculus
\cite{deRoo:1980mm}, but now in four-component form,~\footnote{
  This version of the chiral density formula differs by an overall
  factor of $-\tfrac{1}{2}$ from the usual one \cite{deRoo:1980mm}.
  This arises as a result of the normalization of the component $C$ of
  the chiral multiplet, or equivalently, a different definition of the
  superspace measure.} 
\begin{align} 
  \label{eq:chiral-density} 
  -2 \, e^{-1} \mathcal L_\mathrm{ch} =& \,
  C -\varepsilon^{ij}\, \bar\psi_{\mu i} \gamma^\mu \Lambda_j
  - \tfrac{1}{8}\,\varepsilon^{ij}\varepsilon^{kl}\, \bar \psi_{\mu i}
  T_{ab\,jk}\gamma^{ab}\gamma^\mu \Psi_l
  - \tfrac{1}{16}( T_{ab\,ij} \varepsilon^{ij})^2 \,A \nonumber\\
  &\,
	- \tfrac{1}{2}\,\varepsilon^{ik}\varepsilon^{jl}\,\bar\psi_{\mu
          i}\gamma^{\mu\nu}\psi_{\nu j}\,B_{kl} 
	+ \varepsilon^{ij} \bar \psi_{\mu i}\psi_{\nu j}(F^{-\mu\nu}
		-\tfrac{1}{2} A\, T^{\mu\nu}{}_{kl}\,\varepsilon^{kl} )\nonumber\\
  &\,
	- \tfrac{1}{2} e^{-1} \varepsilon^{\mu\nu\rho\sigma}\,
        \varepsilon^{ij}\varepsilon^{kl}  
		\bar\psi_{\mu i}\psi_{\nu j}
		(\bar\psi_{\rho k}\gamma_\sigma\Psi_{l} +\bar\psi_{\rho k}
			\psi_{\sigma j}\, A)~.
\end{align}
For further details of the superspace geometry, we refer to
Appendix \ref{App:SuperCSG}.

\subsection{Chiral multiplet actions and the kinetic multiplet in
  curved superspace}
In section \ref{sec:ext-chiral-super}, we discussed four types of
actions which could be written down in flat superspace. Each has a
straightforward extension to curved superspace. If we restrict
ourselves to pure conformal supergravity without additional matter
multiplets, there is only a single possible action given by the chiral
superspace integral
\begin{align}
\int \rd^4x\, \rd^4\theta\, \mathcal{E}\, W^{\alpha \beta} W_{\alpha \beta}~,
\end{align}
which we have already discussed at the linearized level in \eqref{eq:flatW2}.

The remaining actions that we discussed in section
\ref{sec:ext-chiral-super} require general chiral multiplets
and vector multiplets, which are contained respectively in chiral
and reduced chiral superfields.  To couple these to
conformal supergravity in superspace requires merely the
covariantization of the chiral constraint 
and the reducibility constraint, respectively.

The simplest action we discussed was the vector multiplet action \eqref{eq:vector-action},
whose curved generalization reads simply
\begin{align}
  \label{eq:two-der-vector}
  \tfrac{1}{2} \int \rd^4x\, \rd^4\theta\, \mathcal{E}\, \cX^2~.
\end{align}
Its component expression was given in \cite{deRoo:1980mm} using the
superconformal multiplet calculus. However, our focus will be on the
curved superspace generalizations of the actions
\eqref{eq:phi-phi-bar} and \eqref{eq:flat-nlk}.

Let us begin with \eqref{eq:phi-phi-bar}.
It generalizes to curved superspace
in a completely straightforward manner:
\begin{align}
  \label{eq:Phi'PhiSuper}
  \int \rd^4x\, \rd^4\theta\, \rd^4\bar\theta\, E\, \Phi' \bar\Phi
	= \int \rd^4x\, \rd^4\theta\, \mathcal{E}\, \Phi' \bar\nabla^4\bar\Phi~.
\end{align}
We have emphasized that the same action can be written using
\eqref{eq:DtoFterm} as a chiral integral of the product of $\Phi'$ and
the kinetic multiplet $\mathbb T(\bar\Phi)$. At the component level,
the Lagrangian is the supersymmetrization of $A' \Box_\mathrm{c}
\Box_\mathrm{c} \bar A$ and was analyzed in \cite{deWit:2010za}.  This
class of higher derivative action admits an obvious generalization in
the presence of several chiral multiplets $\Phi^I$ with weights $w_I$.
Introducing a homogeneous function $\cH(\Phi, \bar\Phi)$ of weight zero,
\begin{align}\label{eq:Hhomo1}
\sum_I w_I \Phi^I \cH_I =  0~,
\end{align}
where $\cH_I := \pa\cH/\pa \Phi^I$,
one can construct a higher derivative action by integrating $\cH$ over the full
superspace,\footnote{Similar structures were considered in the context of low-energy
effective actions in flat space \cite{Henningson:1995eh, deWit:1996kc,
Buchbinder:1999jn, Banin:2002mf, Argyres:2003tg}.}
\begin{align}
\int \rd^4x\, \rd^4\theta\, \rd^4\bar\theta\, E\, \cH~.
\end{align}
By virtue of the formula \eqref{eq:DtoFterm} and its complex conjugate, one can show
that the action is invariant under the K\"ahler-like transformations
\begin{align}
\cH \rightarrow \cH + \L(\Phi) + \bar \L(\bar\Phi)
\end{align}
where the holomorphic function $\L(\Phi)$ must similarly be homogeneous.
It follows that the component action will depend only on the K\"ahler metric
$\cH_{I \bar J}$, which is subject to the homogeneity condition
\begin{align}\label{eq:Hhomo2}
  \sum_I w_I \Phi^I \cH_{I \bar J} = 0~.
\end{align}
The locally supersymmetric version was analyzed in
\cite{deWit:2010za}, with particular attention paid to the special
case where the chiral multiplets were vector multiplets $\cX^I$ with
$w=1$ or the Weyl-squared chiral multiplet $W^{\alpha \beta}
W_{\alpha\beta}$ with $w=2$. This class can be broadened further while
maintaining the K\"ahler structure by considering the chiral
multiplets $\Phi^I$ to be themselves composite in various ways.

It was noted in \cite{deWit:2010za} that a broad class of higher derivative
chiral superspace integrals lift naturally to full superspace integrals
involving functions $\cH$ by stripping away an operator $\bar\nabla^4$ as
in \eqref{eq:Phi'PhiSuper}. However, it turns out that the curved
version of the action \eqref{eq:flat-nlk},
\begin{align}
\int \rd^4x\, \rd^4\theta\, \mathcal{E}\, \Phi' \bar\nabla^4 \ln\bar\Phi ~,
\end{align}
where $\Phi'$ has weight $w'=0$ and $\Phi$ has nonzero weight $w$,
does \emph{not} belong to this class.
At first glance, a naive
application of \eqref{eq:DtoFterm} would seem to indicate
\begin{align}\label{eq:wrongTtoD}
\int \rd^4x\, \rd^4\theta\, \mathcal{E}\, \Phi' \bar\nabla^4 \ln\bar\Phi \stackrel{?}{=}
	\int \rd^4x\, \rd^4\theta\,\rd^4\bar\theta\, E\, \Phi' \ln\bar\Phi
\end{align}
with the full superspace Lagrangian falling into the
class of generic function $\cH$ already considered. However, the proposed
Lagrangian $\cH = \Phi' \ln\bar\Phi$ transforms inhomogeneously under dilatations
and so is \emph{not} permissible; in other words,
$\cH$ does not obey the homogeneity conditions
\eqref{eq:Hhomo1} or \eqref{eq:Hhomo2}.\footnote{This obstruction is specific
for curved superspace. For flat superspace, $\cH$ must be homogeneous
only up to K\"ahler transformations; see e.g. \cite{Buchbinder:1999jn}
where such actions were considered.}
Nevertheless, the left-hand side of \eqref{eq:wrongTtoD} \emph{does} transform
appropriately. This is because the kinetic multiplet
$\mathbb T(\ln\bar\Phi)$ is a conformal primary chiral multiplet of weight $w=2$,
obeying
\begin{align}
\bar \nabla^{\dalpha i} \mathbb T(\ln\bar\Phi) = 0~, \qquad
S_\alpha{}^i \mathbb T(\ln\bar\Phi) = \bar S^{\dalpha}{}_i \mathbb T(\ln\bar\Phi) = 0~.
\end{align}
Both conditions are straightforward enough to check, although they require some
$\rm SU(2)$ index gymnastics.\footnote{The key idea for the first condition is that there
are only four anti-commuting $\bar\nabla^{\dalpha i}$ derivatives, so a product of
five of them must vanish (up to curvatures, which contribute nothing in this case).
The next condition, that it is annihilated
by $S_\alpha{}^i$, is easy enough as that operator anti-commutes with $\bar\nabla^{\dalpha i}$;
checking the last condition, that $\bar S^{\dalpha}{}_i$ similarly gives zero, is
a minor exercise using the algebra of the operators given in Appendix \ref{App:SuperCSG}.}
Now by comparing to the flat space limit, it is obvious that
\begin{align}\label{eq:curved-nlk}
\int \rd^4x\, \rd^4\theta\, \mathcal{E}\, \Phi' \,\bar\nabla^4 \ln\bar\Phi
	= \int \rd^4x\, e\, A'\, \Box_{\rm c} \Box_{\rm c} \ln \bar A +
        \textrm{additional terms}\,. 
\end{align}
The complete expression, which we will present in this paper, corresponds to
a \emph{new chiral supersymmetric invariant}.

This invariant has already appeared in physical applications. In \cite{Banerjee:2011ts},
the $5D$ mixed gauge-gravitational Chern-Simons invariant \cite{Hanaki:2006pj} was
dimensionally reduced, and a characteristic subset of $4D$ terms was obtained
which broke down into three classes.
The first class was easily identified as the usual chiral superspace integral
of a holomorphic function. Another class seemed to coincide with the
full superspace integral of a real function $\cH \sim \Phi' \ln\bar\Phi + \textrm{h.c.}$,
while the remainder, involving terms of the Gauss-Bonnet variety,
could not be identified with any currently known invariant.
It is clear to us now that these latter two classes of terms are actually
contained within the single invariant \eqref{eq:curved-nlk},
which is intrinsically chiral and cannot be decomposed further
in a manifestly superconformal way.

Before setting out to calculate the expression \eqref{eq:curved-nlk}
explicitly, we should make an important observation. In the
introduction, we noted that the non-linear Lagrangian
\eqref{eq:box-box-log} with $\phi^\prime$ constant, must depend on the
field $\ln \bar \phi$ only via total derivative terms.  We expect the
same should hold for its supersymmetrized version, namely that when
$\Phi^\prime$ is constant in \eqref{eq:curved-nlk} the dependence on
$\ln\Phi$ is only in the form of total-derivative terms.  To see this,
suppose we have two such kinetic multiplets built out of the logarithm
of two different anti-chiral superfields $\bar\Phi_1$ and
$\bar\Phi_2$, taken to have the same weight $w$ for simplicity.  The
difference is obviously
\begin{align}
\bar\nabla^4 \ln\bar\Phi_1 - \bar\nabla^4\ln\bar\Phi_2
	= \bar\nabla^4 \ln(\bar\Phi_1 / \bar\Phi_2)\,,
\end{align}
and the quantity under the spinor derivatives on the right-hand side is actually
a proper weight-zero multiplet. It follows that any chiral integrand involving
such a difference can be written as a full superspace integral
and then as an anti-chiral superspace integral, discarding total
derivatives in the equalities. Hence,  
\begin{align}
  \label{eq:NLKdiff}
  \int \rd^4x\, \rd^4\theta\, \mathcal{E}\, \Phi' \,\bar\nabla^4 \ln
  (\bar\Phi_1 / \bar\Phi_2) &= \int \rd^4x\, \rd^4\theta\,
  \rd^4\bar\theta\, E\, \Phi' \ln(\bar\Phi_1 / \bar\Phi_2) \eol &=
  \int \rd^4x\, \rd^4\bar\theta\, \bar{\mathcal{E}}\, (\nabla^4 \Phi')
  \ln(\bar\Phi_1 / \bar\Phi_2)~.
\end{align}
Taking the weight-zero chiral superfield $\Phi'$ to be actually constant,
it follows that the right-hand side of \eqref{eq:NLKdiff} vanishes and therefore 
\begin{align}
  \int \rd^4x\, \rd^4\theta\, \mathcal{E}\, \bar\nabla^4 \ln
  \bar\Phi_1 = \int \rd^4x\, \rd^4\theta\, \mathcal{E}\, \bar\nabla^4
  \ln \bar\Phi_2~.
\end{align}
In other words, the integral
$\int \rd^4x\, \rd^4\theta\, \mathcal{E}\, \bar\nabla^4 \ln \bar\Phi$
is \emph{independent} of the components of $\ln \bar\Phi$ up to total
derivatives. This observation will be an important check that we have
correctly calculated the additional terms in \eqref{eq:curved-nlk}. It
is to this task which we now turn.

\section{The component structure of the kinetic multiplet}
\label{sec:component-kinetic-multiplet}
\setcounter{equation}{0} 
In this section, we proceed to construct the kinetic multiplet
$\mathbb T(\ln\bar\Phi)$ in supergravity along with the corresponding
Lagrangian \eqref{eq:flat-nlk}. The starting point is the formula for
the Q- and S-supersymmetry transformations of a general $N=2$ chiral
multiplet $\Phi$ with Weyl weight $w$ in four-component notation
\cite{deRoo:1980mm,deWit:1980tn,deWit:2010za},
\begin{align}
  \label{eq:conformal-chiral}
  \delta A =&\,\bar\epsilon^i\Psi_i\,, \nonumber\\[.2ex]
  \delta \Psi_i =&\,2\,\Slash{D} A\epsilon_i + B_{ij}\,\epsilon^j +
  \tfrac12   \gamma^{ab} F_{ab}^- \,\varepsilon_{ij} \epsilon^j + 2\,w
  A\,\eta_i\,,  \nonumber\\[.2ex]
  \delta B_{ij} =&\,2\,\bar\epsilon_{(i} \Slash{D} \Psi_{j)} -2\,
  \bar\epsilon^k \Lambda_{(i} \,\varepsilon_{j)k} + 2(1-w)\,\bar\eta_{(i}
  \Psi_{j)} \,, \nonumber\\[.2ex]
  \delta F_{ab}^- =&\,\tfrac12
  \varepsilon^{ij}\,\bar\epsilon_i\Slash{D}\gamma_{ab} \Psi_j+
  \tfrac12 \bar\epsilon^i\gamma_{ab}\Lambda_i
  -\tfrac12(1+w)\,\varepsilon^{ij} \bar\eta_i\gamma_{ab} \Psi_j \,,
  \nonumber\\[.2ex]
  \delta \Lambda_i =&\,-\tfrac12\gamma^{ab}\Slash{D}F_{ab}^-
   \epsilon_i  -\Slash{D}B_{ij}\varepsilon^{jk} \epsilon_k +
  C\varepsilon_{ij}\,\epsilon^j
  +\tfrac14\big(\Slash{D}A\,\gamma^{ab}T_{abij}
  +w\,A\,\Slash{D}\gamma^{ab} T_{abij}\big)\varepsilon^{jk}\epsilon_k
  \nonumber\\*
  &\, -3\, \gamma_a\varepsilon^{jk}
  \epsilon_k\, \bar \chi_{[i} \gamma^a\Psi_{j]} -(1+w)\,B_{ij}
  \varepsilon^{jk}\,\eta_k + \tfrac12 (1-w)\,\gamma^{ab}\, F_{ab}^-
    \eta_i \,, \nonumber\\[.2ex]
    \delta C =&\,-2\,\varepsilon^{ij} \bar\epsilon_i\Slash{D}\Lambda_j
  -6\, \bar\epsilon_i\chi_j\;\varepsilon^{ik}
    \varepsilon^{jl} B_{kl}   \nonumber\\*
  &\, -\tfrac14\varepsilon^{ij}\varepsilon^{kl} \big((w-1)
  \,\bar\epsilon_i \gamma^{ab} {\Slash{D}} T_{abjk}
    \Psi_l + \bar\epsilon_i\gamma^{ab}
    T_{abjk} \Slash{D} \Psi_l \big) + 2\,w \varepsilon^{ij}
    \bar\eta_i\Lambda_j \,.
\end{align}
In this convention the spinors $\epsilon^i$ and $\eta_{i}$ are the positive
chirality spinorial parameters associated with Q- and S-supersymmetry.
The corresponding negative chirality parameters are denoted by
$\epsilon_i$ and $\eta^i$. (In two-component form, the positive
chirality spinors would be denoted by $\epsilon_\alpha{}^i$ and
$\eta_{\alpha i}$, and the negative chirality spinors by
$\bar\epsilon^\dalpha{}_i$ and $\bar\eta^{\dalpha i}$.) 

One can see from \eqref{eq:conformal-chiral} that the highest
component $C$ of a $w=0$ chiral multiplet is anti-chiral and invariant
under S-supersymmetry. This observation allows the construction of the
chiral $w=2$ kinetic multiplet $\mathbb T(\bar\Phi)$ whose lowest
component is $\bar C$.  Although such an analysis was carried out in
components in \cite{deWit:2010za} by consecutively considering
supersymmetry transformations and identifying the higher-theta
components, it could just as easily be carried through in
superspace. The starting point is to take $\mathbb T(\bar\Phi) := -2
\bar\nabla^4 \bar\Phi$ and to identify its components using the curved
superspace version of \eqref{eq:flatPhi_components},
\begin{alignat}{2}
     A\vert_{\mathbb T(\bar\Phi)} &:= -2 \bar\nabla^4
     \bar\Phi\vert_{\theta=0}~, \quad& 
     \Psi_{\alpha i}\vert_{\mathbb T(\bar\Phi)} &:= -2
     \nabla_{\alpha i} \bar\nabla^4 \bar\Phi\vert_{\theta=0}~,
     \nonumber \\
     B_{ij}\vert_{\mathbb T(\bar\Phi)} &:= \nabla_{ij} \bar\nabla^4
     \bar\Phi\vert_{\theta=0}~, \quad& 
     F_{ab}^-\vert_{\mathbb T(\bar\Phi)} &:= \tfrac{1}{2}
     (\sigma_{ab})_\alpha{}^\beta \nabla_\beta{}^\alpha \bar\nabla^4
     \bar\Phi\vert_{\theta=0} ~,\nonumber \\
     \Lambda_{\alpha i}\vert_{\mathbb T(\bar\Phi)} &:= -\tfrac{1}{3}
     \varepsilon^{jk} \nabla_{\alpha k} \nabla_{ji} \bar\nabla^4
     \bar\Phi\vert_{\theta=0}~, \quad& 
     C\vert_{\mathbb T(\bar\Phi)} &:= 4 \nabla^4 \bar\nabla^4
     \bar\Phi \vert_{\theta=0}~.
\label{eq:TPhiCompSuper}
\end{alignat}
Since $\bar\Phi$ is anti-chiral, each of these components may be
evaluated in the usual superspace fashion: (anti)commute each
$\nabla_{\alpha i}$ past the other covariant derivatives until they
annihilate $\bar\Phi$. The resulting expression should be
rearranged (using the superspace commutation relations) so that all
the spinor derivatives $\bar\nabla^{\dalpha i}$ act directly
upon $\bar\Phi$. The calculation is straightforward,
although more and more complicated as the number of spinor derivatives
increases; for such calculations, the (anti)-commutation relations
given in Appendix \ref{App:SuperCSG} are necessary.

Now we wish to construct the non-linear version of the kinetic
multiplet, $\mathbb T(\ln\bar\Phi)$.
As we have already alluded to in section \ref{sec:ext-chiral-super}, 
we will choose to define the components of $\ln\Phi$
using \eqref{eq:conformal-nonlinear-chiral}, which coincides with
using the curved superspace version of \eqref{eq:flatPhi_components}
with $\Phi$ replaced by $\ln\Phi$.
It is straightforward to determine the Q- and S-supersymmetry transformation
rules of these components
\begin{align}
  \label{eq:conformal-nonlinear-chiral transf}
  \delta \hat{A} =&\,\bar\epsilon^i\hat{\Psi}_i\,, \nonumber\\[.2ex]
  \delta \hat{\Psi}_i =&\,2\,\Slash{D} \hat{A}\epsilon_i + 
  \hat{B}_{ij}\,\epsilon^j + \tfrac12   \gamma^{ab} \hat{F}_{ab}^-
  \,\varepsilon_{ij} \epsilon^j + 2\,w\,\eta_i\,,
  \nonumber\\[.2ex]
  \delta \hat{B}_{ij} =&\,2\,\bar\epsilon_{(i} \Slash{D} \hat{\Psi}_{j)}
  -2\, \bar\epsilon^k \hat{\Lambda}_{(i} \,\varepsilon_{j)k} + 
   2\,\bar\eta_{(i} \hat{\Psi}_{j)} \,,
  \nonumber\\[.2ex]
  \delta \hat{F}_{ab}^- =&\,\tfrac12
  \varepsilon^{ij}\,\bar\epsilon_i\Slash{D}\gamma_{ab} \hat{\Psi}_j+
  \tfrac12 \bar\epsilon^i\gamma_{ab}\hat{\Lambda}_i
  -\tfrac12\,\varepsilon^{ij} \bar\eta_i\gamma_{ab} \hat{\Psi}_j \,,
  \nonumber\\[.2ex]
  \delta \hat{\Lambda}_i =&\,-\tfrac12\gamma^{ab}\Slash{D}\hat{F}_{ab}^-
   \epsilon_i  -\Slash{D}\hat{B}_{ij}\varepsilon^{jk} \epsilon_k +
  \hat{C}\varepsilon_{ij}\,\epsilon^j
  +\tfrac14\big(\Slash{D}\hat{A}\,\gamma^{ab}T_{abij}
  +w\,\Slash{D}\gamma^{ab} T_{abij}\big)\varepsilon^{jk}\epsilon_k
  \nonumber\\
  &\, -3\, \gamma_a\varepsilon^{jk}
  \epsilon_k\, \bar \chi_{[i} \gamma^a\hat{\Psi}_{j]} -\,\hat{B}_{ij}
  \varepsilon^{jk}\,\eta_k + \tfrac12 \,\gamma^{ab}\, \hat{F}_{ab}^-
    \eta_i \,, \nonumber\\[.2ex]
    \delta \hat{C} =&\,-2\,\varepsilon^{ij} \bar\epsilon_i\Slash{D}\hat{\Lambda}_j
   -6\, \bar\epsilon_i\chi_j\;\varepsilon^{ik}
    \varepsilon^{jl} \hat{B}_{kl}
	+\tfrac14\varepsilon^{ij}\varepsilon^{kl} 
   \big(\bar\epsilon_i \gamma^{ab} {\Slash{D}} T_{abjk}
    \hat{\Psi}_l - \bar\epsilon_i\gamma^{ab}
    T_{abjk} \Slash{D} \hat{\Psi}_l \big)~.
\end{align}
Comparing these transformation laws to those in
\eqref{eq:conformal-chiral}, one notes the appearance of
non-linearities involving the weight $w$.  Every term is linear in the
components of $\hat \Phi=\ln\Phi$ except for the terms proportional to
$w$, which are independent of $\ln\Phi$.  As discussed earlier, this
arises ultimately from the inhomogeneous transformation of $\ln\Phi$
under dilatations. Note, however, that the covariant derivatives in
\eqref{eq:conformal-nonlinear-chiral transf} do also depend on the
Weyl weight and therefore contain similar terms. For instance,
consider the transformation \eqref{eq:D-A-shift}, which obviously
requires a term $-w\big(b_\mu-\mathrm{i} A_\mu\big)$ in the covariant
derivative $D_\mu\hat A$ which no longer depends on $\ln\Phi$.

As mentioned in section \ref{sec:ext-chiral-super}, the highest component
$\hat C$ of $\ln\Phi$ is a weight 2 conformal primary and (anti)chiral under
Q-supersymmetry. This means we may use $\hat{\bar C}$ as the lowest component
of a chiral multiplet, which will be the kinetic multiplet
$\mathbb T(\ln\bar\Phi)$. Within superspace, we can define its components exactly
as in \eqref{eq:TPhiCompSuper}, with $\bar\Phi$ replaced by $\ln\bar\Phi$,
and the subsequent computational steps are as outlined above, except for the
generation of terms involving $w$.

An alternative procedure is to begin with the condition
$A\vert_{\mathbb{T}(\ln\bar\Phi)} = \hat{\bar C}$ and derive
$\Psi_{i}\vert_{\mathbb{T}(\ln\bar\Phi)}$ by applying a
Q-supersymmetry transformation to both sides. Continuing in this way,
one can build up the entire multiplet. This was the procedure that was
originally applied to the linear kinetic multiplet
$\mathbb{T}(\bar\Phi)$ in \cite{deWit:2010za}, but which is now
considerably more involved. A convenient way of applying the same
strategy is to focus only on the $w$-dependent terms by unpackaging
the full covariant derivatives. Although this sacrifices manifest
covariance, it exploits the high degree of overlap between $\mathbb
T(\ln\bar\Phi)$ and the kinetic multiplet $\mathbb T(\bar\Phi)$
studied in \cite{deWit:2010za}.  

We have followed both lines of approach and confirmed agreement
between them, up to the fermionic terms in
$C\vert_{\mathbb{T}(\ln\bar\Phi)}$; these have passed other
non-trivial checks using S-supersymmetry.  The result is (in four
component notation),
\begin{align}
  \label{eq:T-components}
  A\vert_{\mathbb{T}(\ln\bar\Phi)} =&\, \hat{\bar C} \,, \nonumber\\[.3ex]
  \Psi_i\vert_{\mathbb{T}(\ln\bar\Phi)} =&\,
  - 2\,\varepsilon_{ij} \Slash{D} \hat \Lambda^j
  - 6\, \;\varepsilon_{ik} \varepsilon_{jl} \chi^j \hat B^{kl}
  - \tfrac14\varepsilon_{ij}\varepsilon_{kl} \, \gamma^{ab} T_{ab}{}^{jk}
  \stackrel{\leftrightarrow}{\Slash{D}} \hat \Psi^l \,,  \nonumber\\[.6ex]
  B_{ij}\vert_{\mathbb{T}(\ln\bar\Phi)} =&\,
	- 2\,\varepsilon_{ik}\varepsilon_{jl} \big(\Box_\mathrm{c} + 3\,D\big) \hat B^{kl}
	- 2\, \hat F^+_{ab}\, R(\mathcal{V})^{ab\,k}{}_{i}\, \varepsilon_{jk} \nonumber\\
	 &\, 
	-6\,\varepsilon_{k(i}\,\bar\chi_{j)} \hat \Lambda^k
	+ 3\, \varepsilon_{ik} \varepsilon_{jl} \hat {\bar\Psi}^{(k}\Slash{D}\chi^{l)} \,,
	\nonumber \\[.6ex]
   F_{ab}^-\vert_{\mathbb{T}(\ln\bar\Phi)} =&\,
	- \big(\delta_a{}^{[c} \delta_b{}^{d]}
    - \tfrac12\varepsilon_{ab}{}^{cd}\big)
		\big[4\, D_c D^e \hat F^+_{ed}
		+ (D^e \hat {\bar A}\,D_cT_{de}{}^{ij}
		+ D_c \hat {\bar A} \,D^eT_{ed}{}^{ij})\varepsilon_{ij}
		- w D_c D^e T_{ed}{}^{ij} \varepsilon_{ij}
	\big]  
	\nonumber\\ 
        & \,
	+ \Box_\mathrm{c} \hat {\bar A} \,T_{ab}{}^{ij}\varepsilon_{ij}
	- R(\mathcal{V})^-{}_{\!\!ab}{}^i{}_k \,\hat B^{jk} \,\varepsilon_{ij}
	+ \tfrac1{8} T_{ab}{}^{ij} \,T_{cdij} \hat F^{+cd}
	- \varepsilon_{kl}\,\hat {\bar\Psi}^k\stackrel{\leftrightarrow}{\Slash{D}} R(Q)_{ab}{}^l
        \nonumber\\
        &\, 
	-\tfrac94 \varepsilon_{ij} \,\hat {\bar\Psi}^i\gamma^c\gamma_{ab}D_c \chi^j
	+  3\,\varepsilon_{ij} \bar\chi^i\gamma_{ab} \Slash{D} \hat {\Psi}^j
	+ \tfrac3{8}T_{ab}{}^{ij}\varepsilon_{ij} \bar\chi_k \hat {\Psi}^k \,,\nonumber\\[.6ex] 
        \Lambda_i\vert_{\mathbb{T}(\ln\bar\Phi)} =&\,
	2\,\Box_\mathrm{c}\Slash{D} \hat \Psi^{j}\varepsilon_{ij}
	+ \tfrac1{4}  \gamma^c\gamma_{ab} (2\, D_c
        T^{ab}{}_{ij}\,\hat \Lambda^{j} + T^{ab}{}_{ij} \,D_c \hat \Lambda^{j}) \nonumber\\*
        &\, 
	- \tfrac1{2}\varepsilon_{ij} \big(R(\mathcal{V})_{ab}{}^j{}_k
	+ 2\mathrm{i} \,
        R(A)_{ab}\delta^j{}_k\big)\,\gamma^c\gamma^{ab} D_c\hat \Psi^k
        \nonumber\\*
        &\, 
	+ \tfrac12\,\varepsilon_{ij} \big( 3\, D_b D - 4\mathrm{i} D^a R(A)_{ab}
		+\tfrac{1}{4} T_{bc}{}^{ij}\stackrel{\leftrightarrow}{D_a} T^{ac}{}_{ij}
		\big)\,\gamma^b \hat \Psi^{j} \nonumber\\*
	 &\, 
	- 2\,\hat F^{+ab}\, \Slash{D}R(Q)_{ab}{}_{i}
	+ 6\,\varepsilon_{ij}  D\, \Slash{D} \hat \Psi^{j} \nonumber\\
	 &\,
	+ 3 \,\varepsilon_{ij}\,\big(\Slash{D}\chi_k\,\hat B^{kj}
	+ \Slash{D} \hat {\bar A}\,\Slash{D}\chi^{j}  \big) \nonumber\\
	& \, 
	+ \tfrac32\big( 2\,\Slash{D} \hat B^{kj} \varepsilon_{ik}
	+ \Slash{D} \hat F_{ab}^+ \gamma^{ab} \, \delta^j_{i}
	+ \tfrac14  \varepsilon_{kl} T_{ab}{}^{kl}\,\gamma^{ab}
		\,\Slash{D} \hat {\bar A}\, \delta_i{}^j\big)  \chi_j \nonumber\\
	& \,
	+ \tfrac94 \, (\bar\chi^l\gamma_a\chi_l) \,\varepsilon_{ij}\gamma^a \hat \Psi^{j}
	- \tfrac92 \, (\bar\chi_i\gamma_a\chi^k)
        \,\varepsilon_{kl}\gamma^a \hat \Psi^{l} \nonumber\\
	& \,
	-\tfrac32w\, \varepsilon_{jk}\,  D^a T_{ab}{}^{jk}\gamma^b\chi_i
	\,,  \nonumber\\[.6ex]
C\vert_{\mathbb{T}(\ln\bar\Phi)} =&\, 4 (\Box_{\rm c} + 3 D)
        \Box_{\rm c} \hat {\bar A} 
	+ 6 (D_a D) \, D^a \hat {\bar A}
	- 16 \,D^a \Big(R(D)_{ab}^+ D^b \hat {\bar A}\Big)
	\eol & \,
	- D^a (T_{ab ij} T^{cbij} D_c \hat {\bar A})
	- \tfrac{1}{2} D^a(T_{ab ij} T^{cb ij}) D_c \hat {\bar A}
	- 9 \,\bar \chi_j \gamma^a \chi^j\, D_a \hat {\bar A}
	\eol & \,
	+ \tfrac{1}{2} D_a D^a (T_{bc ij} \hat F^{bc+}) \varepsilon^{ij}
	+ 4 \varepsilon^{ij} D_a \Big(D^b T_{bc ij} \hat F^{ac+} + D^b
        \hat F_{bc}^+ T^{ac}{}_{ij} \Big) 
	\eol & \,
	- \tfrac{9}{2} \varepsilon^{jk} \bar\chi_j \gamma^{ab} \chi_k \hat F_{ab}^+
	+ 9 \bar\chi_j \chi_k \hat {B}^{jk}
	+ \tfrac{1}{16} (T_{ab}{}^{ij} \varepsilon_{ij})^2 \hat {\bar C}
	\eol & \,
	+ 6 D^a D_a \bar \chi_j \hat\Psi^j
	+ 3 \bar \chi_j \slashed{D}\slashed{D} \hat\Psi^j
	+ 3 D_a (\bar \chi_j \gamma^a \slashed{D} \hat\Psi^j)
	+ 9 D \bar\chi_j \hat\Psi^j
	\eol & \,
	- 8 D^a \bar R(Q)_{ab j} D^b \hat\Psi^j
	+ 6 D_b \bar\chi_j \gamma^b \slashed{D} \hat\Psi^j
	\eol & \,
	+ \tfrac{3}{2} D^a T_{ab ij} \bar\chi^i \gamma^b \hat\Psi^j
	+ 3 D^a (T_{ab ij} \bar\chi^i \gamma^b \hat\Psi^j )
	+ \tfrac{3}{2} D^a (T_{abij} \bar\chi^i) \gamma^b \hat\Psi^j
	\eol & \,
	+ 3 \Big(\tfrac{1}{2} R(\cV)^+_{ab}{}^i{}_j
		- R(D)_{ab}^+ \delta^i{}_j \Big) \bar\chi_i \gamma^{ab} \hat\Psi^j
	- 2 R(\cV)^+_{ab}{}^i{}_j \bar R(Q)^{ab}{}_i \hat\Psi^j
	- \tfrac{1}{2} T^{ab}{}_{ij} \bar R(S)^+_{ab}{}^i \hat\Psi^j
	\eol & \,
	+ \tfrac{1}{8} \varepsilon^{ij} T_{ab ij}
		\Big(3 \bar \chi_k \gamma^{ab} \hat \Lambda^k + 2 \bar
                R(Q)^{ab}_k \hat \Lambda^k\Big) 
	\eol & \, 
        + w\Big\{9 \bar\chi_j \slashed{D} \chi^j
	- R(\cV)_{ab}^+{}^i{}_j R(\cV)^{ab+}{}^j{}_i
	- 8  R(D)^+_{ab} R(D)^{ab+}
	\eol &\quad \qquad
	-  D^a T_{ab ij} D_c T^{cb ij}
	-  D^a (T_{abij} D_c T^{cb ij})\Big\} ~.
\end{align}
The result agrees with the corresponding expressions for
the usual kinetic multiplet discussed in \cite{deWit:2010za} by taking
$w=0$. In this limit, the superfield $\ln\bar\Phi$ becomes a normal
$w=0$ anti-chiral multiplet with $\mathbb T(\ln\bar\Phi)$ its
associated kinetic multiplet.

Now we can calculate the component Lagrangian $\mathcal L$
corresponding to the action
\begin{align} 
  \label{eq:SphiT}
  -2 \int \rd^4x\, \rd^4\theta\, \mathcal{E}\, \Phi' \, \mathbb T(\ln \bar\Phi)~.
\end{align}
This is a straightforward application of \eqref{eq:chiral-density} and
the product rule, (C.1) of \cite{deWit:2010za}, or, equivalently, the
direct application of \eqref{eq:chiral-densitySuper}.  We will ignore
all fermions, which significantly simplifies the resulting
expression. Expanding out the covariant d'Alembertians using, for example,
the expression for $f_\mu{}^a$ given in \eqref{eq:f-bos-uncon} leads to
\begin{align}
  \label{eq:Cdegauged}
  C\vert_{\mathbb T(\ln\bar\Phi)}
	=&\, \mathcal{D}_a V^a + \tfrac{1}{16} ( T_{ab\,ij}
        \varepsilon^{ij})^2\, \hat {\bar C} \nonumber\\
        &\, + w\big\{ - 2\, \mathcal{R}^{ab} \mathcal{R}_{ab} +
        \tfrac{2}{3}  \mathcal{R}^2 
	- 6 \,  D^2 + 2 \, R(A)^{ab} R(A)_{ab}
	-  R(\cV)_{ab}^+{}^i{}_j R(\cV)^{ab+}{}^j{}_i \nonumber\\
	 & \qquad
	+ \tfrac{1}{128}  T^{ab ij} T_{ab}{}^{kl} T^{cd}{}_{ij}T_{cd kl}
	+  T^{ac ij} D_a D^b T_{bc ij}\big\} \,,
\end{align}
where $V^a$ is given by
\begin{align}
  \label{eq:Vexpression}
  V^a =&\, 4 \mathcal{D}^a \mathcal{D}^2 \hat{\bar A}
	- 8 \mathcal{R}^{ab} \mathcal{D}_b \hat{\bar A}
	+ \tfrac{8}{3} \mathcal{R} \mathcal{D}^a \hat{\bar A}
	+ 8 D \,\mathcal{D}^a \hat{\bar A}
	- 8 \mathrm{i}\, R(A)^{ab} \mathcal{D}_b \hat{ \bar A} \nonumber\\*
	&\, - 2 T^{ac ij} T_{bc ij} \mathcal{D}^b \hat{\bar A}
	+ \tfrac{1}{2}  \varepsilon^{ij} \mathcal{D}^a T_{bc ij} \hat F^{bc+}
	+ 4 \,\varepsilon^{ij} T^{ac}{}_{ij} \mathcal{D}^b \hat
        F_{bc}^+ \nonumber\\*
	&\, + w\big\{ \tfrac{2}{3} \mathcal{D}^a \mathcal{R}
	- 4 \,\mathcal{D}^a D
	-  \mathcal{D}^b (T^{ac ij} T_{bcij})\big\} ~.
\end{align}
Here the derivatives $\mathcal{D}_a$ are covariant with respect to
the linearly acting bosonic transformations. Hence they do not contain
the connection field or the conformal boosts $f_\mu{}^a$. Note that we
have kept the K-connection $f_\mu{}^a$ within the fully covariant
derivatives in the last term of \eqref{eq:Cdegauged} for later
convenience, but there is no obstacle in extracting it here as well.

Performing a similar decomposition in
$B_{ij}\vert_{\mathbb{T}(\ln\bar\Phi)}$ and
$F_{ab}^-\vert_{\mathbb{T}(\ln\bar\Phi)}$ and dropping a number of
total derivatives, we find
\begin{align}
  \label{eq:NLactionComp}
  e^{-1} \mathcal{L}_{} =&\,
	4\,\mathcal{D}^2 A'\,\mathcal{D}^2 \hat{\bar A}
	+ 8\,\mathcal{D}^a A'\, \big[\mathcal{R}_{ab}
	-\tfrac13 \mathcal R \,\eta_{ab}\big]\mathcal{D}^b \hat{\bar A}
	+ C'\,\hat{\bar C}
	\nonumber \\[.1ex]
	&\,
	- \mathcal{D}^\mu B'_{ij} \,\mathcal{D}_\mu \hat B^{ij}
	+ (\tfrac16 \mathcal{R} +2\,D) \, B'_{ij} \hat B^{ij}
	\nonumber\\[.1ex]
	&\,
	- \big[\varepsilon^{ik}\,B'_{ij} \,\hat F^{+\mu\nu} \,
		R(\mathcal{V})_{\mu\nu}{}^{j}{}_{k}
		+\varepsilon_{ik}\,\hat B^{ij}\,F'^{-\mu\nu} R(\mathcal{V})_{\mu\nu j}{}^k \big]
	\nonumber\\[.1ex]
	&\,
	-8\, D\, \mathcal{D}^\mu A'\, \mathcal{D}_\mu \hat{\bar A}
	+ \big(8\, \mathrm{i}\, R(A)_{\mu\nu}
		+ 2\, T_\mu{}^{cij}\, T_{\nu cij}\big)
		\mathcal{D}^\mu A' \,\mathcal{D}^\nu \hat{\bar A}  \nonumber\\[.1ex]
	&\,
	-\big[\varepsilon^{ij} \mathcal{D}^\mu T_{bc ij}
		\mathcal{D}_\mu A' \,\hat F^{+bc}
		+ \varepsilon_{ij} \mathcal{D}^\mu T_{bc}{}^{ij}
		\mathcal{D}_\mu \hat{\bar A}\,F'^{-bc}\big] \nonumber\\[.1ex]
	&\,
	-4\big[\varepsilon^{ij} T^{\mu b}{}_{ij}\,\mathcal{D}_\mu A'
		\,\mathcal{D}^c \hat F^{+}_{cb}
	+ \varepsilon_{ij} T^{\mu bij}\,\mathcal{D}_\mu \hat{\bar A}
        \,\mathcal{D}^c F'^{-}_{cb}\big] 
	\nonumber\\[.1ex]
	&\,
	+ 8\, \mathcal{D}_a F'^{-ab}\, \mathcal{D}^c \hat F^+_{cb}
	+ 4\,F'^{-ac}\, \hat F^+_{bc}\, \mathcal R_a{}^b
	+\tfrac1{4} T_{ab}{}^{ij} \,T_{cdij} F'^{-ab} \hat F^{+cd} 
	\nonumber\\[.1ex]
	&\,
	+w\,\Big\{ - \tfrac{2}{3}  \mathcal{D}^a A' \,\mathcal{D}_a
        \mathcal{R} + 4  \mathcal{D}^a A'\, \mathcal{D}_a D 
	-  T^{acij} T_{bc ij}\, \mathcal{D}^b \mathcal{D}_a A'
	\nonumber\\[.1ex]
	&\quad \qquad
	- 2 \mathcal{D}^a F_{ab}'^- \,\mathcal{D}_c T^{cb}{}^{ij} \varepsilon_{ij}
	+ \mathrm{i}\,  F'^{-ab} R(A)_{ad}^- \,T_b{}^{dij} \varepsilon_{ij}
	+ F_{ab}^- T^{ab ij} \varepsilon_{ij} (\tfrac{1}{12} \mathcal{R} - \tfrac{1}{2} D)
	\nonumber\\[.1ex]
	&\quad\qquad
	+ A' \,\big[\tfrac{2}{3} \mathcal{R}^2 - 2\, \mathcal{R}^{ab} \mathcal{R}_{ab} - 6\, D^2
		+ 2 \, R(A)^{ab} R(A)_{ab} -  R(\cV)^{+ab}{}^i{}_j\, R(\cV)^+_{ab}{}^j{}_i
		\nonumber \\
                & \quad \qquad\qquad
		+ \tfrac{1}{128}  T^{ab ij} T_{ab}{}^{kl} T^{cd}{}_{ij} T_{cd kl}
		+  T^{ac ij} D_a D^b T_{bc ij}\big]\Big\} \,~.
\end{align}
The above Lagrangian is the central result of this paper
and can be used to construct a large variety of
invariants in the same way as has been done in \cite{deWit:2010za}.
Three brief comments should be made about it.
First, in the limit $w=0$, we recover exactly (4.2) of
\cite{deWit:2010za}.
Second, the $w$-terms appear not only explicitly in the
final four lines of \eqref{eq:NLactionComp} but also implicitly
within the covariant derivatives of $\hat{\bar A}$, as we have already
stressed earlier.
Finally, we argued in section \ref{sec:curved-kinetic} that if $\Phi'$ is set
to a constant, then the action cannot actually depend on the
components of $\ln\bar\Phi$.  This is apparent in
\eqref{eq:NLactionComp} by inspection: only the last two lines survive
in this limit and they depend on the conformal supergravity fields
alone. We note in particular the appearance of the non-conformal part
of the Gauss-Bonnet invariant involving $\tfrac{2}{3} \mathcal{R}^2 -
2 \mathcal{R}^{ab} \mathcal{R}_{ab}$. This confirms our conjecture
that the kinetic multiplet based upon $\ln\bar\Phi$ can be used to
generate the $N=2$ Gauss-Bonnet invariant. This will be the topic of
the next section.

\section{The $N=2$ Gauss-Bonnet invariant in and out of superspace}
\label{sec:GaussBonnet}
\setcounter{equation}{0} 
We now have all of the building blocks necessary to construct the $N=2$
Gauss-Bonnet invariant. Based on our discussion in section \ref{sec:ext-chiral-super},
we were led to postulate the action
\begin{align}\label{eq:N2GB}
S_\chi^- = S_{\rm W}^- + S_{\rm NL}^-
	= - \int \rd^4x\, \rd^4\theta\, \mathcal{E}\, \Big(
		W^{\alpha \beta} W_{\alpha \beta} 
		+ w^{-1} \mathbb T(\ln\bar\Phi)
	\Big)
	= \int \rd^4x\, \Big(\mathcal{L}_{\rm W}^- + \mathcal{L}_{\rm NL}^-\Big)
\end{align}
as the $N=2$ supersymmetric Gauss-Bonnet, based mainly on the form its
component action took in the linearized limit.
Using the results of section \ref{sec:component-kinetic-multiplet}, we
can verify explicitly that its component Lagrangian contains the
combination \eqref{eq:Euler} of curvature-squared terms.
However, the full $N=2$ Gauss-Bonnet must not only include this combination,
but must also be a topological quantity.

We will establish its topological nature in the next two sections
using two complementary methods. First, we will analyze its
component structure, keeping only the bosonic terms, and show
that it indeed reduces to a topological quantity. In principle,
this should be sufficient as it is unlikely that the fermionic
terms would not be a topological invariant if the bosonic terms are.
However, in order to eliminate this possibility, we will subsequently present
a superspace argument which encompasses all terms.

Afterwards, we will comment briefly on an alternative way of formulating
the Gauss-Bonnet in superspace which sheds further light on some of its
features.

\subsection{The $N=2$ Gauss-Bonnet in components}
In section \ref{sec:component-kinetic-multiplet}, we provided the explicit
expressions for the various components of the kinetic multiplet $\mathbb T(\ln\bar\Phi)$.
It is straightforward to put them together to construct the component action
\eqref{eq:N2GB}. To keep the calculation concise, we will again neglect all fermionic terms.

We begin with the density formula for the kinetic multiplet,
\begin{align}
-2 \int \rd^4x\, \rd^4\theta\, \mathcal{E}\, \mathbb T(\ln\bar\Phi) &= \int \rd^4x\, e\, \Big(
	C\vert_{\mathbb T(\ln\bar\Phi)}
   - \tfrac{1}{16} ( T_{ab\,ij} \varepsilon^{ij})^2\, A\vert_{\mathbb T(\ln\bar\Phi)}\Big)
	\equiv 2 w \int \rd^4x \, \mathcal{L}_{\rm NL}^-
\end{align}
where $C\vert_{\mathbb T(\ln\bar\Phi)}$ and $A\vert_{\mathbb T(\ln\bar\Phi)}$ are given
in \eqref{eq:T-components}. We have already discussed how the dependence on the fields
of the anti-chiral multiplet must be limited to total derivative terms, but we would
like to explicitly check this.
Making use of \eqref{eq:Cdegauged}, we easily find
\begin{align}
    \label{eq:LTexpression}
    2 w \,e^{-1} \mathcal{L}_{\rm NL}^- &= \mathcal{D}_a V^a
	- 2 w \mathcal{R}^{ab} \mathcal{R}_{ab} + \tfrac{2}{3} w \mathcal{R}^2
	- 6 w D^2
	\eol & \quad
	+ 2 w R(A)^{ab} R(A)_{ab}
	- w R(\cV)_{ab}^+{}^i{}_j R(\cV)^{ab+}{}^j{}_i
	\eol & \quad
	+ \tfrac{1}{128} w T^{ab ij} T_{ab}{}^{kl} T^{cd}{}_{ij}T_{cd kl}
	+ w T^{ac ij} D_a D^b T_{bc ij}
\end{align}
where the components of the multiplet $\ln\bar\Phi$ are confined to
the covariant term $V^a$ given in \eqref{eq:Vexpression}.

The well-known conformal supergravity invariant constructed from the square
of the superconformal Weyl tensor is
\begin{align}\label{eq:CsgAction}
e^{-1} \mathcal{L}_{\rm W}^-
	&= \tfrac{1}{2} C^{abcd} C_{abcd} - \tfrac{1}{2} C^{abcd} \tilde C_{abcd}
	- 2 R(A)_{ab}^- R(A)^{ab-}
	+ \tfrac{1}{2} R(\cV)_{ab}^-{}^i{}_j R(\cV)^{ab-}{}^j{}_i
	\eol * & \quad
	+ 3 D^2 
	- \tfrac{1}{2} T^{ac ij} D_a D^b T_{bcij}
	- \tfrac{1}{256} T^{ab ij} T_{ab}{}^{kl} T^{cd}{}_{ij}T_{cd kl}~.
\end{align}
Combining the expressions \eqref{eq:LTexpression} and \eqref{eq:CsgAction} with
the appropriate coefficients leads to
\begin{align}
  e^{-1} \mathcal{L}_\chi^- = e^{-1} \mathcal{L}_{\rm W}^- + e^{-1}
  \mathcal{L}_{\rm NL}^- 
	&=  \tfrac{1}{2} C^{abcd} C_{abcd} - \mathcal{R}^{ab}
        \mathcal{R}_{ab} + \tfrac{1}{3} \mathcal{R}^2
	- \tfrac{1}{2} C^{abcd} \tilde C_{abcd}
	\eol & \quad
	+ R(A)_{ab} \tilde R(A)^{ab}
	- \tfrac{1}{2} R(\cV)_{ab}{}^i{}_j \tilde R(\cV)^{ab}{}^j{}_i
	+ \tfrac{1}{2} w^{-1} \mathcal{D}_a V^a~.
\end{align}
As required, $\mathcal{L}_\chi^-$ is a topological invariant. It
involves respectively the Euler density, the Pontryagin density, the
SU(2) and U(1) topological invariants, and an explicit total covariant
derivative. It is interesting (although perhaps coincidental) that the
specific combination of U(1) and SU(2) curvatures appearing in the
above expression can be rewritten purely in terms of the U(2)
curvature.

\subsection{The $N=2$ Gauss-Bonnet is topological in superspace}
Next, we give a purely superspace argument that the action \eqref{eq:N2GB}
is topological -- that is, it is independent (up to a total derivative)
of the choice of $\bar\Phi$ and of the fields of conformal supergravity.
We have already shown it is independent of the components of $\bar\Phi$
via a simple argument in section \ref{sec:curved-kinetic}.
Proving invariance under the supergravity fields is much more involved.
In principle, the superspace connections depend in a very complicated way
on the $N=2$ conformal supergravity prepotential, which is a real
scalar superfield $H$.\footnote{The references \cite{RivellesTaylor, HST} 
showed that the linearized $\cN=2$ Weyl multiplet can be described by a real unconstrained 
prepotential $H$, in agreement with the supercurrent analysis of \cite{Sohnius:1978pk}.
The origin of such a  prepotential in the harmonic superspace
approach to $\cN=2$ supergravity (see \cite{GIOS} and references therein)
was revealed in \cite{Siegel-curved} at the linearized level,
and in \cite{KT} at the fully nonlinear level.} Then applying a small deformation
$\delta H$ to the prepotential, the action shifts to first order,
$S \rightarrow S + \delta S$, where
\begin{align}\label{eq:deltaS}
\delta S = \int \rd^4x\, \rd^4\q\, \rd^4\bar\q\, E\, \delta H \frac{\delta S}{\delta H}~.
\end{align}
The quantity $\delta S / \delta H$ is
the supercurrent multiplet provided $\delta H$ is defined correctly;
we will elaborate on this shortly.
If the action is topological, then $\delta S / \delta H = 0$.
Our goal will be to prove this last condition for the Gauss-Bonnet invariant.

To make these manipulations a bit more concrete, we consider first the
second order Weyl action $S_{\rm W}^-$ given by the space-time integral of
\eqref{eq:flatW2} involving the linearized super-Weyl
tensor $W_{\alpha\beta}$. This superfield is given in terms of the
prepotential $H$ as
\begin{align}
W_{\alpha \beta} = \bar D^4 D_{\alpha\beta} H~,
\end{align}
which satisfies the Bianchi identity
$D^{\alpha \beta} W_{\alpha \beta} = \bar D_{\dalpha \dbeta} \bar W^{\dalpha \dbeta}$.
The prepotential $H$ contains the linearized connections and covariant fields
of the Weyl multiplet. Applying a small deformation $H \rightarrow H + \delta H$,
one finds the action changes by
\begin{align}\label{eq:varyCSGlin}
\delta S_{\rm W}^{-}
	= - 2 \int \rd^4x\, \rd^4\theta\, \rd^4\bar\theta\, \delta H \,D^{\alpha \beta} W_{\alpha\beta}~.
\end{align}
The quantity $\cJ = -2 D^{\alpha \beta} W_{\alpha \beta}$ is the (linearized)
$N=2$ supercurrent for this action. One can check that it satisfies the
constraint \cite{Sohnius:1978pk}
\begin{align}\label{eq:DJ}
D_{ij} \cJ = \bar D^{ij} \cJ = 0~,
\end{align}
which is a consequence of the fact that $H$ is defined only up to the gauge
transformation \cite{KT, BK:N2sc}
\begin{align}\label{eq:deltaH}
\delta_\Omega H = \tfrac{1}{12} D_{ij} \Omega^{ij} + \tfrac{1}{12} \bar D^{ij} \bar \Omega_{ij}
\end{align}
for an unconstrained complex superfield $\Omega^{ij}$.

These manipulations were rather simple because of the linearized nature
of the super-Weyl tensor. In a generic curved background, there will be some elaborations. 
For instance, because $H$ is a prepotential, it generically appears non-polynomially
in the definitions of the connections and the curvature $W_{\alpha\beta}$,
and so there is some ambiguity in how one should define its variation.
Nevertheless, one expects that just as one can introduce small covariant deformations
to the component fields,
\begin{align}\label{eq:varyConnections}
\delta e_\mu{}^a = e_\mu{}^b h_b{}^a~, \qquad
\delta \psi_\mu{}^{\alpha i} = e_\mu{}^b \varphi_b{}^{\alpha i}~, \qquad \textrm{etc.},
\end{align}
it should be possible to introduce a similar small covariant deformation
$\cH$ to the prepotential.\footnote{For an
  extensive pedagogical discussion of this procedure for $N=1$ supergravity, we
  refer the reader to the standard textbook references
  \cite{GGRS,Buchbinder:1998qv}.
  The generalization to a manifestly
  superconformal setting was obtained in \cite{Butter:BFF}. There
  seems to be no particular obstruction to implementing an analogous
  procedure for $N=2$ conformal supergravity, but this has not yet
  been done.}
Here the key idea is that one is deforming around
an \emph{arbitrary curved} background.
The corresponding variation of the action would be
\begin{align}
\delta S = \int \rd^4 x\,\rd^4\theta\, \rd^4\bar\theta\, E\, \cH \cJ
\end{align}
where the deformation $\cH$ and the supercurrent $\cJ$
are both covariant conformal primary superfields, generalizing our
previous formula \eqref{eq:deltaS}. At the component level, this formula
would simply amount to
\begin{align}\label{eq:varyScomp}
\delta S = \int \rd^4 x\, e\, \Big(
	h^{ba} T_{ba} + \varphi^{b\alpha i} J_{b \alpha i} + \cdots
	+ \delta D J_D
	\Big)
\end{align}
where $T_{ba}$ is the stress-energy tensor, $J_{b\alpha i}$ is the
supersymmetry current, and so on up through $J_D$, which is the
variation of the action with respect to the field $D$.
By comparing with the linearized case, we can deduce that $\cH$
must have Weyl weight $w=-2$ and so $\cJ$ must be weight $w=2$;
it follows that the variation $\delta D$ appears only in the highest component of $\cH$,
and so $\cJ \vert_{\theta=0} = -\tfrac{1}{4} J_D$, with the normalization
given by matching to the linearized case.
A gauge transformation of the component fields corresponds
to a superfield gauge transformation\footnote{In the $\rm SU(2)$ superspace formulation of conformal
supergravity \cite{KLRT},
the gauge transformation \eqref{eq:deltaHCurv} coincides
with the transformation given in \cite{BK:AdS}.}
\begin{align}\label{eq:deltaHCurv}
\delta_\Omega \cH = \tfrac{1}{12} \nabla_{ij} \Omega^{ij} + \tfrac{1}{12} \bar \nabla^{ij} \bar \Omega_{ij}~,
\end{align}
which is the curved generalization of \eqref{eq:deltaH}.
One can check that this respects the S-supersymmetry invariance
of $\cH$ provided $\Omega^{ij}$ has $w=-3$ and $c=-1$.
For this choice, the variation of the action is zero, $\delta S = 0$,
so it follows that the supercurrent $\cJ$ must obey the current conservation equations
\begin{align}\label{eq:Jconseqn}
\nabla_{ij} \cJ = \bar\nabla^{ij} \cJ = 0~,
\end{align}
which are the curved generalizations of \eqref{eq:DJ}.
These conditions are invariant under S-supersymmetry precisely when $\cJ$ has $w=2$.

Now let us return to the case of interest. The naive covariantization of
\eqref{eq:varyCSGlin} is
\begin{align}
\delta S_{\rm W}^-
	= -2 \int \rd^4x\, \rd^4\theta\, \rd^4\bar\theta\, E\, \cH \,\nabla^{\alpha \beta} W_{\alpha\beta}
\end{align}
and so $\cJ_{\rm W} = -2 \nabla^{\alpha\beta} W_{\alpha\beta}$.
In principle there could be additional covariant corrections on the right-hand side,
but it is easy to see that no such corrections exist. The $N=2$ supercurrent
must be a real conformal primary $w=2$ superfield,
and the \emph{unique} such superfield one may construct in conformal supergravity is
$\nabla^{\alpha\beta} W_{\alpha\beta}$.\footnote{This statement is a little too
strong. In principle, one could have terms like
$(\nabla^{\gamma \delta} W_{\gamma\delta})^2 / |W^{\alpha \beta} W_{\alpha\beta}|$.
The correct statement is
that so long as our component action has a regular Minkowski limit, we expect that
the supercurrent should also have a regular Minkowski limit, and so we may exclude such terms.}
Moreover, $\cJ_{\rm W}$ must also
obey the constraint \eqref{eq:Jconseqn}, which one can check is indeed satisfied
for $\cJ_{\rm W} \propto \nabla^{\alpha\beta} W_{\alpha\beta}$.

Remarkably, we may now apply the same argument to the variation of the
kinetic multiplet action $S_{\rm NL}^-$. Taking
\begin{align}
\delta S_{\rm NL}^-
	= \int \rd^4x\, \rd^4\theta\, \rd^4\bar\theta\, E\, \cH \,\cJ_{\rm NL}~,
\end{align}
we observe that $\cJ_{\rm NL}$ \emph{cannot} depend on $\ln\bar\Phi$
since the original action does not actually depend on it. Thus,
$\cJ_{\rm NL}$ can only depend on the conformal supergravity fields.
But we have just argued that this leaves only one option:
$\cJ_{\rm NL} \propto \nabla^{\alpha\beta} W_{\alpha\beta}$.
This means that there must be \emph{some} combination of
$\mathbb T(\ln\bar\Phi)$ and $W^{\alpha\beta} W_{\alpha\beta}$ that is topological.
As we already know that the combination
$W^{\alpha \beta} W_{\alpha \beta} + w^{-1} \mathbb T(\ln\bar\Phi)$
yields a topological action if we turn off all fields except the vierbein
and the lowest component of $\ln\bar\Phi$, we must conclude that
\begin{align}
\cJ_{\rm NL} = 2 \nabla^{\alpha\beta} W_{\alpha\beta}~.
\end{align}
It follows that
\begin{align}
\delta S_\chi^- = \delta S_{\rm W}^- + \delta S_{\rm NL}^- = 0~,
\end{align}
and therefore this combination is indeed topological for a generic supergravity background.

\subsection{The $N=2$ Gauss-Bonnet in an alternative superspace}
\label{sec:GaussBonnetc}
We close this section by elaborating upon alternative formulations of
the $N=2$ Gauss-Bonnet in superspace. The formulation in \eqref{eq:N2GB}
is very close in spirit
to the component formulation \eqref{eq:Lchi'a} constructed from conformal gravity
coupled to a scalar field. A natural question to ask is what
the superspace analog of \eqref{eq:Lchi'b} should be, where the
scalar field has been gauge-fixed to unity and conformal gravity reduced
to Poincar\'e gravity.

This question is naturally addressed in the superspace formulation for $N=2$ 
conformal supergravity given in \cite{KLRT} 
where only the Lorentz and SU(2) transformations are 
explicitly gauged, while the remaining local superconformal symmetries are realized as super-Weyl transformations.\footnote{This formulation makes use of the superspace geometry originally proposed in \cite{Grimm} without any connection with conformal supergravity.}
We refer to this conformal supergravity formulation 
as SU(2) superspace. 
The covariant superspace derivatives are given by
\begin{align}
E_M{}^A \mathcal{D}_A = \pa_M - \tfrac{1}{2} \Omega_M{}^{ab} M_{ab} -
\tfrac{1}{2} \cV_M{}^i{}_j I^j{}_i 
\end{align}
and the algebra of superspace covariant derivatives depends not
only on the superfield $W_{\alpha\beta}$, which contains the Weyl
multiplet, but also on additional torsion superfields $S_{ij}$ and
$Y_{\alpha\beta}$, which are both complex and symmetric in their indices,
as well as the real superfield $G_a$. The latter torsion superfields
give direct access to the Ricci tensor (as opposed to merely the Weyl
tensor), which is an advantage of using this formulation as opposed
to conformal superspace.

It turns out there is a straightforward mapping between conformal and SU(2)
superspace, which can be accomplished by adopting the K- and S-gauge $B_M=0$
and extracting the U(1), K- and S-connections from the covariant
derivative.\footnote{More precisely, one recovers U(2) superspace \cite{Howe}
(see also \cite{KLRT-M2})
in this manner, which can be further reduced to SU(2) superspace by an
additional super-Weyl gauge-fixing \cite{KLRT-M2}.}
These turn out to contain the multiplet associated with the
Ricci tensor. Just as adopting the gauge $b_\mu=0$ in conformal gravity allows
the decomposition
\begin{align}\label{eq:BoxBoxphiDG}
\Box_{\rm c} \Box_{\rm c} \ln\phi
	&= \cD^2 \cD^2 \ln \phi
	+ \mathcal{D}^a \Big(
		\tfrac{2}{3} \mathcal{R} \mathcal{D}_a \ln\phi - 2 \mathcal{R}_{ab} \mathcal{D}^b \ln\phi\Big)
	\eol & \qquad
	- \tfrac{1}{2} w\mathcal{R}^{ab}\mathcal{R}_{ab} + \tfrac{1}{6} w
        \mathcal{R}^2 + \tfrac{1}{6} w \cD^2 \mathcal{R}~, 
\end{align}
performing the same procedure in superspace allows
\begin{align}\label{eq:NLKinSU2}
\bar\nabla^4 \ln\bar\Phi &=
	\bar\Delta \ln\bar\Phi
	- \tfrac{1}{2} w \mathbb T_0
\end{align}
where
\begin{align}
  \label{eq:SU2Delta}
  \bar\Delta \ln \bar\Phi := \tfrac{1}{96} \varepsilon_{ik}
  \varepsilon_{jl} \bar {\mathcal{D}}^{ij} \bar{\mathcal{D}}^{kl} \ln
  \bar\Phi - \tfrac{1}{96} \bar{\mathcal{D}}_{\dalpha \dbeta} \bar
  {\mathcal{D}}^{\dalpha \dbeta} \ln \bar\Phi + \tfrac{1}{6}
  \varepsilon_{ik} \varepsilon_{jl}\bar S^{ij} \bar{\mathcal{D}}^{kl}
  \ln \bar\Phi + \tfrac{1}{6} \bar Y_{\dalpha \dbeta} \bar
  {\mathcal{D}}^{\dalpha \dbeta} \ln \bar\Phi
\end{align}
is the chiral projection operator of SU(2) superspace \cite{Muller:1988ux, KT-M:DReps}
and
\begin{align}
\mathbb T_0 := - \bar Y_{\dalpha \dbeta} \bar Y^{\dalpha \dbeta}
	- \varepsilon_{ik} \varepsilon_{jl} \bar S^{ij} \bar S^{kl}
	-\tfrac{1}{6} \varepsilon_{ik} \varepsilon_{jl} \bar{\mathcal{D}}^{ij} \bar S^{kl}
\end{align}
is a combination of torsion superfields which is independent of
$\bar\Phi$. The term $\bar\Delta \ln\bar\Phi$ of \eqref{eq:NLKinSU2}
corresponds to the first line of \eqref{eq:BoxBoxphiDG} while the
second term involving $\mathbb T_0$ corresponds
to the three $w$-dependent curvature terms.
Moreover, the combination $\mathbb T_0$ is actually chiral
since both $\bar\nabla^4\ln\bar\Phi$ and $\bar\Delta\ln\bar\Phi$ are
chiral in SU(2) superspace. This is quite a non-trivial statement
since none of the individual terms are chiral, nor can the expression be
written as the chiral projection of some covariant term. In other
words, $\mathbb T_0$ is an \emph{additional} non-trivial chiral
invariant in SU(2) superspace, which contains the second line of
\eqref{eq:BoxBoxphiDG} as its highest component.

The analogy we have drawn between \eqref{eq:BoxBoxphiDG} and \eqref{eq:NLKinSU2}
is not superficial. In the component expression \eqref{eq:BoxBoxphiDG},
the first line possesses an inhomogeneous contribution under a Weyl transformation
which is precisely balanced by the second line. The same property holds for
\eqref{eq:NLKinSU2}. Using the super-Weyl transformation introduced in \cite{KLRT},
one finds
\begin{align}
  \delta_\S \bar\Delta \ln \bar\Phi = 2 \S \bar\Delta \ln \bar\Phi + w \bar\Delta
  \bar \S~, \qquad 
  \delta_\S \mathbb T_0 = 2 \S \mathbb T_0 + 2 \bar\Delta \bar \S~,
\end{align}
with chiral parameter $\Sigma$.
It follows that $\delta_\S \bar\nabla^4 \ln\bar\Phi = 2
\,\S \,\bar\nabla^4 \ln\bar\Phi$.

Let us now consider the action for the kinetic multiplet in SU(2)
superspace, where it becomes 
\begin{align}
    \label{eq:NLKinSU2action}
  \int \rd^4x\, \rd^4\theta\, \mathcal{E}\, \Phi' \mathbb T(\ln\bar\Phi)
	&= -2 \int \rd^4x\, \rd^4\theta\, \mathcal{E}\, \Phi' \bar\nabla^4 \ln\bar\Phi \nonumber \\
	&= -2 \int \rd^4x\, \rd^4\theta\, \rd^4\bar\theta\, E\, \Phi' \ln\bar\Phi
	+ w \int \rd^4x\, \rd^4\theta\, \mathcal{E}\, \Phi' \mathbb T_0
\end{align}
after using the chiral projection operator $\bar\Delta$ to rewrite a
chiral superspace integral in terms of a full superspace integral. It
is easy to see that the part of the action involving $\bar\Phi$
vanishes when $\Phi'$ is a constant. The pure curvature contributions
to the Gauss-Bonnet are isolated in the remaining term $\mathbb T_0$,
which is the \emph{intrinsic} part of the kinetic multiplet and is
explicitly independent of the components of $\bar\Phi$. In fact, there
is no obstruction to performing a super-Weyl transformation to
explicitly fix $\bar\Phi$ to a constant; then its contribution to the
action vanishes completely.

Just as we proposed the action
\begin{align}
  \label{eq:GammaCSG}
  S_\chi^- &= \int \rd^4x\, \rd^4\theta\, \mathcal{E} \,\Big(
	- W^{\alpha \beta} W_{\alpha \beta}
	+ 2 w^{-1} \bar\nabla^4 \ln\bar\Phi \Big)
  \nonumber \\ 
	&= \int \rd^4x\, e\, \Big(
		\tfrac{1}{2} C^{abcd} C_{abcd} - \tfrac{1}{2} C^{abcd} \tilde C_{abcd}
		+  2 w^{-1} \Box_{\rm c} \Box_{\rm c} \bar A
		+ \cdots	\Big)
\end{align}
as the Gauss-Bonnet in conformal superspace, we can similarly now exhibit the
Gauss-Bonnet in SU(2) superspace as
\begin{align}
    \label{eq:GammaSU2}
    S_\chi^- &= -\int \rd^4x\, \rd^4\theta\, \mathcal{E} \,\Big(
	W^{\alpha \beta} W_{\alpha \beta} +  \mathbb T_0\Big) \nonumber \\
    &= \int \rd^4x\, \rd^4\theta\, \mathcal{E} \,\Big(
	- W^{\alpha \beta} W_{\alpha \beta}
    + \bar Y_{\dalpha \dbeta} \bar Y^{\dalpha \dbeta}
    + \varepsilon_{ik} \varepsilon_{jl} \bar S^{ij} \bar S^{kl}
    + \tfrac{1}{6} \varepsilon_{ik} \varepsilon_{jl} \bar{\mathcal{D}}^{ij} \bar S^{kl}
    \Big) \nonumber \\
    &= \int \rd^4x\, e\, \Big(
    \tfrac{1}{2} C^{abcd} C_{abcd} - \tfrac{1}{2} C^{abcd} \tilde C_{abcd}
		- \,\mathcal{R}^{ab}\mathcal{R}_{ab}
		+ \tfrac{1}{3} \mathcal{R}^2
		+ \tfrac{1}{3} \cD^2 \mathcal{R} + \cdots
	\Big)~.
\end{align}
These two actions correspond respectively to the supersymmetric versions
of the actions \eqref{eq:Lchi'a} and \eqref{eq:Lchi'b}.
In both cases, the details of the elided terms can be reconstructed using
the explicit results for the kinetic multiplet. We further observe that
because the imaginary part of the chiral superspace integral of
$W^{\alpha\beta} W_{\alpha\beta}$ is a total derivative, the supersymmetric
Pontryagin term, the same must hold for $\mathbb T_0$.

It is actually possible to cast the conformal superspace action 
into the same form as \eqref{eq:NLKinSU2action}. Suppose we have some
chiral multiplet $\Phi_0$  which is nowhere vanishing. For simplicity,
let us take its weight to be $w_0=1$, although any nonzero weight will
suffice. It is trivial to rewrite
\begin{align}
\mathbb T(\ln \bar \Phi) = \mathbb T(\ln (\bar \Phi / (\bar\Phi_0)^w)) +
w \mathbb T(\ln\bar\Phi_0) 
\end{align}
where $\mathbb T(\ln (\bar \Phi / (\bar\Phi_0)^w))$ is a usual kinetic multiplet
of a weight zero anti-chiral superfield and the non-linear behavior has
been isolated within
$\mathbb T_0 := \mathbb T(\ln\bar\Phi_0)$.
It follows that
\begin{align}
\int \rd^4x\, \rd^4\theta\, \mathcal{E}\, \Phi' \mathbb T(\ln\bar\Phi)
	&= -2 \int \rd^4x\, \rd^4\theta\, \rd^4\bar\theta\, E\, \Phi' \ln (\bar\Phi / (\bar\Phi_0)^w)
	+ w \int \rd^4x\, \rd^4\theta\, \mathcal{E}\, \Phi' \mathbb T_0~.
\end{align}
If one then chooses to work in the gauge where $\bar\Phi_0 = 1$, one
recovers \eqref{eq:NLKinSU2action}. Note however that the action is
actually independent of the choice of $\bar\Phi_0$.

\section{Summary and conclusions}
\label{sec:summary-conclusions}
\setcounter{equation}{0}
The main goal of this paper was to establish the existence of a new
class of higher-derivative $N=2$ supersymmetric invariants based on
the non-linear extension of the kinetic multiplet. Now that we have
obtained its explicit form, we can address in more detail the two issues
mentioned in the introduction.

In the recent paper \cite{Banerjee:2011ts}, the $5D$ mixed
gauge-gravitational Chern-Simons term constructed originally in
\cite{Hanaki:2006pj} was reduced to four dimensions.  The resulting
$4D$ Lagrangian, denoted $\mathcal{L}_{\rm vww}$, could not be
completely classified in terms of known supersymmetric invariants.  In
particular, there appeared to be three sets of terms.  The first set
was easily identified as arising from a known chiral invariant based
on a holomorphic function; the second and third sets were more
puzzling.  One seemed to belong to the class based on the kinetic
multiplet that had already been constructed in \cite{deWit:2010za},
corresponding, as discussed in section \ref{sec:curved-kinetic}, to a
full superspace integral of a K\"ahler potential
\begin{align}\label{eq:nlkH}
\mathcal{H} \propto \mathrm{i} \,c_A (t^A \ln \bar \cX^0 - \bar t^A \ln \cX^0)~,
\end{align}
where the coefficients $c_A$ were real constants, the field $\cX^0$
was the Kaluza-Klein vector multiplet, and the fields $t^A = \cX^A /
\cX^0$ were the ratio of vector multiplets.  The other set of terms
involved curvature bilinears such as $c_A t^A \,\mathcal{R}^{ab}
\mathcal{R}_{ab}$ and $c_A t^A\, R(\mathcal{V})_{ab}^+{}^i{}_j
R(\mathcal{V})^{+ab}{}^j{}_i$.

Based on the results of this paper, it has become clear to us that
the second and third sets of terms actually arise from a single
invariant based on the non-linear version of the kinetic multiplet.
The key point is that the higher-derivative Lagrangian constructed in
\cite{deWit:2010za} depended on a K\"ahler metric $\cH_{I \bar J}$
with the additional homogeneity condition $\cX^I \cH_{I \bar J} = 0$.
The proposed function \eqref{eq:nlkH} does not obey this condition;
however, it seems that one can
relax slightly the homogeneity condition and
``patch up'' the component Lagrangian by including certain curvature-squared
combinations, such as $c_A t^A \,\mathcal{R}^{ab} \mathcal{R}_{ab}$,
exactly of the sort found in \cite{Banerjee:2011ts}.
The resulting higher-derivative action arises not by using the full superspace action
of \cite{deWit:2010za}, but rather the non-linear kinetic multiplet action
\begin{align}\label{eq:nlkAction}
\mathrm{i} \,c_A \int \rd^4x\, \rd^4\q\, \cE\, t^A \,\mathbb T(\ln\bar \cX^0) + \textrm{h.c.}
\end{align}
constructed in this paper.
This single action appears to contain the second and third sets of
terms identified in \cite{Banerjee:2011ts}.
In other words, it seems that the $4D$ Lagrangian $\cL_{\rm vww}$ contains
only two supersymmetric invariants: one based on a holomorphic function
of chiral multiplets and the other based on \eqref{eq:nlkAction}.
At the present time, we cannot be more definitive, as the analysis of
\cite{Banerjee:2011ts} was based on a few characteristic terms only,
with the goal of reconstructing what the $4D$ invariant should be.
Now that we are confident in our identification of these
terms, we plan to revisit the analysis of \cite{Banerjee:2011ts} to
ensure full equality.

It is an interesting question whether the new $4D$ invariants we have
constructed also arise from reduction of other $5D$ invariants.
Recently, the dilaton-Weyl formulation of $5D$ conformal supergravity
has been used to construct all of the $5D$ $R^2$ invariants \cite{Ozkan:2013nwa},
in addition to the Gauss-Bonnet combination \cite{Ozkan:2013uk}, building on the
work of \cite{Bergshoeff:2011xn}.
It is probable that an off-shell dimensional reduction of these actions would produce 
$4D$ invariants equivalent to the ones under consideration, but such
explicit reductions have not yet been undertaken.

The second question has to do with black hole entropy.
Originally the first calculation of the entropy of BPS black holes involving higher-derivative
couplings was based on the supersymmetric extension of the square of
the Weyl tensor \cite{LopesCardoso:1998wt,LopesCardoso:2000qm}. More
precisely \eqref{eq:flatW2} was generalized to a holomorphic and
homogeneous function $F$ of weight two, depending on $W^2=
W_{\alpha\beta} W^{\alpha\beta}$ and the vector multiplets $\cX^I$, i.e.
\begin{equation}
  \label{eq:higher-order-Weyl}
  S \propto \int \mathrm{d}^4x\, \mathrm{d}^4\theta\; \mathcal{E}\,F(\cX^I, W^2) +
  \mathrm{h.c.}
\end{equation}
A somewhat surprising result was that the actual contribution from the
higher-derivative terms did not originate from the square of the Weyl
tensor, but from the terms $T^{ac ij} D_a D^b T_{bc ij}$ required by
supersymmetry.  Some time later, in a specific model
\cite{Sen:2005iz}, the entropy was calculated by replacing the square
of the Weyl tensor by the Gauss-Bonnet combination
\begin{align}
  \label{eq:SenReplacement}
  C^{abcd} C_{abcd} \Longrightarrow C^{abcd} C_{abcd} - 2
  \mathcal{R}^{ab} \mathcal{R}_{ab} + \tfrac{2}{3} \mathcal{R}^2~,
\end{align}
keeping the same coefficient in front of $C^2$ term.  Since the
supersymmetrization of the Gauss-Bonnet term was not known, no
additional terms were included. The surprising result was that this
pure Gauss-Bonnet coupling gave rise, at least in this model, to the
same result as \cite{LopesCardoso:1998wt,LopesCardoso:2000qm}.

With the results of this paper it is now straightforward to analyze
the reasons behind this unexpected match, which holds even when
including all the terms required by supersymmetry.  The relevant terms
in the supersymmetrization \eqref{eq:higher-order-Weyl} of the Weyl
tensor squared are
\begin{align}
  \label{eq:LwRel}
  e^{-1} A^\prime \,\mathcal{L}_{\rm W}^- \sim A^\prime \Big\{&
  \tfrac{1}{2} C^{abcd} C_{abcd} - 
  \tfrac{1}{2} C^{abcd} \tilde C_{abcd} \nonumber\\
  &- \tfrac{1}{2} T^{ac ij} D_a D^b T_{bcij} 
	- \tfrac{1}{256} T^{ab ij} T_{ab}{}^{kl} T^{cd}{}_{ij}T_{cd kl}\Big\}\,,
\end{align}
where $A^\prime$ denotes the scalar associated with the ratio of two
vector multiplets.  As already mentioned, the sole contribution to the
BPS black hole entropy in the original calculation came from the third
term above. The reason is that the Wald entropy follows in this
particular case from varying the action with respect to
$\mathcal{R}_{ab}{}^{cd}$ and subsequently restricting the background
to ensure that the near-horizon horizon is fully supersymmetric (for
further details we refer to \cite{LopesCardoso:1998wt,
  LopesCardoso:2000qm}).  In this near-horizon background both the
Weyl tensor and the Ricci scalar vanish, so that the term quadratic in
the Weyl tensor cannot give a contribution to the entropy. However, it
turns out that the square of the (conformally) covariant derivatives
acting on $T_{bcij}$ involve terms linear in the Ricci tensor, while
the tensor fields $T$ are non-vanishing so that this term determines
the entropy.

Let us now give the relevant terms in the non-linear kinetic
multiplet, which can be added to \eqref{eq:LwRel} to carry out the
replacement \eqref{eq:SenReplacement} in the fully supersymmetric
context, 
\begin{align}
  \label{eq:phi-NL-lagr}
  e^{-1} A^\prime\,\mathcal{L}_{\rm NL}^- &\sim A^\prime \Big\{-
  \mathcal{R}^{ab} \mathcal{R}_{ab} + \tfrac{1}{3} \mathcal{R}^2 +
  \tfrac{1}{2} T^{ac ij} D_a D^b T_{bc ij} + \tfrac{1}{256} T^{ab ij}
  T_{ab}{}^{kl} T^{cd}{}_{ij}T_{cd kl}\Big\}\,.
\end{align}
Here the first and the third term do both contribute to the entropy,
but as it turns out their contribution cancels in the near-horizon
geometry by virtue of the relation $\mathcal{R}_{ab} = -\tfrac{1}{8}
T_{a}{}^{cij} T_{bc ij}$. Hence it follows that the replacement
\eqref{eq:SenReplacement} \emph{at the fully supersymmetric level}
does not affect the result for the BPS black hole
entropy.\footnote{ 
  Similarly, $\cL_{\rm NL}^-$ contributes nothing to the electric
  charges of BPS black holes.} 
Moreover, the terms depending on the tensor fields cancel in the sum
of \eqref{eq:LwRel} and \eqref{eq:phi-NL-lagr}, so that in the
calculation based on the Gauss-Bonnet term the supersymmetric
completion will not contribute, just as indicated by the result of
\cite{Sen:2005iz}.  

In addition one may also consider the actual value of the two
invariants in the supersymmetric near-horizon background. This is the
reason why we also included the $T^4$ terms in \eqref{eq:LwRel} and
\eqref{eq:phi-NL-lagr}, as they are the only other terms that can
generate additional contributions to the action in the near-horizon
geometry. Working out this particular contribution, we find that
\eqref{eq:phi-NL-lagr} vanishes, and furthermore that the
$T$-dependent terms vanish in the sum of \eqref{eq:LwRel} and
\eqref{eq:phi-NL-lagr}. Hence the supersymmetric completion does not
contribute to the Gauss-Bonnet coupling, and the value of the action
will not change under the replacement \eqref{eq:SenReplacement} at the
fully supersymmetric level. We should add that this last result has a
bearing on the evaluation of the logarithmic corrections to the BPS
entropy in \cite{Sen:2011ba}. There the square of the Weyl tensor and
the Gauss-Bonnet invariant were equated and their contributions summed
without further information of the possible supersymmetric completion
of the coupling to a Gauss-Bonnet term. This was necessary in order to
obtain quantitative agreement when comparing two methods for calculating
the logarithmic corrections. Our above analysis thus confirms and
clarifies the earlier observations in
\cite{Sen:2005iz,Sen:2011ba}.

We have showed for this case that the non-linear version of the
kinetic multiplet vanishes at supersymmetric field configurations and
it does not contribute to the entropy of a BPS black hole.  A
more complete analysis, establishing the existence of a BPS
non-renormalization theorem in a more general Lagrangian, would
proceed along the same lines as in \cite{deWit:2010za}, which
established that Lagrangians involving the usual kinetic multiplet
$\mathbb T(\bar\Phi)$ will vanish for a supersymetric background and
also their first derivative with respect to fields or parameters will
vanish in such a background. The latter would imply in particular that
they cannot contribute to the BPS black hole entropy or to the
electric charges. The proof was based on the fact that weight-zero
chiral superfields must be proportional to a constant in the
supersymmetric limit. For the non-linear version of the kinetic
multiplet $\mathbb T(\ln\bar\Phi)$ considered here, there is a marked
difference because $\Phi$ is a chiral multiplet of non-zero
weight. Its supersymmetric value is therefore not necessarily
proportional to a constant, which makes the corresponding BPS analysis
significantly more involved, with constraints imposed on the
supergravity background as well as the chiral multiplet itself.  We
intend to give a more thorough analysis of these features in the near
future.

\section*{Acknowledgements}
This work is supported in part by the ERC Advanced
Grant no. 246974, {\it ``Supersymmetry: a window to non-perturbative physics''}.
The work of DB and SMK   was supported in part by the Australian Research Council,
project No. DP1096372.  

\begin{appendix}
%
\section{Notations and conventions}
\label{App:NC}
\setcounter{equation}{0}
In this paper, we have used in parallel both superspace, which is
conventionally written in two-component notation, and multiplet
calculus, which is usually carried out in four-component notation. To
aid the reader in translating any given formula between the two
notations, we summarize our conventions for both.

Space-time indices are denoted $\mu,\nu,\ldots$, Lorentz indices are denoted
$a,b,\ldots$, and SU(2) indices are denoted $i,j,\ldots$.
The Lorentz metric is $\eta_{ab} = \textrm{diag}(-1,1,1,1)$ and the
antisymmetric tensor $\veps_{abcd}$ is imaginary, with $\veps_{0123} = -\ri$.
Our two-component conventions follow mainly \cite{Wess:1992cp}
with the following modification: the spinor matrices are given by
$\sigma^a = (-\mathbf 1, -\tau^i)$ with $\tau^i$ the Pauli matrices,
so that the matrix in the Dirac conjugate can be written $\ri \gamma^0$
as in \cite{deWit:2010za}.
A generic four-component Dirac fermion $\Psi$ decomposes
into spinors $\psi_\alpha$ and $\bar\chi^\dalpha$, which
are respectively left-handed and right-handed two-component 
spinors. The Dirac conjugate $\bar\Psi = \ri \Psi^\dag \gamma^0$
has components $\chi^\alpha = (\bar\chi^\dalpha)^*$
and $\bar\psi_\dalpha = (\psi_\alpha)^*$.
Spinor indices can be raised and lowered using the antisymmetric tensor
$\eps_{\alpha \beta}$:
\begin{gather}\label{eq:epsLorentz}
\psi^\beta = \eps^{\beta \alpha} \psi_\alpha~, \qquad \psi_\alpha = \eps_{\alpha \beta} \psi^\beta~,\qquad
\eps_{\alpha \beta} \eps^{\beta \gamma} = \delta_\alpha^\gamma~, \qquad
\eps^{12} = \eps_{21} = 1~.
\end{gather}
Similar equations pertain for $\eps_{\dalpha \dbeta}$ and dotted spinors.
We define
\begin{align}
(\bsigma^a)^{\dalpha \alpha} := \eps^{\dalpha \dbeta} \eps^{\alpha \beta} (\sigma^a)_{\beta \dbeta}~, \qquad
\bsigma^a = (\sigma^0, -\sigma^i)
\end{align}
so that
\begin{align}
(\sigma^a \bsigma^b + \sigma^b \bsigma^a)_\alpha{}^\beta = -2 \eta^{ab} \delta_\alpha^\beta~, \qquad
(\bsigma^a \sigma^b + \bsigma^b \sigma^a)^\dalpha{}_\dbeta = -2 \eta^{ab} \delta^\dalpha_\dbeta~.
\end{align}
The four-component $\gamma$ matrices, which differ from those of \cite{Wess:1992cp},
are built out of the $\sigma$ matrices and obey
\begin{gather}
\gamma^a =
\left(\begin{array}{cc}
0 & \ri \,(\sigma^a)_{\alpha \dbeta} \\
\ri \,(\bsigma^a)^{\dalpha \beta} & 0
\end{array}\right)~, \qquad (\gamma^a)^\dag = \gamma_a~, \qquad \{\gamma^a, \gamma^b\} = 2 \eta^{ab}~, \nonumber \\
\gamma_5 = -\ri \gamma^0 \gamma^1 \gamma^2 \gamma^3 = 
\begin{pmatrix}
\delta_\alpha{}^\beta & 0 \\
0 & -\delta^\dalpha{}_\dbeta
\end{pmatrix}~.
\end{gather}
We define antisymmetric combinations of $\gamma$ and $\sigma$ matrices as
\begin{align}
(\sigma^{ab})_\alpha{}^\beta &:= \tfrac{1}{4} (\sigma^a \bsigma^b - \sigma^b \bsigma^a)_\alpha{}^\beta ~, \qquad
(\bsigma^{ab})^\dalpha{}_\dbeta := \tfrac{1}{4} (\bsigma^a \sigma^b - \bsigma^b \sigma^a)^\dalpha{}_\dbeta~, \nonumber \\
\gamma^{ab} &:= \tfrac{1}{2} [\gamma^a, \gamma^b] =
\begin{pmatrix}
-2 (\sigma^{ab})_\alpha{}^\beta & 0 \\
0 & -2 (\bsigma^{ab})^\dalpha{}_\dbeta
\end{pmatrix}~.
\end{align}
One can check that $(\sigma^{ab})_{\alpha\beta} = \eps_{\beta \gamma} (\sigma^{ab})_\alpha{}^\gamma$
is \emph{symmetric} in its spinor indices
and similarly for
$(\bar\sigma^{ab})_{\dalpha\dbeta} = \eps_{\dalpha \dgamma} (\bar\sigma^{ab})^\dgamma{}_\dbeta$.
These obey the duality properties
\begin{align}
\tfrac{1}{2} \veps_{abcd} \sigma^{cd} = -\sigma_{ab}~, \qquad
\tfrac{1}{2} \veps_{abcd} \bsigma^{cd} = +\bsigma_{ab}~, \qquad
\tfrac{1}{2} \veps_{abcd} \gamma^{cd} = -\gamma_5 \gamma_{ab}~.
\end{align}

The main difference between four-component and two-component notation (aside from
the use of $\gamma$- versus $\sigma$-matrices) is
that the latter usually yields more direct information about the Lorentz group
representation of the field in question. For example, in four-component calculations,
one must remember the chirality of all spinor quantities. This is accomplished
in $N=2$ multiplet calculus by using the location of the $\rm SU(2)$ index
to distinguish between the left-handed and right-handed fields;
for example, $\psi_\mu{}^i$ and $\psi_{\mu i}$ are always, respectively,
the left-handed and right-handed gravitinos
while $\phi_\mu{}^i$ and $\phi_{\mu i}$ are always, respectively, the right-handed
and left-handed S-supersymmetry connections.
In two-component notation, the first pair are written as
$\psi_\mu{}_\alpha{}^i$ and $\bar\psi_\mu{}^\dalpha{}_i$ and
the second pair by
$\bar\phi_\mu{}^{\dalpha i}$ and $\phi_\mu{}_{\alpha i}$ with the explicit
spinor index denoting the chirality, so one can in principle raise or lower
the SU(2) index using the antisymmetric tensor $\veps_{ij}$.
However, we will avoid doing this to maintain maximum compatibility with
four-component notation.

Similarly, vectors and tensors can be written with spinor indices to explicitly
indicate their properties under the Lorentz group.
A vector $V^a$ is associated with a field $V_{\alpha \dalpha}$ with one
dotted and one undotted index via
\begin{align}
V_{\alpha \dalpha} = (\sigma^a)_{\alpha \dalpha} V_a~, \qquad
V_a = -2 (\bsigma_a)^{\dalpha \alpha} V_{\alpha \dalpha}~.
\end{align}
Similarly, an antisymmetric two-form $F_{ab}$ is associated with
symmetric bi-spinors $F_{\alpha \beta}$ and $F_{\dalpha \dbeta}$
corresponding to its anti-selfdual and selfdual parts,
\begin{gather}
F_{ab}^- = (\sigma_{ab})_\alpha{}^\beta F_\beta{}^\alpha~, \qquad
F_{ab}^+ = (\bar\sigma_{ab})^\dalpha{}_\dbeta F^\dbeta{}_\dalpha~, \nonumber \\
F_{ab}^\pm = \tfrac{1}{2} (F_{ab} \pm \tilde F_{ab})~, \qquad
\tilde F_{ab} = \tfrac{1}{2} \veps_{abcd} F^{cd}~, \qquad
\tilde F_{ab}^\pm = \pm F_{ab}^\pm~. \label{eq:Fselfdual}
\end{gather}
If $F_{ab}$ is real, then $(F_{\alpha\beta})^* = -F_{\dalpha \dbeta}$.
We always apply symmetrization and antisymmetrization with unit strength,
so that $F_{[ab]} = F_{ab}$ and $F_{(\alpha \beta)} = F_{\alpha \beta}$.

Finally, we remind the reader that $\rm SU(2)$ indices are swapped by
complex conjugation, $(T_{abij})^* = T_{ab}{}^{ij}$, and we make use
of the invariant $\rm SU(2)$ tensor $\veps^{ij}$ and $\veps_{ij}$
defined as $\veps^{\1\2} = \veps_{\1\2} = 1$ with $\veps^{ij}
\veps_{kj} = \delta^i_k$.  As already stated, unlike in the superspace
approaches \cite{Butter:2011sr, KLRT}, we do \emph{not} raise or lower
SU(2) indices with the $\veps_{ij}$ tensor.

\section{Multiplet calculus formulation of $N=2$ conformal
  supergravity}
\setcounter{equation}{0}
\label{App:SC}
In this appendix, we present the transformation rules for the $N=2$
conformal supergravity (or Weyl) multiplet and their relation to the
superconformal algebra. Recall that the superconformal algebra
comprises the generators of the general-coordinate, local Lorentz,
dilatation, special conformal, chiral $\mathrm{SU}(2)$ and
$\mathrm{U}(1)$, supersymmetry (Q) and special supersymmetry (S)
transformations.  The gauge fields associated with general-coordinate
transformations ($e_\mu{}^a$), dilatations ($b_\mu$), R-symmetry
($\mathcal{V}_\mu{}^i{}_j$ and $A_\mu$) and Q-supersymmetry
($\psi_\mu{}^i$) are independent fields.  The remaining gauge fields
associated with the Lorentz ($\omega_\mu{}^{ab}$), special conformal
($f_\mu{}^a$) and S-supersymmetry transformations ($\phi_\mu{}^i$) are
composite objects \cite{deWit:1980tn,deWit:1984pk,deWit:1984px}.  The
multiplet also contains three other fields: a Majorana spinor doublet
$\chi^i$, a scalar $D$, and a selfdual Lorentz tensor $T_{abij}$,
which is anti-symmetric in $[ab]$ and $[ij]$. The Weyl and chiral
weights have been collected in table \ref{table:weyl}.
%
\begin{table}[t]
\begin{tabular*}{\textwidth}{@{\extracolsep{\fill}}
    |c||cccccccc|ccc||ccc| }
\hline
 & &\multicolumn{9}{c}{Weyl multiplet} & &
 \multicolumn{2}{c}{parameters} & \\[1mm]  \hline \hline
 field & $e_\mu{}^{a}$ & $\psi_\mu{}^i$ & $b_\mu$ & $A_\mu$ &
 $\mathcal{V}_\mu{}^i{}_j$ & $T_{ab}{}^{ij} $ &
 $ \chi^i $ & $D$ & $\omega_\mu^{ab}$ & $f_\mu{}^a$ & $\phi_\mu{}^i$ &
 $\epsilon^i$ & $\eta^i$
 & \\[.5mm] \hline
$w$  & $-1$ & $-\tfrac12 $ & 0 &  0 & 0 & 1 & $\tfrac{3}{2}$ & 2 & 0 &
1 & $\tfrac12 $ & $ -\tfrac12 $  & $ \tfrac12  $ & \\[.5mm] \hline
$c$  & $0$ & $-\tfrac12 $ & 0 &  0 & 0 & $-1$ & $-\tfrac{1}{2}$ & 0 &
0 & 0 & $-\tfrac12 $ & $ -\tfrac12 $  & $ -\tfrac12  $ & \\[.5mm] \hline
 $\gamma_5$   &  & + &   &    &   &   & + &  &  &  & $-$ & $ + $  & $
 -  $ & \\ \hline
\end{tabular*}
\vskip 2mm
\renewcommand{\baselinestretch}{1}
\parbox[c]{\textwidth}{\caption{\label{table:weyl}{\footnotesize
Weyl and chiral weights ($w$ and $c$) and fermion
chirality $(\gamma_5)$ of the Weyl multiplet component fields and the
supersymmetry transformation parameters.}}}
\end{table}

Under Q-supersymmetry, S-supersymmetry and special conformal
transformations the Weyl multiplet fields transform
as
\begin{eqnarray}
  \label{eq:weyl-multiplet}
  \delta e_\mu{}^a & =& \bar{\epsilon}^i \, \gamma^a \psi_{ \mu i} +
  \bar{\epsilon}_i \, \gamma^a \psi_{ \mu}{}^i \, , \nonumber\\
  \delta \psi_{\mu}{}^{i} & =& 2 \,\mathcal{D}_\mu \epsilon^i - \tfrac{1}{8}
  T_{ab}{}^{ij} \gamma^{ab}\gamma_\mu \epsilon_j - \gamma_\mu \eta^i
  \, \nonumber \\
  \delta b_\mu & =& \tfrac{1}{2} \bar{\epsilon}^i \phi_{\mu i} -
  \tfrac{3}{4} \bar{\epsilon}^i \gamma_\mu \chi_i - \tfrac{1}{2}
  \bar{\eta}^i \psi_{\mu i} + \mbox{h.c.} + \Lambda^a_{\rm K} e_{\mu a} \, ,
  \nonumber \\
  \delta A_{\mu} & =& \tfrac{1}{2} \mathrm{i} \bar{\epsilon}^i \phi_{\mu i} +
  \tfrac{3}{4} \mathrm{i} \bar{\epsilon}^i \gamma_\mu \, \chi_i +
  \tfrac{1}{2} \mathrm{i}
  \bar{\eta}^i \psi_{\mu i} + \mbox{h.c.} \, , \nonumber\\
  \delta \mathcal{V}_\mu{}^{i}{}_j &=& 2\, \bar{\epsilon}_j
  \phi_\mu{}^i - 3
  \bar{\epsilon}_j \gamma_\mu \, \chi^i + 2 \bar{\eta}_j \, \psi_{\mu}{}^i
  - (\mbox{h.c. ; traceless}) \, , \nonumber \\
  \delta T_{ab}{}^{ij} &=& 8 \,\bar{\epsilon}^{[i} R(Q)_{ab}{}^{j]} \,
  , \nonumber \\
  \delta \chi^i & =& - \tfrac{1}{12} \gamma^{ab} \, \Slash{D} T_{ab}{}^{ij}
  \, \epsilon_j + \tfrac{1}{6} R(\mathcal{V})_{\mu\nu}{}^i{}_j
  \gamma^{\mu\nu} \epsilon^j -
  \tfrac{1}{3} \mathrm{i} R_{\mu\nu}(A) \gamma^{\mu\nu} \epsilon^i + D
  \epsilon^i +
  \tfrac{1}{12} \gamma_{ab} T^{ab ij} \eta_j \, , \nonumber \\
  \delta D & =& \bar{\epsilon}^i \,  \Slash{D} \chi_i +
  \bar{\epsilon}_i \,\Slash{D}\chi^i \, .
\end{eqnarray}
Here $\epsilon^i$ and $\epsilon_i$ denote the spinorial parameters of
Q-supersymmetry, $\eta^i$ and $\eta_i$ those of S-supersymmetry, and
$\Lambda_{\rm K}{}^a$ is the transformation parameter for special conformal
boosts.  The full superconformally covariant derivative is denoted by
$D_\mu$, while $\mathcal{D}_\mu$ denotes a covariant derivative with
respect to Lorentz, dilatation, chiral $\mathrm{U}(1)$, and
$\mathrm{SU}(2)$ transformations,
\begin{equation}
  \label{eq:D-epslon}
  \mathcal{D}_{\mu} \epsilon^i = \big(\partial_\mu - \tfrac{1}{4}
    \omega_\mu{}^{cd} \, \gamma_{cd} + \tfrac1{2} \, b_\mu +
    \tfrac{1}{2}\mathrm{i} \, A_\mu  \big) \epsilon^i + \tfrac1{2} \,
  \mathcal{V}_{\mu}{}^i{}_j \, \epsilon^j  \,.
\end{equation}

The covariant curvatures are given by
\begin{align}
  \label{eq:curvatures}
  R(P)_{\mu \nu}{}^a  = & \, 2 \, \partial_{[\mu} \, e_{\nu]}{}^a + 2 \,
  b_{[\mu} \, e_{\nu]}{}^a -2 \, \omega_{[\mu}{}^{ab} \, e_{\nu]b} -
  \tfrac1{2} ( \bar\psi_{[\mu}{}^i \gamma^a \psi_{\nu]i} +
  \mbox{h.c.} ) \, , \nonumber\\[.2ex]
  R(Q)_{\mu \nu}{}^i = & \, 2 \, \mathcal{D}_{[\mu} \psi_{\nu]}{}^i -
  \gamma_{[\mu}   \phi_{\nu]}{}^i - \tfrac{1}{8} \, T^{abij} \,
  \gamma_{ab} \, \gamma_{[\mu} \psi_{\nu]j} \, , \nonumber\\[.2ex]
  R(A)_{\mu \nu} = & \, 2 \, \partial_{[\mu} A_{\nu ]} - \mathrm{i}
  \left( \tfrac12
    \bar{\psi}_{[\mu}{}^i \phi_{\nu]i} + \tfrac{3}{4} \bar{\psi}_{[\mu}{}^i
    \gamma_{\nu ]} \chi_i - \mbox{h.c.} \right) \, , \nonumber\\[.2ex]
  R(\mathcal{V})_{\mu \nu}{}^i{}_j =& \, 2\, \partial_{[\mu}
  \mathcal{V}_{\nu]}{}^i{}_j +
  \mathcal{V}_{[\mu}{}^i{}_k \, \mathcal{V}_{\nu]}{}^k{}_j  +  2 (
    \bar{\psi}_{[\mu}{}^i \, \phi_{\nu]j} - \bar{\psi}_{[\mu j} \,
    \phi_{\nu]}{}^i )
  -3 ( \bar{\psi}_{[\mu}{}^i \gamma_{\nu]} \chi_j -
    \bar{\psi}_{[\mu j} \gamma_{\nu]} \chi^i ) \nonumber\\
& \, - \delta_j{}^i ( \bar{\psi}_{[\mu}{}^k \, \phi_{\nu]k} -
  \bar{\psi}_{[\mu k} \, \phi_{\nu]}{}^k )
  + \tfrac{3}{2}\delta_j{}^i (\bar{\psi}_{[\mu}{}^k \gamma_{\nu]}
  \chi_k - \bar{\psi}_{[\mu k} \gamma_{\nu]} \chi^k)  \, , \nonumber\\[.2ex]
  R(M)_{\mu \nu}{}^{ab} = & \,
  \, 2 \,\partial_{[\mu} \omega_{\nu]}{}^{ab} - 2\, \omega_{[\mu}{}^{ac}
  \omega_{\nu]c}{}^b
  - 4 f_{[\mu}{}^{[a} e_{\nu]}{}^{b]}
  + \tfrac12 (\bar{\psi}_{[\mu}{}^i \, \gamma^{ab} \,
  \phi_{\nu]i} + \mbox{h.c.} ) \nonumber\\
& \, + ( \tfrac14 \bar{\psi}_{\mu}{}^i   \,
  \psi_{\nu}{}^j  \, T^{ab}{}_{ij}
  - \tfrac{3}{4} \bar{\psi}_{[\mu}{}^i \, \gamma_{\nu]} \, \gamma^{ab}
  \chi_i
  - \bar{\psi}_{[\mu}{}^i \, \gamma_{\nu]} \,R(Q)^{ab}{}_i
  + \mbox{h.c.} ) \, , \nonumber\\[.2ex]
  R(D)_{\mu \nu} = & \,2\,\partial_{[\mu} b_{\nu]} - 2 f_{[\mu}{}^a
  e_{\nu]a}
  - \tfrac{1}{2} \bar{\psi}_{[\mu}{}^i \phi_{\nu]i} + \tfrac{3}{4}
    \bar{\psi}_{[\mu}{}^i \gamma_{\nu]} \chi_i
    - \tfrac{1}{2} \bar{\psi}_{[\mu i} \phi_{\nu]}{}^i + \tfrac{3}{4}
  \bar{\psi}_{[\mu i} \gamma_{\nu]} \chi^i \,,  \nonumber\\[.2ex]
  R(S)_{\mu\nu}{}^i  = \,&  2\,{\cal D}_{[\mu}\phi_{\nu]}{}^i
  -2 f_{[\mu}{}^a\gamma_a\psi_{\nu]}{}^i
  -\ft18 \Slash{D} T_{ab}{}^{ij}\gamma^{ab}\gamma_{[\mu} \psi_{\nu]\, j}
     -\tfrac32 \gamma_a\psi_{[\mu}{}^i\,\bar\psi_{\nu]}{}^j\gamma^a{\chi}_j
     \nonumber\\
     \,& +\ft14 R({\cal V})_{ab}{}^i{}_j\gamma^{ab}
     \gamma_{[\mu}\psi_{\nu]}{}^j
     +\ft12 \mathrm{i}
     R(A)_{ab}\gamma^{ab}\gamma_{[\mu}\psi_{\nu]}{}^i
     \,,\nonumber\\[.2ex]
     R(K)_{\mu\nu}{}^a = \,& 2 \,{\cal D}_{[\mu} f_{\nu]}{}^a
     -\ft14\big(\bar{\phi}_{[\mu}{}^i\gamma^a\phi_{\nu]i}
     +\bar{\phi}_{[\mu i} \gamma^a\phi_{\nu]}{}^i\big)  \nonumber\\
     &\,
     +\tfrac14\big(\bar{\psi}_{\mu }{}^iD_b T^{ba}{}_{ij}\psi_{\nu}{}^j
     -3\, e_{[\mu}{}^a\psi_{\nu]}{}^i\Slash{D}\chi_i +\ft32
     D\,\bar{\psi}_{[\mu}{}^i\gamma^a\psi_{\nu]j}
     -4\,\bar{\psi}_{[\mu}{}^i\gamma_{\nu]}D_b R(Q)^{ba}{}_i
     +\mbox{h.c.}\big)\,. \nonumber\\
     &{~}
\end{align}
The connections $\omega_{\mu}{}^{ab}$, $\phi_\mu{}^i$ and $f_{\mu}{}^a$
are algebraically determined by imposing the conventional constraints
\begin{gather}
  R(P)_{\mu \nu}{}^a =  0 ~, \quad
  \gamma^\mu R(Q)_{\mu \nu}{}^i + \tfrac32 \gamma_{\nu}
  \chi^i = 0 ~, \nonumber \\
  e^{\nu}{}_b \,R(M)_{\mu \nu a}{}^b - \mathrm{i} \tilde{R}(A)_{\mu a} +
  \tfrac1{8} T_{abij} T_\mu{}^{bij} -\tfrac{3}{2} D \,e_{\mu a} = 0
  \,.  \label{eq:conv-constraints}
\end{gather}
Their solution is given by
\begin{align}
  \label{eq:dependent}
  \omega_\mu{}^{ab} =&\, -2e^{\nu[a}\partial_{[\mu}e_{\nu]}{}^{b]}
     -e^{\nu[a}e^{b]\sigma}e_{\mu c}\partial_\sigma e_\nu{}^c
     -2e_\mu{}^{[a}e^{b]\nu}b_\nu   \nonumber\\
      &\, -\ft{1}{4}(2\bar{\psi}_\mu^i\gamma^{[a}\psi_i^{b]}
     +\bar{\psi}^{ai}\gamma_\mu\psi^b_i+{\rm h.c.}) \,,\nonumber\\
     \phi_\mu{}^i  =& \, \tfrac12 \left( \gamma^{\rho \sigma} \gamma_\mu -
    \tfrac{1}{3} \gamma_\mu \gamma^{\rho \sigma} \right) \left(
    \mathcal{D}_\rho
    \psi_\sigma{}^i - \tfrac{1}{16} T^{abij} \gamma_{ab} \gamma_\rho
    \psi_{\sigma j} + \tfrac{1}{4} \gamma_{\rho \sigma} \chi^i \right)
    \,,  \nonumber\\
    f_\mu{}^{\mu}  =& \, \tfrac{1}{6} R(\omega,e) - D - \left(
      \tfrac1{12} e^{-1}
    \varepsilon^{\mu \nu \rho \sigma} \bar{\psi}_\mu{}^i \, \gamma_\nu
    \mathcal{D}_\rho \psi_{\sigma i} - \tfrac1{12} \bar{\psi}_\mu{}^i
    \psi_\nu{}^j T^{\mu \nu}{}_{ij} - \tfrac1{4} \bar{\psi}_\mu{}^i
    \gamma^\mu \chi_i +
    \mbox{h.c.} \right) \, .
\end{align}
We will also need the bosonic part of the expression for the
uncontracted connection $f_\mu{}^a$,
\begin{equation}
  \label{eq:f-bos-uncon}
  f_\mu{}^a= \tfrac12 R(\omega,e)_\mu{}^a -\tfrac14 \big(D+\tfrac13
  R(\omega,e)\big) e_\mu{}^a -\tfrac12\mathrm{i}\tilde R(A)_\mu{}^a +
  \tfrac1{16} T_{\mu b} {}^{ij} T^{ab}{}_{ij} \,,
\end{equation}
where $R(\omega,e)_\mu{}^a= R(\omega)_{\mu\nu}{}^{ab} e_b{}^\nu$ is
the non-symmetric Ricci tensor, and $R(\omega,e)$ the corresponding
Ricci scalar. The curvature $R(\omega)_{\mu\nu}{}^{ab}$ is associated
with the spin connection field $\omega_\mu{}^{ab}$.

\section{Superspace formulation of $N=2$ conformal supergravity}
\setcounter{equation}{0}
\label{App:SuperCSG}
We summarize in this appendix the structure of conformal superspace,
whose component reduction reproduces the superconformal multiplet
calculus. Relative to \cite{Butter:2011sr}, we have made several
changes of normalization of various operators, connections and
curvatures so that the matching with tensor calculus is as transparent
as possible.  With the explicit results given here, one can (with some
effort) reproduce the component results of section
\ref{sec:component-kinetic-multiplet}.

Recall that $\cN=2$ superspace is a supermanifold parametrized by
local coordinates $z^M = (x^\mu, \theta^{\mathfrak{m} \imath},
\bar\theta_{\dot{\mathfrak{m}} \imath})$.  Together with
superdiffeomorphisms, we equip the superspace with additional symmetry
generators -- the Lorentz transformations ($M_{ab}$), Weyl dilatations
($\mathbb D$), chiral $\rm U(1)$ rotations ($\mathbb A$), $\rm SU(2)$
transformations ($I^i{}_j$), special conformal transformations
($K_a$), and S-supersymmetry ($S_\alpha{}^i$ and $\bar
S^\dalpha{}_i$).  One introduces connection one-forms associated with
each of these generators, including a vielbein $E_M{}^A$ associated
with covariant diffeomorphisms, and defines the covariant derivative
$\nabla_A$ as in \eqref{eq:defCD}.  It transforms under Lorentz,
dilatation and $\rm SU(2) \times U(1)$ transformations as
\begin{gather} [M_{ab}, \nabla_c] = -\eta_{bc} \nabla_a + \eta_{ac}
  \nabla_b ~,\quad [M_{ab}, \nabla_{\gamma i}] =
  -{(\sigma_{ab})_\gamma}^{\beta} \nabla_{\beta i}~, \quad
  [M_{ab}, \bar \nabla^{\dgamma i}] =
  -{(\bsigma_{ab})^\dgamma}_{\dbeta} \bar \nabla^{\dbeta i}~,
  \nonumber \\ 
  [\mathbb D,\nabla_a] = \nabla_a, \quad [\mathbb D, \nabla_{\alpha
    i}] = \tfrac{1}{2} \nabla_{\alpha i}, \quad [\mathbb D, \bar
  \nabla^{\dalpha i}] = \tfrac{1}{2} \bar \nabla^{\dalpha i}~, \eol{}
  [\mathbb A, \nabla_{\alpha i}] = \tfrac{1}{2} \ri \,\nabla_{\alpha
    i},\quad [\mathbb A, \bar \nabla^{\dalpha i}] = -\tfrac{1}{2} \ri
  \,\bar \nabla^{\dalpha i}~, \eol{} [I^j{}_i, \nabla_{\alpha k}] =
  \delta_k^j \nabla_{\alpha i} - \tfrac{1}{2} \delta_i^j
  \nabla_{\alpha k}, \quad [I^j{}_i, \bar \nabla^{\dalpha k}] =
  -\delta_k^i \bar \nabla^{\dalpha j} + \tfrac{1}{2} \delta^i_j \bar
  \nabla^{\dalpha k}~.
\end{gather}
The non-trivial algebraic relations involving $S$, $\bar S$ and $K$ are
\begin{gather}
  \{S_{\alpha}{}^i, \bar S_{\dalpha j} \} = -\ri \,\delta^i_j
  \,(\sigma^a)_{\alpha \dalpha} \,K_a~, \qquad [K_a, \nabla_b] = -
  \eta_{ab} \mathbb D - M_{ab}~, \eol{} \{S_{\alpha}{}^i,
  \nabla_{\beta j}\} = -\delta^i_j \epsilon_{\alpha \beta} \mathbb D +
  2 \delta^i_j M_{\alpha\beta} - \ri \,\delta^i_j \epsilon_{\alpha
    \beta} \mathbb A + 2 \eps_{\alpha \beta} I^i{}_j~, \eol{} \{\bar
  S^{\dalpha}{}_i, \bar \nabla^{\dbeta j} \} = -\delta_i^j
  \epsilon^{\dalpha \dbeta} \mathbb D + 2 \delta_i^j \bar M^{\dalpha
    \dbeta} + \ri \,\delta_i^j \epsilon^{\dalpha \dbeta} \mathbb A - 2
  \eps^{\dalpha \dbeta} I^j{}_i~, \eol{} [K_a, \nabla_{\alpha i}] =
  \ri \,(\sigma_a)_{\alpha \dbeta} \,\bar S^{\dbeta}{}_i, \;\;\; [K_a,
  \bar \nabla^{\dalpha i}] = \ri \,(\bsigma_a)^{\dalpha \beta}
  \,S_{\beta}{}^i ~,\eol{} [S_{\alpha}{}^i, \nabla_a] = -\tfrac{1}{2}
  \ri \,(\sigma_a)_{\alpha \dbeta} \,\bar \nabla^{\dbeta i},\;\;\;
  [\bar S^{\dalpha}{}_i, \nabla_a] = -\tfrac{1}{2} \ri
  \,(\bsigma_a)^{\dalpha \beta} \,\nabla_{\beta i}.
\end{gather}
Above we have used $M_{\alpha\beta}$ and $\bar M^{\dbeta \dalpha}$ as
the anti-selfdual and selfdual parts of $M_{ab}$.

We have not yet specified the (anti-)commutation relations of the
covariant derivatives.  These involve non-vanishing torsion and
curvature tensors, but they are all built out of the covariant Weyl
superfield $W_{\alpha \beta}$, which is a chiral primary superfield
obeying the Bianchi identity $\nabla^{\alpha \beta} W_{\alpha \beta} =
\bar \nabla^{\dalpha \dbeta} \bar W_{\dalpha \dbeta}$.  The algebra of
the spinor derivatives is the simplest:
\begin{gather}
\{\nabla_{\alpha i}, \bar\nabla_{\dbeta}{}^j\} = -2\ri \,\delta_i^j \,
(\sigma^c)_{\alpha \dbeta} \nabla_c  = -2\ri \,\delta_i^j \,
\nabla_{\alpha \dbeta}~, \nonumber \\ 
\{\nabla_{\alpha i}, \nabla_{\beta j}\} = -2 \, \veps_{ij} \eps_{\alpha \beta} \bar \cW~, \qquad
\{\bar\nabla^{\dalpha i}, \bar\nabla^{\dbeta j}\} = -2 \, \veps^{ij} \eps^{\dalpha \dbeta} \cW \nonumber\\
\cW := W^{\alpha \beta} M_{\alpha \beta}
	- \tfrac{1}{2} \nabla^{\beta}{}_j W_\beta{}^\alpha \,S_{\alpha}{}^j
	- \tfrac{1}{2} \nabla^{\dalpha \beta} W_\beta{}^\alpha K_{\alpha \dalpha}~, \qquad
\bar \cW = (\cW)^*~.
\end{gather}
The commutator of the spinor and vector derivatives is
\begin{align}
[\nabla_{\alpha i}, \nabla_{\beta \dbeta}] &= -2 \,\eps_{\alpha\beta} \bar\cW_{\dbeta i}~,\qquad
[\bar\nabla_{\dalpha}{}^i, \nabla_{\beta \dbeta}] = -2 \,\eps_{\dalpha\dbeta } \cW_{\beta}{}^i
\end{align}
where
\begin{align}
\veps_{ij} \cW_{\alpha}{}^j
	&= \tfrac{1}{2} \ri\, W_\alpha{}^\gamma \nabla_{\gamma i}
     - \tfrac{1}{4} \ri\, \nabla^{\phi}{}_i W_{\phi \alpha} \mathbb D
     + \tfrac{1}{4} \nabla^{\phi}{}_i W_{\phi \alpha} \mathbb A
     + \tfrac{1}{2} \ri\, \nabla^{\phi}{}_j W_{\phi \alpha} I^{j}{}_i
     - \tfrac{1}{2} \ri\, \nabla^{\beta}{}_i W^\gamma{}_\alpha M_{\beta \gamma}
     \eol & \quad
     + \tfrac{1}{4} \ri\, \nabla_{\alpha i} \nabla^{\phi}{}_k W_\phi{}^\gamma S_{\gamma}{}^k
     - \tfrac{1}{2} \nabla_\dbeta{}^\phi W_{\phi \alpha} \bar S^{\dbeta}{}_i
     + \tfrac{1}{4} \ri\, \nabla_{\alpha i} \nabla_\dbeta{}^\phi W_{\phi \beta} K^{\dbeta \beta}~,
\end{align}
and $\bar\cW_{\dalpha j} = (\cW_\alpha{}^j)^*$.
Finally, the commutator of the vector derivatives can be written
\begin{align}\label{eq:DefCurv}
[\nabla_a, \nabla_b] &= -T_{ab}{}^c \nabla_c
	- T_{ab}{}^{\gamma j} \nabla_{\gamma j}
	- T_{ab}{}_{\dgamma j} \bar\nabla^{\dgamma j}
	\eol & \quad
	- \tfrac{1}{2} \hat R(M)_{ab}{}^{cd} M_{cd}
     - \tfrac{1}{2} R(\cV)_{ab}{}^i{}_j I^j{}_i
     - R(D)_{ab} \,\mathbb D
     - R(A)_{ab} \,\mathbb A
	\eol & \quad
	- \tfrac{1}{2} R(S)_{ab}{}^{\gamma}{}_j S_{\gamma}{}^j
	- \tfrac{1}{2} R(S)_{ab}{}_{\dgamma}{}^j \bar S^{\dgamma}{}_j
     - \hat R(K)_{ab}{}^c K_c~.
\end{align}
We have placed circumflexes on the Lorentz curvature and K-curvature because their
lowest components will differ from the corresponding curvatures in tensor
calculus in a way we will soon describe.
The anti-selfdual parts of the torsion and curvature tensors are
\begin{align}
T_{{\alpha\beta}}{}^c &= 0~, \qquad
T_{{\alpha\beta}}{}^{\gamma i} =
     - \tfrac{1}{4} \veps^{ij} \nabla^{\gamma}_j W_{\alpha \beta}~, \qquad
T_{{\alpha \beta}}{}_{\dgamma i} = 0~, \nonumber \\
R(\cV)_{{\alpha \beta}}{}^{i}{}_j &= \tfrac{1}{4} \veps^{ik} \nabla_{jk} W_{\alpha \beta}~, \qquad
R(D)_{{\alpha \beta}} = -\ri R(A)_{{\alpha \beta}} =
     \tfrac{1}{16} \nabla_{\alpha}{}^\gamma W_{\gamma \beta}
	+\tfrac{1}{16} \nabla_{\beta}{}^\gamma W_{\gamma \alpha}~, \nonumber \\
\hat R(M)_{{\alpha\beta}}{}^{cd} M_{cd}
     &= \tfrac{1}{4} \nabla^{\gamma \delta} W_{\alpha\beta} M_{\delta \gamma}
     - \tfrac{1}{4} \nabla_{\phi\gamma} W^{\phi \gamma} M_{\alpha \beta}
     -  W_{\alpha \beta} \bar W_{\dgamma \ddelta} \bar M^{\ddelta \dgamma}~, \nonumber \\
R(S)_{{\alpha \beta}}{}^{\gamma}{}_i &=
	\tfrac{1}{24} \veps^{jk} \nabla_{ij} \nabla_{\beta k} W_\alpha{}^\gamma
     + \tfrac{1}{24} \veps^{jk} \nabla_{ij} \nabla_{\alpha k} W_\beta{}^\gamma~, \nonumber \\
R(S)_{{\alpha \beta}}{}_{\dgamma}{}^i &=
     - \tfrac{1}{4} \ri \, \veps^{ij} \nabla_{\beta j} \nabla_\dgamma{}^\gamma W_{\gamma \alpha} 
     - \tfrac{1}{4} \ri\, \veps^{ij} \nabla_{\alpha j} \nabla_\dgamma{}^\gamma W_{\gamma \beta} 
     + \tfrac{1}{2} W_{\alpha \beta} \bar \nabla_{\dphi}{}^i \bar W^\dphi{}_\dgamma~, \nonumber \\
\hat R(K)_{{\alpha \beta}}{}^c &=
     - \tfrac{1}{16} \nabla_{\alpha \beta} \nabla^{\dgamma \delta} W_{\delta}{}^\gamma (\sigma^c)_{\gamma \dgamma}
     + \tfrac{1}{4} W_{\alpha \beta} \nabla_{\gamma \dphi} \bar W^\dphi{}_\dgamma (\bsigma^c)^{\dgamma \gamma}~.
\label{eq:ssTorsion}
\end{align}
The selfdual parts can be found by complex conjugation. These algebraic relations
completely determine the superspace geometry.

The component structure of any superspace theory can be found by identifying
the independent components of the superfields and taking the $\theta=\bar\theta=0$ limit,
which we denote by $\phantom{}\vert_{\theta=0}$. For the connections, we identify
\begin{gather}
e_\mu{}^a \equiv E_\mu{}^a\vert_{\theta=0}, \quad
\psi_\mu{}^{\alpha i} \equiv 2 E_\mu{}^{\alpha i} \vert_{\theta=0}, \quad
\bpsi_\mu{}_{\dalpha i} \equiv 2 E_\mu{}_{\dalpha i} \vert_{\theta=0} \eol
A_\mu \equiv A_\mu\vert_{\theta=0}, \quad
b_\mu \equiv B_\mu\vert_{\theta=0}, \quad
\omega_\mu{}^{ab} \equiv \Omega_\mu{}^{ab} \vert_{\theta=0}, \quad
\cV_\mu{}^i{}_j \equiv \cV_\mu{}^i{}_j\vert_{\theta=0} \eol
\hat f_\mu{}^a \equiv F_\mu{}^a\vert_{\theta=0}, \quad
\phi_\mu{}^{\alpha}{}_i \equiv \Phi_\mu{}^{\alpha}{}_i \vert_{\theta=0}, \quad
\bphi_\mu{}_{\dalpha}{}^i \equiv \bar\Phi_\mu{}_{\dalpha}{}^i \vert_{\theta=0}~. \label{eq:componentForms}
\end{gather}
The covariant components of the Weyl multiplet are found within the superfield
$W_{\alpha \beta}$. The tensor $T_{ab}{}^{ij}$, spinor $\chi_\alpha{}^i$ and
scalar $D$ are given by
\begin{align}
T_{ab}{}^{ij} &:= 2 \veps^{ij} (\sigma_{ab})_\beta{}^\alpha W_\alpha{}^\beta \vert_{\theta=0}~, \qquad
\chi_{\alpha}{}^i := -\tfrac{1}{3} \nabla^{\beta i} W_{\beta \alpha} \vert_{\theta=0}, \qquad
D := \tfrac{1}{12} \nabla^{\alpha \beta} W_{\beta \alpha} \vert_{\theta=0}~.
\end{align}
One can define the component covariant derivative by $\hat D_a = \nabla_a \vert_{\theta=0}$,
leading to
\begin{align}
e_\mu{}^a \hat D_a &=
	\pa_\mu
	- \tfrac{1}{2} \psi_\mu{}^{\alpha i} Q_{\alpha i}
	- \tfrac{1}{2} \bar\psi_\mu{}_{\dalpha i} \bar Q^{\dalpha i}
	- \tfrac{1}{2} \omega_\mu{}^{ab} M_{ab}
	- b_\mu \mathbb D - A_\mu \mathbb A - \tfrac{1}{2} \cV_\mu{}^i{}_j I^j{}_i
	\eol & \qquad \qquad
	- \tfrac{1}{2} \phi_\mu{}^{\alpha}{}_i \,S_{\alpha}{}^i
	- \tfrac{1}{2} \bar\phi_\mu{}_{\dalpha}{}^i \,\bar S^{\alpha}{}_i
	- \hat f_\mu{}^a K_a~,
\end{align}
where we identify $Q_{\alpha i} := \nabla_{\alpha i} \vert_{\theta=0}$ as the supersymmetry
transformation on a component field.

The covariant derivative $D_a$ used in multiplet calculus differs
slightly from $\hat D_a$. They are related by a redefinition of the
K-connection,\footnote{The difference in the $K$-connection
  corresponds to a slight modification of the third conventional
  constraint \eqref{eq:conv-constraints}.}
\begin{align}
D_a = \hat D_a + \tfrac{3}{4} D \,K_a~, \qquad f_\mu{}^a = \hat
f_\mu{}^a - \tfrac{3}{4} e_\mu{}^a~. 
\end{align}
The component curvatures are
\begin{align}\label{eq:DDcurv}
[D_a, D_b] &=
	- \tfrac{1}{2} R(Q)_{ab}{}^{\gamma j} Q_{\gamma j}
	- \tfrac{1}{2} R(Q)_{ab}{}_{\dgamma j} \bar Q^{\dgamma j}
	\eol & \quad
	- \tfrac{1}{2} R(M)_{ab}{}^{cd} M_{cd}
	- \tfrac{1}{2} R(\cV)_{ab}{}^i{}_j \,I^i{}_j
	- R(D)_{ab} \,\mathbb D
	- R(A)_{ab} \,\mathbb A
	\eol & \quad
	- \tfrac{1}{2} R(S)_{ab}{}^\gamma{}_j \,S_\gamma{}^j
	- \tfrac{1}{2} R(S)_{ab}{}_\dgamma{}^j \,\bar S^\dgamma{}_j
	- R(K)_{ab}{}^c  K_c~.
\end{align}
These are related to the superspace curvatures by
\begin{gather}
R(Q)_{ab}{}^{\alpha i} = 2 \,T_{ab}{}^{\alpha i} \vert_{\theta=0}~,\qquad
R(M)_{ab}{}^{cd} = \hat R(M)_{ab}{}^{cd}\vert_{\theta=0} + 3 D \,\delta_a^{[c} \delta_b^{d]}~, \nonumber \\
R(K)_{ab}{}^c = \hat R(K)_{ab}{}^c\vert_{\theta=0} - \tfrac{3}{2} D_{[a} D \, \,\delta_{b]}{}^c~,
\label{eq:ssCurvToTC}
\end{gather}
while the other curvatures in \eqref{eq:DDcurv} are the $\theta=0$
projections of the corresponding superspace curvatures.

Component gauge transformations can be derived directly from how their
corresponding superfields transform. One may explicitly rederive \eqref{eq:weyl-multiplet}, for example,
by taking the transformations with Q-supersymmetry parameters $\xi^{\alpha i} = \eps^{\alpha i}$ and
$\bar \xi_{\dalpha i} = \bar \eps_{\dalpha i}$, S-supersymmetry parameters
$\eta_{\alpha}{}^i$ and $\bar\eta^\dalpha{}_i$,
and special conformal parameter $\Lambda_{\rm K}^a$. For example, if $\Psi$ is some
covariant superfield (e.g. $\Phi$, $\nabla_{\alpha i} \Phi$, etc.)
\begin{align}\label{eq:deltaPsiTC}
\delta \Psi\vert_{\theta=0} = \Big(\eps^{\alpha i} \nabla_{\alpha i} \Psi
	+ \bar \eps_{\dalpha i} \bar \nabla^{\dalpha i} \Psi
	+ \eta^\alpha{}_i\, S_\alpha{}^i \Psi
	+ \bar\eta_\dalpha{}^i \,\bar S^\dalpha{}_i \Psi
	+ \Lambda_{\rm K}^a K_a \Psi\Big)\vert_{\theta=0}~.
\end{align}
As a simple example, let us consider $T_{ab}{}^{ij}$:
\begin{align}
\delta T_{ab}{}^{ij}
	&= - 2 \,\veps^{ij} (\sigma_{ab})^{\beta \alpha} \eps^{\gamma k} \nabla_{\gamma k} W_{\alpha \beta} \vert_{\theta=0}
	= 8\, \eps^{\gamma [i} R(Q)_{ab \,\gamma}{}^{j]}~,
\end{align}
using \eqref{eq:ssTorsion} and \eqref{eq:ssCurvToTC}.
The transformation rules for $\chi_\alpha{}^i$ and $D$ can be derived similarly.
For the connections \eqref{eq:componentForms}, the transformation rules follow
from covariant diffeomorphisms and gauge transformations for the superspace
connections.\footnote{See, for example, the recent discussion in \cite{Butter:2013goa}.}

\section{Gauss-Bonnet invariant in $N=1$ conformal supergravity}
\setcounter{equation}{0}
In the main body of the paper, we have constructed the $N=2$ Gauss-Bonnet
using conformal supergravity, corresponding to the approach taken in
the introduction for the non-supersymmetric case. Because the $N=1$ Gauss-Bonnet
is not usually described in this way, it is reasonable to give a
brief discussion showing how the same construction proceeds in
that case. As it is somewhat out of the main line of presentation
of the paper, we have placed the discussion in this brief appendix.

Recall that all invariants in $N=1$ superspace can
be written either as integrals over the full superspace or over chiral superspace,
\begin{align}
\int \rd^4x\, \rd^2\theta\, \rd^2\bar\theta\, E\, \mathscr{L}~, \qquad
\int \rd^4x\, \rd^2\theta\, \mathcal{E}\, \mathscr{L}_{\mathrm {ch}}~.
\end{align}
In the superspace associated with conformal supergravity, the
covariant derivative $\nabla_A$ is constructed with a connection
associated with each generator in the $N=1$ superconformal
algebra. The algebra of covariant derivatives is given in
\cite{Butter:2009cp} and is constrained so that all curvatures depend
only on the weight 3/2 chiral superfield $W_{\alpha\beta\gamma}$,
which contains the $N=1$ Weyl multiplet.\footnote{ 
  The normalization conventions in \cite{Butter:2009cp} were
  originally chosen to coincide with \cite{Wess:1992cp}, but in this
  appendix we follow the normalization conventions of
  \cite{Buchbinder:1998qv}.  This requires that we rescale the
  supersymmetric Weyl tensor as $W_{\alpha \beta \gamma} \rightarrow 2
  W_{\alpha \beta \gamma}$.} 

The single invariant action one can construct in pure $N=1$ conformal
supergravity involves the chiral superspace integral
\begin{align}
  \int \rd^4x\, \rd^2\theta\, \mathcal{E}\, W^{\alpha \beta \gamma}
  W_{\alpha \beta \gamma} = \int \rd^4x\, e\, \Big(\tfrac{1}{4}
  C_{abcd} C^{abcd} - \tfrac{1}{4} C_{abcd} \tilde C^{abcd} +
  \textrm{additional terms}\Big)~.
\end{align}

To construct the additional terms in \eqref{eq:Lchi'a} requires a
compensator field. The simplest possibility is a chiral superfield
$\Phi$ of weight $w$. It is easy to see that in flat superspace
\begin{align}
-\frac{1}{64} \int \rd^2\theta\, \bar D^2 D^2 \bar D^2 \ln \bar\Phi
	= 
-\frac{1}{4} \int \rd^2\theta\, \bar D^2 \Box \ln \bar\Phi
	= \Box \Box \ln\bar A~.
\end{align}
The generalization of this chiral integrand to conformal superspace turns out to be
its naive covariantization:
$\mathbb S(\ln\bar\Phi) = -\frac{1}{64} \bar\nabla^2 \nabla^2 \bar\nabla^2 \ln\bar\Phi$.
One can check that $\mathbb S$ is a covariant conformal primary chiral multiplet of weight 3.
The proposed chiral invariant corresponding to the $N=1$ Gauss-Bonnet is
\begin{align}
\Gamma &:= W^{\alpha \beta \gamma} W_{\alpha \beta \gamma} 
	+ w^{-1} \mathbb S(\ln\bar\Phi)~, \eol
\int \rd^4x\, \rd^2\theta\, \mathcal{E}\, \Gamma
	&= \int \rd^4x\, e\, \Big(\tfrac{1}{4} C^{abcd} C_{abcd}
	- \tfrac{1}{4} C^{abcd} \tilde C_{abcd}
	+ w^{-1} \Box_{\rm c} \Box_{\rm c} \ln \bar A + \cdots\Big)
\label{eq:N1ConfGB}
\end{align}
where $\Box_c := D^a D_a$ for $D_a = \nabla_a\vert_{\q=0}$, the supercovariant derivative
of $N=1$ conformal supergravity, and we have kept only the relevant terms.

The $N=1$ Gauss-Bonnet is usually formulated in Wess-Zumino
superspace.\footnote{The details of Wess-Zumino superspace are covered in
the standard references \cite{GGRS, Wess:1992cp, Buchbinder:1998qv}, and its
auxiliary field structure corresponds to old minimal supergravity.
The Gauss-Bonnet invariant may be equally well constructed in
new minimal \cite{FSV:GBnewmin} or $\rm U(1)$ supergravity \cite{LeDu:GBU1}.}
To compare our expression to the usual one, we must
rewrite the conformally covariant derivatives $\nabla_A$ in terms of
the Wess-Zumino covariant derivatives $\cD_A$.
The result of the degauging process is
\begin{align}\label{eq:T1phib}
\mathbb S(\ln\bar\Phi) &=
	\Delta \ln\bar\Phi + w \,\mathbb S_0~,\nonumber \\
\Delta\ln\bar\Phi &:= 
	-\tfrac{1}{64} (\bar \cD^2 - 4 R) \Big(
	\cD^2 \bar \cD^2 \ln\bar\Phi
	+ 8 \cD^\alpha (G_{\alpha \dalpha} \bar \cD^\dalpha \ln\bar\Phi)
	\Big)~, \nonumber \\
\mathbb S_0 &:= - \tfrac{1}{4} (\bar\cD^2 - 4 R) \Big(
		2 R \bar R + G^a G_a - \tfrac{1}{4} \cD^2 R
	\Big)~,
\end{align}
where here and below we use the usual $N=1$ abbreviation $\cD^2 = \cD^\alpha \cD_\alpha$
and $\bar \cD^2 = \bar\cD_\dalpha \bar\cD^\dalpha$.
In this expression, the additional torsion superfields $R$ and $G_a$ of
Wess-Zumino superspace appear; these contain respectively the Ricci scalar and
the Einstein tensor. The equation \eqref{eq:T1phib} may be compared both to
the analogous $N=0$ result \eqref{eq:BoxBoxphiDG} and to the $N=2$ result
\eqref{eq:NLKinSU2}.
Under a super-Weyl transformation \cite{HT, Siegel:1978fc} involving a chiral parameter $\Sigma$,
the spinor covariant derivative and the curvature superfields transform as
\begin{gather}
\delta_\Sigma \cD_\alpha = (\bar \Sigma - \tfrac{1}{2} \Sigma) \cD_\alpha + \cD^\beta \Sigma\, M_{\beta \alpha}~,\qquad
	\delta_\Sigma W_{\alpha \beta \gamma} = \tfrac{3}{2} \S \,W_{\alpha\beta\gamma}~, \nonumber \\
\delta_\Sigma R = \tfrac{1}{4} \bar \cD^2 \bar\Sigma + (2 \Sigma - \bar \Sigma) R~, \qquad
\delta_\Sigma G_{\alpha \dalpha} = \tfrac{1}{2} (\Sigma + \bar \Sigma) G_{\alpha \dalpha}
	+ \ri\, \cD_{\alpha \dalpha} (\Sigma - \bar\Sigma)~, 
\end{gather}
while $\Phi$ transforms as $\delta \Phi = w \Sigma \, \Phi$. One can check that
\begin{align}
\delta_\Sigma\,\Delta \ln\bar\Phi = 3 \Sigma \, \Delta\ln\bar\Phi + w \Delta \bar\Sigma~, \qquad
\delta_\Sigma \mathbb S_0 = 3 \Sigma \,\mathbb S_0 - \Delta\bar\Sigma~,
\end{align}
which ensures that $\mathbb S(\ln\bar\Phi)$ transforms homogeneously,
$\delta_\Sigma \mathbb S(\ln\bar\Phi) = 3 \Sigma\, \mathbb S(\ln\bar\Phi)$.

In the form \eqref{eq:T1phib}, it is easy to see that the chiral superspace integral of
$\mathbb S(\ln\bar\Phi)$ depends on the superfield $\bar\Phi$
only via a total covariant derivative; discarding any explicit total derivatives, one finds
\begin{align}
\int \rd^4x\, \rd^2\theta\, \mathcal{E} \,\Gamma &=
	\int \rd^4x\, \rd^2\theta\, \mathcal{E} \,W^{\alpha \beta \gamma} W_{\alpha \beta\gamma}
	+ \int \rd^4x\, \rd^2\theta\,\rd^2\bar\theta \,E \,\Big(2 R \bar R + G^a G_a \Big) \nonumber \\
	&= \int \rd^4x\, \Big(\tfrac{1}{4} \cL_\chi - \tfrac{1}{4} \cL_{\rm P} + \textrm{total derivative}\Big)~.
\end{align}
This is the $N=1$ Gauss-Bonnet in old minimal supergravity \cite{Ferrara:1977mv, TovN},
whose real and imaginary parts correspond respectively to the usual Gauss-Bonnet
invariant and the Pontryagin term.
Its topological nature in superspace was first demonstrated in \cite{Buchbinder:1988tj}
(see also \cite{Buchbinder:1998qv}).
The full component expression appeared first in \cite{FV:GBoldmin}.

There is a curious feature of the operator $\Delta$ which deserves comment.
Taking two chiral multiplets $\Phi$ and $\Phi'$, both now of weight zero,
it is natural to define
\begin{align}
\mathbb S(\bar\Phi) := -\tfrac{1}{64} \bar\nabla^2 \nabla^2 \bar\nabla^2 \bar\Phi
	\equiv \Delta \bar\Phi~.
\end{align}
One can check that
\begin{align}\label{eq:N1PR}
\delta_\S \Delta \bar\Phi = 3 \Sigma \, \Delta \bar\Phi
\end{align}
and so $\Delta$ can be viewed as a super-Weyl covariant mapping from a
weight-zero anti-chiral multiplet to a weight-3 chiral multiplet.
This is the $N=1$ generalization of the Fradkin-Tseytlin
operator discussed in section \ref{sec:introduction}.
One finds\footnote{More details regarding the component expression
can be found in \cite{Fradkin:1985am}.}
\begin{align}
\int \rd^4x\, \rd^2\theta\, \mathcal{E}\, \Phi' \mathbb S(\bar\Phi)
	&= \int \rd^4x\, \rd^2\theta\, \mathcal{E}\, \Phi' \Delta \bar\Phi
	= \int \rd^4x\, e\, A' \Box_{\rm c} \Box_{\rm c} \bar A + \cdots \nonumber \\*
	&= \int \rd^4x\, e\, \Big(\cD^a \cD_a A' \,\cD^b \cD_b \bar A
	+ \mathcal{D}^a A' \Big(
		2 \mathcal R_{ab} - \tfrac{2}{3} \eta_{ab} \mathcal{R}
		\Big) \mathcal{D}^b \bar A + \cdots \Big) ~.
\end{align}
One may equally write
\begin{align}\label{eq:superRiegertAction}
\int \rd^4x\, \rd^2\theta\, \mathcal{E}\, \Phi' \Delta \bar\Phi = \frac{1}{16} \int \rd^4x\, \rd^4\theta\, E\,\Big(
	\cD^2 \Phi' \bar \cD^2 \bar\Phi
	- 8 \cD^\alpha \Phi' \,G_{\alpha \dalpha} \bar \cD^\dalpha \bar\Phi\Big)~.
\end{align}
The expression on the left is super-Weyl invariant as a consequence of
\eqref{eq:N1PR}, while this property is obscured for the expression on the right.
However, one can check that it does transform into
\begin{align}
&-\frac{1}{8} \int \rd^4x\, \rd^4\theta\, E\,\Big(
	\cD^\alpha \Sigma \,\cD_\alpha \Phi \bar\cD^2 \bar\Phi
	+ 4 \ri\, \cD_{\alpha \dalpha} \Sigma \,\cD^\alpha \Phi \bar \cD^\dalpha \bar\Phi
	+ \textrm{c.c.}\Big) \nonumber \\
&= -\frac{1}{8} \int \rd^4x\, \rd^4\theta\, E\,\Big(
	\bar\Phi \bar\cD^2 (\cD^\alpha \Sigma \cD_\alpha \Phi)
	+ 4 \ri\, \bar\Phi \bar \cD^\dalpha (\cD_{\alpha \dalpha} \Sigma \,\cD^\alpha \Phi)
	+ \textrm{c.c.}\Big) = 0~.
\end{align}

One might wonder whether a simpler version of 
this operator could be constructed. The obvious
proposal of $\Phi \nabla^2 \bar\nabla^2 \bar\Phi$ integrated over the
full superspace is unfortunately not a conformal primary;
equivalently, there is no conformally covariant
anti-chiral (or chiral) d'Alembertian in Wess-Zumino superspace.
This is in agreement with the discussion in \cite{BK:Nonlocal}
that the analysis of \cite{Riegert:1984kt} is not directly applicable in
superspace. Rather, one requires the (higher dimension) operator $\Delta$.
The $N=1$ supersymmetric generalization of the construction of \cite{Riegert:1984kt}
is given in \cite{Butter:2013ura}.

We should also mention that the $N=2$ version of the Fradkin-Tseytlin operator
can be constructed in the context of the SU(2) superspace discussed
in section \ref{sec:GaussBonnetc}. It is simply the covariant chiral projector $\bar\Delta$,
and the actions analogous to \eqref{eq:superRiegertAction} are
\begin{align}
\int \rd^4x\, \rd^4\theta\, \mathcal{E}\, \Phi' \bar\Delta \bar\Phi
	= \int \rd^4x\, \rd^4\theta\, \rd^4\bar\theta\, E\, \Phi' \bar\Phi~,
\end{align}
where $\Phi$ and $\Phi'$ are weight-zero chiral superfields and both
integrands are manifestly super-Weyl covariant. This is exactly the
action considered in \cite{deWit:2010za}.

\end{appendix}

\providecommand{\href}[2]{#2}

\end{document}